\documentclass[%
 aip,
 amsmath,amssymb,
reprint,%
]{revtex4-1}

\usepackage{graphicx}% Include figure files
\usepackage{dcolumn}% Align table columns on decimal point
\usepackage{bm}% bold math
\usepackage{mathtools}
\usepackage[utf8]{inputenc}
\usepackage[T1]{fontenc}
\usepackage{mathptmx}
\usepackage{etoolbox}
\usepackage{subcaption}
\usepackage{array}
\usepackage{color}
\usepackage[dvipsnames]{xcolor}
\usepackage{siunitx}  %Einheiten
\usepackage{multirow}
\usepackage[export]{adjustbox}

\makeatletter
\def\@email#1#2{%
 \endgroup
 \patchcmd{\titleblock@produce}
  {\frontmatter@RRAPformat}
  {\frontmatter@RRAPformat{\produce@RRAP{*#1\href{mailto:#2}{#2}}}\frontmatter@RRAPformat}
  {}{}
}%
\makeatother
\begin{document}

%\preprint{AIP/123-QED}

\title[Model-based computational study of reactive fluid systems using transfer operator methods]{Model-based computational study of reactive fluid systems using transfer operator methods}

\author{Anja G\"obel}
\author{Anna Kl\"unker}%
\affiliation{Applied Mathematics, Institute for Mathematics and its Didactics, Leuphana University L\"uneburg, 21335 L\"uneburg, Germany%\\This line break forced% with \\
}%

\author{Alexandra von Kameke}
 %\email{alexandra.vonkameke@haw-hamburg.de}
\affiliation{%
Heinrich-Blasius-Institute, Faculty of Engineering and Computer Science, Hamburg University of Applied Sciences, 20099 Hamburg, Germany%\\This line break forced% with \\
}%

\author{Kathrin Padberg-Gehle}%

\affiliation{Applied Mathematics, Institute for Mathematics and its Didactics, Leuphana University L\"uneburg, 21335 L\"uneburg, Germany%\\This line break forced% with \\
}%
%\email{kathrin.padberg-gehle@leuphana.de}

\date{\today}% It is always \today, today,

\begin{abstract}
    In many industrial applications one aims for the design of efficient chemical reactions. This is often done using experimental analysis of the reactors in combination with numerical twins. However, the simultaneous modeling of flow, mixing and chemical reaction remains a complicated task because the computational costs are still significant and oftentimes prohibitive.
    Here, we model and analyze mixing processes of chemical substances by a transfer operator method using the Lagrangian trajectories of passive particles. In practice, we represent the spatial distributions of the chemical species concentrations by density vectors. These are evolved by means of a stochastic transition matrix, which is obtained as a numerical approximation of a transfer operator. In a further step, we model the reaction of the fluids by specific update rules of the density vectors. The method is applicable for closed and open example systems and different chemical processes such as competitive consecutive reactions. We demonstrate that the dynamic processes are reflected well by this method. Furthermore, we evaluate the interdependence of the reaction fronts with Lagrangian coherent structures which we define using set-oriented methods that also work well on three dimensional data.
    
\end{abstract}
\maketitle
\section{Introduction}
In many industrial applications one aims for efficient chemical reactions and their optimization is a central component of research and development work in modern chemistry and process engineering. The aim is to use resources more efficiently, minimize waste and side products, and maximize the yield of the desired product. For process engineers the question arises as to the extent to which the transport and mixing processes in a chemical reactor affect the efficiency of the reaction. Research in this area focuses on theoretical modeling and analysis as well as on optimizing the underlying flow processes. A key objective is to understand and identify different flow regimes, for example in order to specifically improve mixing and reaction yield in industrial devices such as stirred tank reactors (STR) or microfluidic systems \cite{Aref1984,Mathew2002,PhysRevFluids.7.073904, WEILAND2023100448,ROSSEBURG2018208}.

The outcome of a chemical reaction depends on the mixing pattern within the reactor. Van de Vusse \cite{VANDEVUSSE1964994} demonstrated in a theoretical work that, for a complex reaction scheme, the ideal reactor design that maximizes production of the desired product is not obvious. He examined a competitive consecutive reaction and found that a non-mixing plug-flow reactor (PFR) favors the production of the desired product but also produces the waste product, whereas a perfectly mixing continuous stirred tank reactor (CSTR) reduces the production of the waste product but also decreases the production of the desired product. He compared the two ideal reactor types to determine which is preferable under which conditions. Various studies show that sometimes different reactors (semi-batch reactor \cite{BROOKS19901589} and PFR with a recycle stream \cite{GILLESPIE1966472,DEVERA1979163}) or a periodic inflow of the reactant into a CSTR \cite{SKEIRIK19821015,TANDIONNE19882549} have a positive effect on the production of a desired product. The Van de Vusse reaction scheme is just one example of a chemical reaction that is very time-consuming and costly to investigate in practice. New methods and techniques allow for more efficient and cost-effective investigations to increase the efficiency of chemical reactions.

Our goal is to improve our understanding of the efficiency of chemical reactions by analyzing the distribution of the desired product within the reactor using a model-based approach. One significant advantage of model-based techniques is their efficiency in terms of time and materials \cite{Taylor23}. Model-based optimization of chemical reactions uses computer models to identify the optimal combination of experimental conditions to achieve the desired results. These models can range from data-driven approaches such as machine learning to metaheuristic algorithms inspired by natural phenomena \cite{ZHANG202577} to guide experiments and reduce the time and cost of traditional trial-and-error methods.

Here we develop a computational method to approximate the chemical reactions dynamics in a reactor based on the trajectory data of flow following, passive particles deployed in the liquid. This new method allows us to study the yield and selectivity for different pairings of reaction paths and reactor designs at moderate computational costs.
To mimic reactions of chemical fluids in a reactor we use a set-oriented numerical approach and investigate the dynamics under the influence of complex flow structures to understand how flow and diffusion affect production levels. Set-oriented numerical methods provide a computational framework for the global analysis of classical dynamical systems~\cite{DellnitzHohmann1997,Dellnitz2000}. In this context, transfer operator (or Perron-Frobenius operator) approaches \cite{Dellnitz1999,Froyland14} play a major role. A transfer operator is a linear infinite-dimensional operator that describes the evolution of densities under the action of the nonlinear dynamical systems \cite{Lasota}. Finite-rank approximations of such operators have been frequently used for studying transport and mixing in fluid flows \cite{froyland_prl_2007,vansebille2012,klunker2022open}.

In this work, we extend the transfer operator based transport and mixing studies by including reaction processes of chemical substances. The focus lies on coupling passive transport through a time-dependent flow field with local chemical reactions inside a reactor. Our analysis shows the intricate interaction between transport and reactions.
To evaluate the performance of our method, we consider three different underlying flows fields that describe the motion of passive tracers in fluids: the popular two-dimensional Double Gyre (DG) \cite{SHADDEN2005271}, the three-dimensional Arnold–Beltrami–Childress (ABC) flow\cite{Dombre1986}, and a lab-scale STR model with simulated trajectory data from Weiland et al.~\cite{WEILAND2023100448}.

We discretize the domains of the example flows into small boxes and use Ulam's method \cite{Ulam1964} to construct a finite-rank approximation of the Perron-Frobenius operator. The transition probabilities between boxes are calculated from the trajectories of sample points per box, using the software package GAIO (Global Analysis of Invariant Objects) \cite{Dellnitz2000}. The resulting numerical transfer operator describes the evolution of densities and can be interpreted as a transition matrix of a Markov chain with transition probabilities between boxes as entries.

The Markov chain-based approach is then used to model and analyze mixing and simultaneous chemical reactions. We consider closed and open example systems and study different chemical processes such as a competitive consecutive reaction. The open system includes inflow and outflow dynamics, while the closed system conserves total mass. In each time step, densities are first transported and subsequently react. In practice, we represent the spatial distributions of the chemical species concentrations by density vectors, which are evolved by means of the stochastic transition matrix \cite{klunker2022open}. This approach is similar in spirit to the mapping method \cite{mappingmethod2008}, where a distribution matrix maps the concentration distribution from the inlet to the outlet of a micromixer. In a further step, we model the reactions by specific update rules of the density vectors, based on the corresponding reaction kinetics. 

The progress of reactions under the influence of chaotic or turbulent flow and corresponding mass transport has been a matter of study in different disciplines . The ratio of the flow and the reaction time scales, the dimensionless Damk\"ohler number $Da = \tau_{f}/\tau_{r}$, has been found to play a crucial role in the progression of liquid reactions \cite{KoudellaNeufeld2004,OTTINO19944005,SudhaRohde2025,ShahKostiukKresta2012}. Oftentimes, this dependence is only implicitly investigated \cite{doi:10.1021/acs.iecr.7b01324} by the evaluation of a segregation measure that reflects the homogeneity of the reactant distributions and has been found to depend on the micromixing timescale \cite{baldyga1999turbulent}. Whenever the mixing process has one or more typical timescales, the mass transport and the mixing can be analyzed on this specific timescale by the concept of Lagrangian coherent structures (LCS). LCS determine transport and mixing in fluid flows. Attracting and repelling material lines and surfaces form the boundaries of LCS\cite{haller2015,Haller_2023}, similar to unstable and stable manifolds of hyperbolic objects in classical dynamical systems. Over the past almost three decades, various Lagrangian methods have been proposed to identify these structures (see references \onlinecite{Allshouse}, \onlinecite{Hadjighasem2017} and \onlinecite{Badza2023} for an overview and comparison). All LCS definitions, however, require the selection of an integration time-scale for the evaluation of the mixing that can be freely adjusted to derive specific information, e.g. it can be matched to the reaction time. This can be used to understand the relationship between mixing and reaction rates, which is a topic of current research interest \cite{KAMEKE2019317}. In this work, we also want to investigate how the underlying flow structures influence the reaction progress and yield and take advantage of LCS for this purpose.

Pratt et al.\cite{10.1063/1.4914467} demonstrated the importance of LCS, heuristically identified by means of ridges in finite-time Lyapunov (FTLE) fields, in bringing together reactive scalar filaments. They showed that LCS act as underlying patterns that determine where reaction zones form. They also found that coherent structures in fluid flows can substantially increase reaction rates over intermediate time scales, sometimes by several orders of magnitude, exceeding the steady-state rate. More recent experimental results also report the same observation \cite{10.1063/1.5004649}. This effect is caused by the continued stretching imposed by an attracting LCS, which pulls the scalars into thin filaments, resulting in high concentration gradients and segregation measures that increase the reaction rate.
Gowen and Solomon\cite{10.1063/1.4918594} investigated the usefulness of FTLE fields as a tool for analyzing the propagation of reaction fronts in advection-reaction-diffusion systems. Using theories on burning invariant manifolds (BIM)\cite{Mahoney_2012}, they explain reaction front patterns. They calculated the burning FTLE field based on a three-dimensional system of ordinary differential equations for reaction fronts  and found a clear correlation between the structures in the burning FTLE sections and the BIM intersections.

In the context of set-oriented numerical methods, an entropy-based approach has been proposed as an alternative to FTLE to measure a stretching quantity\cite{FROYLAND20121612}: Finite-time entropy (FTE) fields are a simple and flexible construction that can be directly obtained from a numerical transfer operator. We will use this approach in our present work. Similarly to the results of the FTLE-based studies\cite{10.1063/1.4914467,10.1063/1.4918594}, we hypothesize that reactions occur preferentially along attracting coherent structures, corresponding to regions with high FTE values in backward time (b-FTE). We will test this hypothesis by comparing b-FTE fields and reaction fronts in all example systems.

Our work aims to investigate the interaction between mixing and reaction in chemical reactors to understand how mixing influences the desired reaction yield. Our results should be seen as a first step towards aiding to optimize reaction yields through targeted mixing in the future.

The remainder of this work is organized as follows: Section 2 introduces the set-oriented description of transport and reactions, first reviewing the transfer operator based transport of densities and then discussing selected chemical reactions as example systems. Section 3 introduces the three flow fields that will serve as reactor models, namely the DG, ABC, and STR flows. Section 4 presents the numerical studies of the selected chemical reactions for the different reactor models. We conclude the article with an outlook in section 5. The appendix contains numerical results for the DG reactor modeled with an advection-diffusion-reaction equation.
\section{Set-oriented description of transport and reactions}
In our set-oriented approach, we combine fluid mixing and local chemical reactions inside different reactor models to analyze transport and reactions of fluid particles.
First, we will review the set-oriented framework for the description of transport and mixing in fluid flows. In a second step, we will link this with chemical reaction kinetics.

\subsection{Transport of densities}

We assume that fluid particles (tracers) move passively according to
\begin{equation}
  \dot{\bm{x}}(t)=\bm{u}(\bm{x}(t),t),   \label{eq:ode}
\end{equation}
with trajectories $\bm{x}(t)\in \mathbb{R}^d$ ($d=2$ or $d=3$). Let $\Phi:M \rightarrow M$ be the flow map that maps a tracer $\bm{x}_0=\bm{x}(t_0)$ to its new position $\Phi(\bm{x}_0)=\bm{x}(t_0+\tau)$ over the time span $[t_0,t_0+\tau]$. Note that $\bm{x}(t)$ is a solution of the differential equation \eqref{eq:ode} for the initial value $\bm{x}_0=\bm{x}(t_0)$.

In practice, we represent fluids by density vectors, which are evolved by means of a numerical transfer operator and are subsequently updated according to an underlying chemical reaction scheme. We model a chemical reactor as a compact domain $M \subset \mathbb{R}^d$ ($d=2$ or $d=3$) discretized into $n$ disjoint, connected sets (boxes) $B_i$, $i=1,...n$~\cite{Dellnitz1999}. Here $M$ corresponds to the respective domain of the underlying fluid flow. 
Each box $B_i$ contains $N$ uniformly distributed sample points of chemical substances $x_{0,k}^i$, $k=1,...N$. We use Ulam's method~\cite{Ulam1964}
for a finite-rank approximation of an infinite-dimensional Perron-Frobenius operator as a matrix $\bm{P} \in \mathbb{R}^{n,n}$ as follows:
\begin{equation}
    P_{ij}=\frac{\#\{k:\Phi(x_{0,k}^i)\in B_j\}}{N},\label{eq:Pmatrix}
\end{equation}
i.e. the entry $P_{ij}$ is estimated as the proportion of the fluid particles that gets mapped from box $B_i$ to box $B_j$ under the action of $\Phi$. We can interpret the row-stochastic matrix $\bm{P}$ as the transition matrix of a Markov chain on a finite state space, where the boxes $B_i$ represent the states. Without chemical reaction, a fluid density vector $\bm{V}_t\in\mathbb{R}^{n}$ evolves according to
$$\bm{V}_{t+\tau}=\bm{V}_t\bm{P}.$$
Note that in nonautonomous systems $\bm{P}$ depends explicitly on time, i.e.\ $\bm{P}=\bm{P}_t^{t+\tau}$ but we drop the indices in the following for convenience.
 
To simulate a diffusion process into neighboring boxes, we construct a diffusion matrix $\bm{D}$ by choosing $K$ test points $x_{i,k}\in B_i$, $k=1,...K$, in each box $B_i$ and representing the diffusion over an $\epsilon$-ball centered at $x_{i,k}$~\cite{Froyland14}. A ball of radius $\epsilon$ is approximated as a square or cube with a side length of $2\epsilon$. Therefore, we choose another grid of points $y_{i,k,k'}$, $k'=1,...K'$ uniformly spread over $B_{\epsilon}(x_{i,k})$ to estimate the diffusion matrix~\cite{Froyland14}
\begin{equation}
    D_{ij}=\frac{\#\{y_{i,k,k'}\in B_j\}}{K K'}
\end{equation}
and obtain the diffused transition matrix
\begin{equation}
    \bm{P}_D=\bm{P} \bm{D}.
\end{equation}
Finite-time entropy (FTE) has been proposed as a nonlinear stretching measure that is comparable to the FTLE, but that can be directly computed from the numerical transfer operator\cite{FROYLAND20121612}. For this, let $\mu$ be a probability measure assigning mass to the partition elements $B_i$, $i=1,...,n$. The entropy of $\mu$ is defined by
\begin{equation}
    H(\mu)=-\sum_{i=1}^{n}\mu(B_i)\log \mu(B_i),
\end{equation}
which reduces to
\begin{equation}
    H(p)=-\sum_{i=1}^{n}p_i \log p_i
\end{equation}
when writing $p_i=\mu(B_i)$.
The pushforward of $\mu$ under the flow map $\Phi$ is given by $\Phi_*\mu=\mu\circ \Phi^{-1}$. In the discrete setting, $\Phi_*\mu(B_j)$ is approximated by $\sum_{i=1}^{n}p_i {P}_{ij}=(p\bm{P})_j$, where $\bm{P}$ denotes the numerical transition matrix. Throughout, the elements of the partition $\{B_1,...,B_n\}$ have equal volume.

Let $\bm{e}_i$ denote the $n$-vector with a single nonzero entry 1 at position $i$. The FTE associated with partition element $B_i$ is defined as~\cite{FROYLAND20121612}
\begin{equation}
    FTE(B_i,T)=\frac{1}{|T|}H(e_i \bm{P}^{(T)})=-\frac{1}{|T|}\sum_{j=1}^{n}P_{ij}^{(T)}\log P_{ij}^{(T)},
\end{equation}
where $\bm{P}^{(T)}$ is the discretized transfer operator corresponding to $\Phi^T$.
As we are interested in attracting flow structures, we will have to consider FTE fields computed in backward time (b-FTE).
To obtain a suitable backward-time transition matrix without explicitly constructing a discretization of $\Phi^{-1}$, one may exploit the following simple relation: if the initial and final domains coincide, then an appropriate backward-time matrix, $\tilde{\bm{P}}$, is obtained by transposing $\bm{P}$ and normalizing the rows to ensure unit row sums. For autonomous (time-independent) systems with respect to $\tilde{\bm{P}}^{(T)}$, we denote the backward time FTE by $FTE_b(\cdot,T)$.

For nonautonomous (time-dependent) systems, let $\bm{P}^{(T_n)}(t)$ denote the composite operator that captures the flow over the full finite-time horizont $T_n$, where $\bm{P}^{(T_n)}(t)$ is obtained by the composition of $N_{step}$ successive transition matrices, such that $T_n=N_{step}\tau$. For $\bm{P}^{(T_n)}(t)$, we define~\cite{FROYLAND20121612}
\begin{eqnarray}
    FTE(B_i,t,T_n)&=&\frac{1}{|T_n|}H(e_i \bm{P}^{(T_n)}(t)) \\
    &=&-\frac{1}{|T_n|}\sum_{j=1}^{n}P^{(T_n)}(t)_{ij}\log P^{(T_n)}(t)_{ij}.
\end{eqnarray}
With respect to $\tilde{\bm{P}}^{(T_n)}(t)$, we denote the b-FTE by $FTE_b(\cdot,t,T_n)$.

To approximate the transfer operator, we use GAIO~\cite{Dellnitz2000}: By successively bisecting $M \subset \mathbb{R}^d$ into smaller rectangles ($d=2$) or cuboids ($d=3$) and using a binary tree for storage, the partition elements $B_i$ are obtained. In each time step $\tau$, we move the fluids based on the transition matrix $\bm{P}$ (or $\bm{P}_{D}$) and then update the change in concentrations due to the chemical reaction kinetics.

Let $\bm{C}_{S_t} \in \mathbb{R}^{n}$ represent a chemical substance $S$ at time $t$.
In what follows we denote by 
\begin{equation}
  \hat{\bm{C}}_{S_t}:= \bm{C}_{S_t} \bm{P}  \label{eq:evolvedC}
\end{equation}
the time-evolved concentration vector for a chemical substance $S$ on the time span $[t, t+\tau]$. We obtain $\bm{C}_{S_{t+\tau}}$ from $\hat{\bm{C}}_{S_t}$ by a simulation of the chemical reaction as outlined below.

\subsection{Example reaction systems}
Now that we have described fluid transport, we will explain how we calculate the local concentration density in each box $B_i$ based on the underlying chemical reaction rule.
Our example systems include a simple chemical reaction in closed reactor models and van de Vusse's competitive consecutive reaction\cite{VANDEVUSSE1964994} as an example of a complex chemical reaction in an open reactor model.

\subsubsection{Simple chemical reaction}\label{sec:simplereaction}
For the simple chemical reaction, we consider a closed system as a reactor model that is filled with reactants in selected boxes at time $t_0$.
We consider the elementary chemical reaction involving reactants A and B that form product C depending on the reaction rate constant, $k_r$ [\unit{\l\per\mol\per\s}]. \qty{1}{\mol} of A and \qty{1}{\mol} of B form \qty{1}{\mol} of C. If A and B are in box $B_i$, they react with each other.
$$
A+B \xlongrightarrow{k_r} C
$$
The rate constant $k_r$ depends on temperature and reaction conditions. The reaction proceeds quickly if $k_r$ is large. If $k_r$ is small, the reaction proceeds slowly.
The reaction rate $r_S$ describes how quickly the concentrations of substance S change \cite{Levenspiel1999}:
$$
-r_A=-r_B=r_C=k_r c_A c_B
$$
The concentrations change over time. A and B decrease at the same rate (because they react 1:1):
\begin{eqnarray*}
\dot{c_A} &=& - k_r c_{A}c_{B} \\
\dot{c_B} &=& - k_r c_{A}c_{B}.
\end{eqnarray*}
C increases at exactly the same rate as A (or B) decreases, because {\qty{1}{\mol} of each reactant produce \qty{1}{\mol} of product. 
Since product C is formed, the rate of change in the concentration of C is positive.
$$
\dot{c_C} = + k_r c_{A} c_{B}
$$
The total amount [\unit{mol}] therefore does not remain the same in this example. The change in concentrations is always given by the stoichiometric ratio, in this case: 1:1:1.

Assuming we have plenty of A and only a little B, the maximum concentration of B that can react is consumed at a 1:1 ratio. In this case, B is the limiting reactant. The limiting reactant determines when the reaction stops. The reaction in a box automatically stops as soon as the concentration of one of the reactants in that box $B_i$ reaches \qty{0}{\mol}.

The Euler method is used to simulate the reaction process over the (rather) short time intervals of length $\tau$. Therefore, this simple numerical approach provides an efficient approximation of the differential equations of the reaction kinetics.  

% By choosing a sufficiently small time step $h \leq \tau$ and $\frac{\tau}{h} \in \mathbb{N}$ steps, the Euler method yields results in good agreement with the analytical solution. The concentration of a substance in box $B_i$ at time $t+h$ is determined by the concentration in that box after transport and subsequent reaction. We need the reaction rate $r_A$ of A to determine the time-dependent concentrations of A, B, and C.

% In each time step $h$, the change in concentrations in the respective concentration vectors $\bm{C}_{A_t}$, $\bm{C}_{B_t}$, $\bm{C}_{C_t} \in \mathbb{R}^n$ is determined as follows.
% Let 
% $
% \hat{\bm{C}}_{S_t}:= \bm{C}_{S_t} \bm{P}
% $
% denote the time-evolved concentration vector for a chemical substance $S\in \{A,B,C\}$.
Using the time-evolved concentration vectors from \eqref{eq:evolvedC} the  vectorized Euler scheme for the reaction process is then given as 
\begin{eqnarray}
    \hat{\bm{C}}_{A_{t+h}} &=& \hat{\bm{C}}_{A_{t}} - h k_r \hat{\bm{C}}_{A_{t}}\circ  \hat{\bm{C}}_{B_{t}} \notag\\
    \hat{\bm{C}}_{B_{t+h}} &=& \hat{\bm{C}}_{B_{t}} - h k_r \hat{\bm{C}}_{A_{t}} \circ  \hat{\bm{C}}_{B_{t}} \label{eq:simplereact}\\
    \hat{\bm{C}}_{C_{t+h}} &=& \hat{\bm{C}}_{C_{t}} + h k_r \hat{\bm{C}}_{A_{t}}\circ  \hat{\bm{C}}_{B_{t}}\notag
\end{eqnarray}
Here $\circ$ denotes the Hadamard product (element-wise multiplication of the vector entries) and $h =\frac{\tau}{N_h}$, $N_h \in\mathbb{N}$ the internal time step of the Euler scheme. We iterate $N_h$ steps and finally set
$$
\bm{C}_{S_{t+\tau}}:=\hat{\bm{C}}_{S_{t+\tau}},
$$ 
which is the concentration vector for the chemical substance $S \in \{A,B,C\}$ at time $t+\tau$ after advection and reaction.  
The reaction is simulated using  $k_r=0.1 \unit{\l\per\mol\per\s}$ for the reaction rate constant.

\subsubsection{Competitive consecutive reaction} \label{sec:vandevusse}
For the competitive consecutive reaction in an open reactor model, we continuously feed the reactant into selected input area boxes and collect the chemical products from selected output area boxes.
Van de Vusse~\cite{VANDEVUSSE1964994} examined the competitive consecutive reaction
\begin{eqnarray*}
    A &\xlongrightarrow{k_1}& B \xlongrightarrow{k_2} C \\
    2A &\xlongrightarrow{k_3}& D
\end{eqnarray*}
with constant reaction rates
\begin{eqnarray*}
    &k_1& [\unit{\per\s}] \text{ (first order reaction)}, \\
    &k_2& [\unit{\per\s}] \text{ (first order reaction) and} \\
    &k_3& [\unit{\l\per\mol\per\s}] \text{ (second order reaction)}.
\end{eqnarray*}
to maximize the production of the desired product B.

This reaction can be described as competitive, in that on the one hand, the reactant A becomes the desired product B, and on the other hand, 2A becomes the waste product D. Furthermore, it can be characterized as consecutive, in that after A becomes B, B becomes the waste product C.

For each time step, we calculate the current reaction rates per box $B_i$, where
\begin{eqnarray*}
    -r_{A1} = r_{B1} &=& k_1 c_A, \\
    -r_{B2} &=& k_2 c_B \text{ and} \\
    -r_{A3} &=& k_3 c_A^2,
\end{eqnarray*}
and get the reaction rates for each fluid in that box:
\begin{eqnarray*}
    -r_A = -r_{A1}-r_{A3} &=& k_1 c_A+k_3 c_A^2 \\
    r_B = r_{B1}+r_{B2} &=& k_1 c_A-k_2 c_B \\
    r_C = -r_{B2} &=& k_2 c_B \\
    r_D &=& \frac{1}{2} k_3 c_A^2
\end{eqnarray*}
We simulate Van de Vusse’s competitive consecutive reaction in an open reactor model.
The constant feed concentration $c_{A_0}$ of pure A is continuously fed into the input area boxes $B_{\text{input}}$ of the open reactor (volume $V$ per box, flow rate $v_0$) and the concentrations of all products are pumped out of the output area boxes $B_{\text{output}}$. We assume that the inflow and outflow velocities are identical. The calculation of concentrations varies depending on the location of the box $B_{i}$ in the open reactor model. For each box $B_i \notin B_{input} \cup B_{output}$, the change of concentrations evolve according to
\begin{eqnarray*}
    \dot{c_A} &=& - k_1 c_A - k_3 c_A^2 \\
    \dot{c_B} &=& k_1 c_A - k_2 c_B \\
    \dot{c_C} &=& k_2 c_B  \\
    \dot{c_D} &=& \frac{1}{2} k_3 c_A^2.
\end{eqnarray*}
% where the concentration updates are simulated as 
% \begin{eqnarray*}
%     \bm{C}_{A_{t+h}} &=& \bm{C}_{A_{t}} - h k_1 \bm{C}_{A_{t}} - h k_3 \bm{C}_{A_{t}}\circ \bm{C}_{A_{t}}, \\
%     \bm{C}_{B_{t+h}} &=& \bm{C}_{B_{t}} + h k_1 \bm{C}_{A_{t}} - h k_2 \bm{C}_{B_{t}}, \\
%     \bm{C}_{C_{t+h}} &=& \bm{C}_{C_{t}} + h k_2 \bm{C}_{B_{t}} \text{ and} \\
%     \bm{C}_{D_{t+h}} &=& \bm{C}_{D_{t}} + h \frac{1}{2} k_3 \bm{C}_{A_{t}}\circ \bm{C}_{A_{t}}.
% \end{eqnarray*}
In each box $B_i \in B_{\text{input}}$, the change of concentrations $c_A$ is calculated differently by
$$
\dot{c_A} = \frac{v_0}{V} c_{A0} - k_1 c_A - k_3 c_A^2,
$$
% where the concentration update is simulated as
% $$
% \bm{C}_{A_{t+h}}= \bm{C}_{A_{t}} + h \frac{v_0}{V} C_{A0} - h k_1 \bm{C}_{A_{t}} - h k_3 \bm{C}_{A_{t}}\circ \bm{C}_{A_{t}}.
% $$
whereas for each box $B_i \in B_{\text{output}}$, the changes in concentrations are given by
\begin{eqnarray*}
\dot{c_A} &=& -\frac{v_0}{V} c_A - k_1 c_A - k_3 c_A^2 \\
\dot{c_B} &=& -\frac{v_0}{V} c_B + k_1 c_A - k_2 c_B \\
\dot{c_C} &=& -\frac{v_0}{V} c_C + k_2 c_B \\
\dot{c_D} &=& -\frac{v_0}{V} c_D + \frac{1}{2} k_3 c_A^2.
\end{eqnarray*}

Let $\bm{I}\in\mathbb{R}^n$ denote an indicator vector of the inflow boxes where $\bm{I}_i=1$ if $B_i \in B_{\text{input}}$ and $0$ otherwise, and similarly $\bm{O}\in\mathbb{R}^n$ the indicator vector for the outflow boxes $B_{\text{output}}$. In accordance with the approach proposed for the simple reaction (see sect.\ \ref{sec:simplereaction}), the vectorized Euler scheme for the reaction of the substances based on the time-evolved concentration vectors from \eqref{eq:evolvedC} is defined as  
\begin{eqnarray}
    \hat{\bm{C}}_{A_{t+h}} &=& \hat{\bm{C}}_{A_{t}} + h\left(\frac{v_0}{V} C_{A0}\bm{I} -\frac{v_0}{V} \hat{\bm{C}}_{A_{t}}\circ \bm{O}\right. \notag \\ && \left.  - k_1 \hat{\bm{C}}_{A_{t}} - k_3 \hat{\bm{C}}_{A_{t}}  \circ \hat{\bm{C}}_{A_{t}}\right)\notag\\
    \hat{\bm{C}}_{B_{t+h}} &=& \hat{\bm{C}}_{B_{t}} +h\left( -\frac{v_0}{V} \hat{\bm{C}}_{B_{t}} \circ \bm{O} + k_1 \hat{\bm{C}}_{A_{t}} - k_2 \hat{\bm{C}}_{B_{t}}\right)\notag \\
    \hat{\bm{C}}_{C_{t+h}} &=& \hat{\bm{C}}_{C_{t}} +h\left( -\frac{v_0}{V} \hat{\bm{C}}_{C_{t}}\circ \bm{O} + k_2 \hat{\bm{C}}_{B_{t}}\right) \label{eq:vandevusse}\\
    \hat{\bm{C}}_{D_{t+h}} &=& \hat{\bm{C}}_{D_{t}} +h\left(-\frac{v_0}{V} \hat{\bm{C}}_{D_{t}}\circ \bm{O} + \frac{1}{2} k_3 \hat{\bm{C}}_{A_{t}} \circ \hat{\bm{C}}_{A_{t}}\right).\notag
\end{eqnarray}
The concentration vector for each chemical substance $S \in \{A,B,C, D\}$ at time $t+\tau$ after advection and reaction is again obtained as 
$\bm{C}_{S_{t+\tau}}:=\hat{\bm{C}}_{S_{t+\tau}}$ when iterating the Euler scheme for $N$ time steps of length $h=\frac{\tau}{N}$.

The competitive consecutive reaction in an open reactor model is simulated using reaction rate constants $k_1 = \frac{1}{72} \unit{\per\s}$, $k_2 = \frac{1}{36} \unit{\per\s}$ and $k_3 = \frac{1}{360} \unit{\l\per\mol\per\s}$. The example reaction systems are simulated and analyzed in the reactor models described below.

\section{Reactor models}

We examine three reactor models: a two-dimensional DG flow~\cite{SHADDEN2005271}, a three-dimensional ABC flow~\cite{Dombre1986}, and a three-dimensional STR~\cite{WEILAND2023100448}. These models demonstrate the effects of fluid particle transport within a reactor on the production of the desired product.
The chemical reactor corresponds to the domain of the flow $M$, where $M \subset \mathbb{R}^2$ is a rectangle and $M \subset \mathbb{R}^3$ is a cube or cylinder. We distinguish between closed and open reactor models. In the closed system, $M$ is invariant. This means no particles starting in $M$ can leave $M$ and no particles starting in $\mathbb{R}^d \setminus M$ ($d=2$ or $d=3$) can enter $M$.
In some cases, a real reaction vessel functions more like a semi-batch reactor than a batch reactor (closed STR), as the chemical reaction already takes place during the filling phase. In our closed reactor models, we do not represent a filling process and start a reaction process with the assumed initial distribution of the chemical substances.
In a closed system, the initial distribution of fluid particles changes due to mixing and reactions.

In the open system, we implement inflow and outflow dynamics in defined inflow and outflow areas of the reactor. There, particles enter and leave $M$ as in a CSTR. We use the explicit Euler method to approximate continuous inflow and outflow, as described in sect.~\ref{sec:vandevusse}. In an open system, only a neutral substance required for mixing is initially present in the reactor, not any reacting tracers. Due to the continuous inflow of the reactants, the concentrations of the resulting products initially increase.
The three reactor models are explained below, starting with the two-dimensional DG reactor model.

\subsection{DG reactor model}
The DG flow is simple enough mathematically to be easily handled analytically and numerically. However, it is also complex enough dynamically to simulate realistic phenomena, such as chaotic mixing, transport barriers, and LCS.
The DG mixes the fluids over the domain $[0,2]\times[0,1]$ with the velocity field $\bm{u}$ defined by~\cite{SHADDEN2005271}
\begin{eqnarray}
    \dot{x} &=& -0.5 \pi \sin (\pi f(x,t)) \cos (\pi y) \notag \\
    \dot{y} &=& 0.5 \pi \cos (\pi f(x,t)) \sin (\pi y) \frac{df}{dx} (x,t), \text{where} \notag \\
    f(x,t) &=& \delta \sin(2 \pi t) x^2 + (1-2\delta \sin(2 \pi t)) x.\label{eq:doublegyre}
\end{eqnarray}
We differentiate between an autonomous ($\delta=0$) and a time-dependent system with $\delta = 0.1$ or $\delta = 0.5$ (see fig.~\ref{DG_Flow}).
For better comparison with physically scaled reactor models, the nondimensional DG was dimensionalized by assigning one spatial unit to $0.1\unit{\m}$ and one time unit to $1\unit{\s}$. Consequently, the computational domain corresponds to $0.2\unit{\m}\times0.1\unit{\m}$. For the autonomous system, the average flow velocity is $U_0\approx 0.105\unit{\m\per\s}$, which implies a flow time scale $\tau_{f_0}\approx \frac{0.2\unit{\m}}{0.105\unit{\m\per\s}}\approx 1.9\unit{\s}$.
For the time-dependent systems, the average flow velocities are $U_{0.1}\approx 0.105\unit{\m\per\s}$ and $U_{0.5}\approx 0.102\unit{\m\per\s}$, implying the flow time scales $\tau_{f_{0.1}}\approx 1.9\unit{\s}$ and $\tau_{f_{0.5}}\approx 2.0\unit{\s}$.
\begin{figure}[ht]
\begin{center}
    \begin{tabular}{lll}  
    {\text{a)}} & {\text{b)}} & {\text{c)}} \\
    \includegraphics[width=0.32\linewidth]{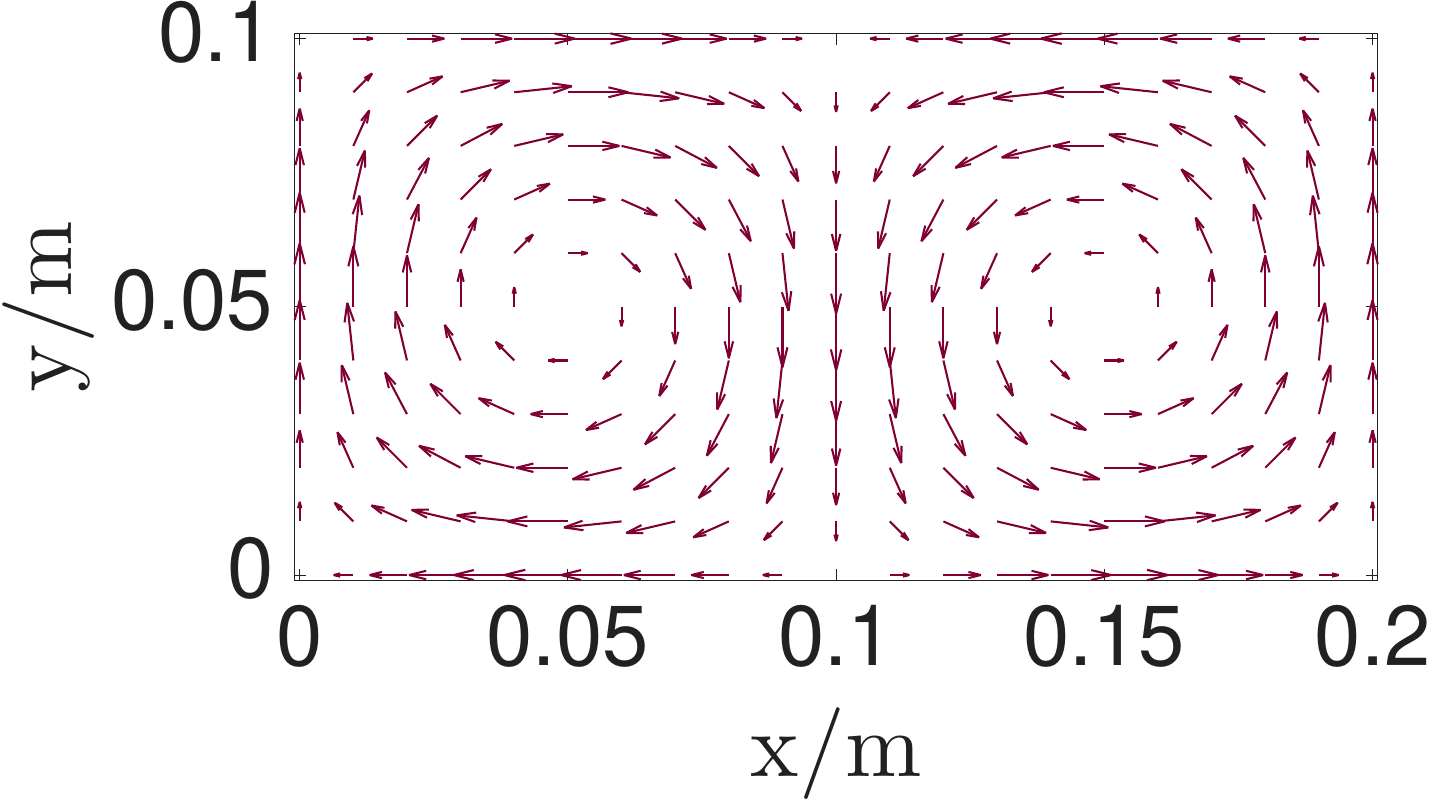}&
    \includegraphics[width=0.32\linewidth]{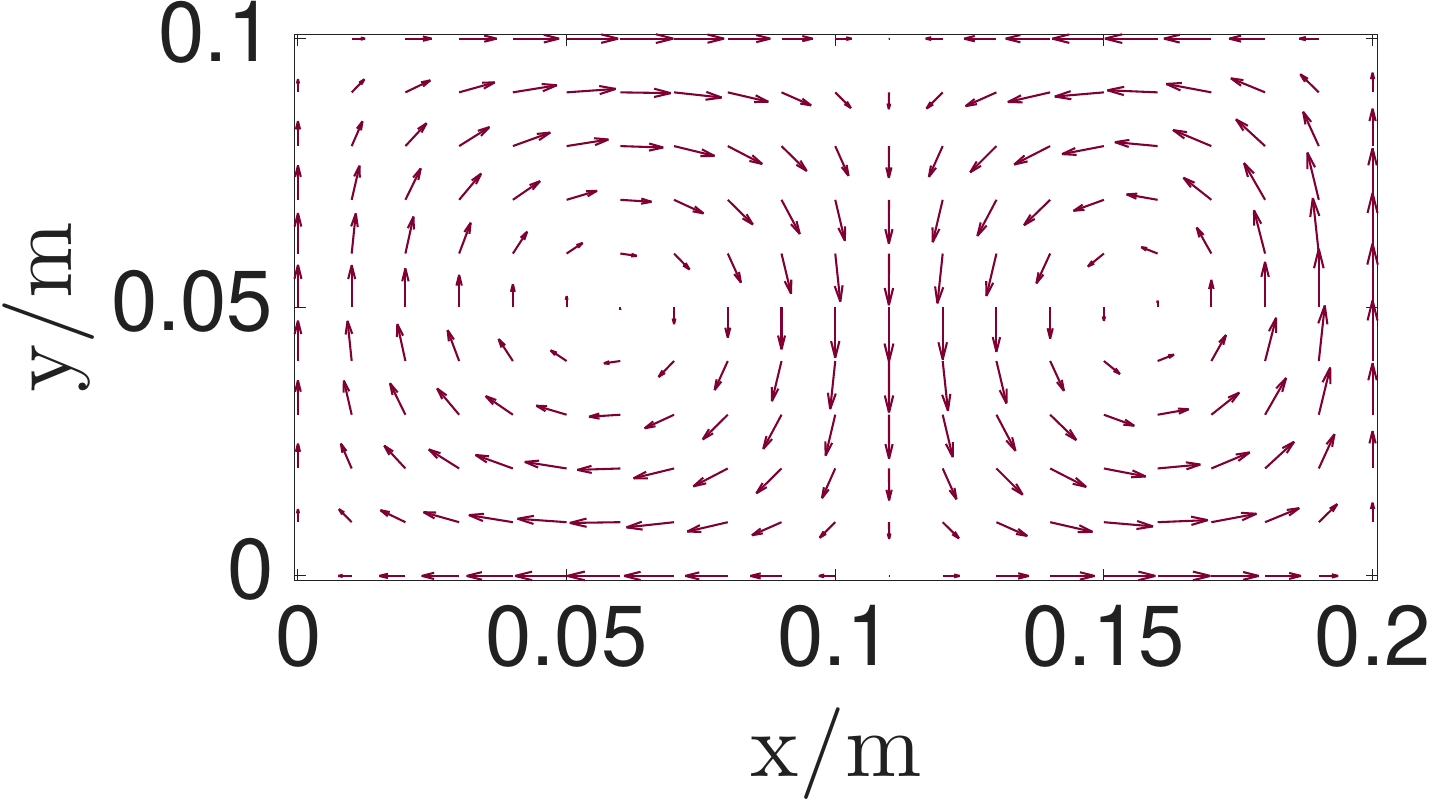} & \includegraphics[width=0.32\linewidth]{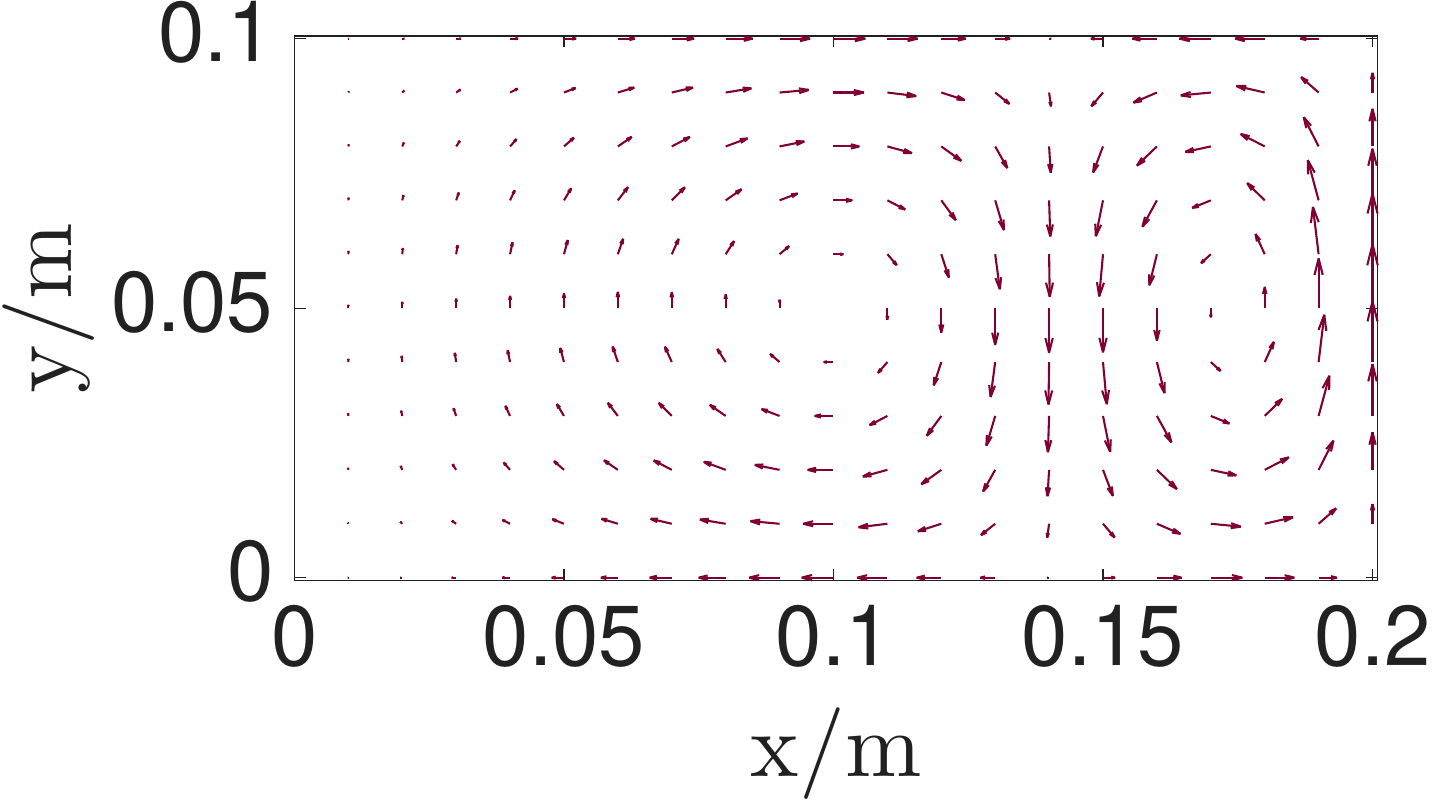}
    \end{tabular}
    \end{center}
        \caption{Flow velocity field of the DG in an autonomous system with $\delta=0$ (a) and in a time-dependent system with $\delta=0.1$ (b) and $\delta=0.5$ (c) at $t=0.25\unit{\s}$.}
    \label{DG_Flow}
\end{figure}
\subsection{ABC flow reactor model}
The ABC flow is a three-dimensional steady flow that is exact solution of Euler's equation~\cite{Dombre1986}. It involves three real parameters $a$, $b$ and $c$.
The Eulerian flow is $2 \pi$-periodic in $x$, $y$, and $z$, relative to rectangular Cartesian coordinates (see fig.~\ref{fig:ABC_mixing}), and has velocity $\bm{u} = (\dot{x}, \dot{y}, \dot{z})$, where
\begin{eqnarray*}
\dot{x} &=& a \sin z + c \cos y,\\
\dot{y} &=& b \sin x + a \cos z,\\
\dot{z} &=& c \sin y + b \cos x.
\end{eqnarray*} 
% For real $\bm{x}(0)$, where $\bm{x}$ is in the periodic torus $T^3$, there exists a unique solution $\bm{x}(t)$ for all real $t$. 
The map $\bm{x}(0) \mapsto \bm{x}(t)$ defines a conservative dynamical system. $M=\mathbb{T}^3$ is a periodic cube (torus) and its coordinates are defined modulo $2 \pi$:
$$
0 \leq x \leq 2 \pi, 0 \leq y \leq 2 \pi, 0 \leq z \leq 2 \pi.
$$
Early numerical experiments by Hénon~\cite{Henon1966} have proven chaos in the special case $a=\sqrt{3}$, $b = \sqrt{2}$ and $c = 1$, which we will consider in our numerical study.
For better comparison with physically scaled reactor models, the nondimensional ABC flow was dimensionalized by assigning one spatial unit to $0.1\unit{\m}$ and one time unit to $1\unit{\s}$. Consequently, the computational domain corresponds to $0.2\pi \unit{\m}\times0.2\pi \unit{\m}\times0.2\pi \unit{\m}$. The average flow velocity of our ABC flow reactor model is $U\approx 0.233 \unit{\m\per\s}$. This implies the flow time scale $\tau_{f}=\frac{0.2\pi\unit{\m}}{0.233\unit{\m\per\s}}\approx 2.7\unit{\s}$.

\subsection{STR model}
The final reactor model we consider is an STR, which provides a realistic representation of fluid mixing and reaction processes. Here, rather than employing a velocity field description, we utilize given trajectory data obtained from a Lattice–Boltzmann simulation of a \qty{2.8}{\l} lab-scale vessel by Weiland et al.~\cite{WEILAND2023100448}. The reactor is cylindrical, with a base radius of $r=\qty{0.0655}{\m}$ and a height of $l=\qty{0.24}{\m}$ (see fig.~\ref{fig:STR_mixing}). The rotational speed of the mixer is $252$ rotations per minute. The dataset comprises approximately 355,000 particle trajectories spanning about \qty{2.4}{\s}, resolved into 239 frames. The average flow velocity of the STR model is $U\approx 0.062 \unit{\m\per\s}$. This implies the flow time scale $\tau_{f}=\frac{0.24\unit{\m}}{0.062\unit{\m\per\s}}\approx 3.9\unit{\s}$.
We distinguish between a closed and an open STR model representing a CSTR with input and output dynamics (see fig.~\ref{fig:STR_VDV}).

\section{Computational studies}
In this study, we simulate both non-reactive mixing and a simple chemical reaction within the DG, ABC flow, and closed STR reactor models, as well as a competitive consecutive reaction in an open STR model.
We evolve reactants A and B both with and without additional diffusion in the different closed reactor models and compare the product distributions.
Furthermore, we examine the correlation between backward-time FTE fields and the spatial distribution of the reaction front. Finally, for the open STR model, we analyze the total yield and the internal distribution of the desired product.
\subsection{Mixing and reaction in a DG reactor model}\label{sec:mixreactDG}
We discretize the DG reactor model into $2^{15}$ boxes.
Each box $B_i$ is a square with a side length of $d=\num{7.8e-4}\unit{m}$ and contains 100 test points on a uniform inner grid.
The initial distribution at time $t_0=0\unit{\s}$ is $c_{A} = 1\unit{\mol\per\l}$ per box in the upper half of the reactor and $c_{B} = 1\unit{\mol\per\l}$ per box in the lower half (see fig.~\ref{fig:DG_mixing}).
We simulate diffusion by approximating an $\epsilon$-ball for each test point with radius $\epsilon=0.002\unit{\m}$ by 25 grid points, so that $\epsilon \approx 5r$, with a radius $r=\frac{1}{2}d=\num{3.9e-4}\unit{\m}$ per box $B_i$.
For the DG, we calculate the transition matrix $\bm{P}$ (or $\bm{P}_{D}$) for $\tau=0.1\unit{\s}$. For the periodic systems with $\delta = 0.1$ and $\delta = 0.5$, we thus calculate ten transition matrices, i.e.\ on $[0, 0.1]\unit{\s}, [0.1, 0.2]\unit{\s}, \ldots [0.9, 1]\unit{\s}$.

When we mix substance A for each time step $\tau=0.1\unit{s}$ with substance B and compare the product distribution at time $t=1\unit{\s}$, we see a pronounced difference in mixing without and with diffusion (see fig.~\ref{fig:DG_mixing}). Without diffusion, at $t=1\unit{\s}$, there are predominantly boxes containing only A or only B. With diffusion, there are predominantly boxes that contain both A and B.
\begin{figure}
    \begin{center}
    \begin{tabular}{lll}
    & {\text{a)}} & \\
    & \includegraphics[width=0.31\linewidth]{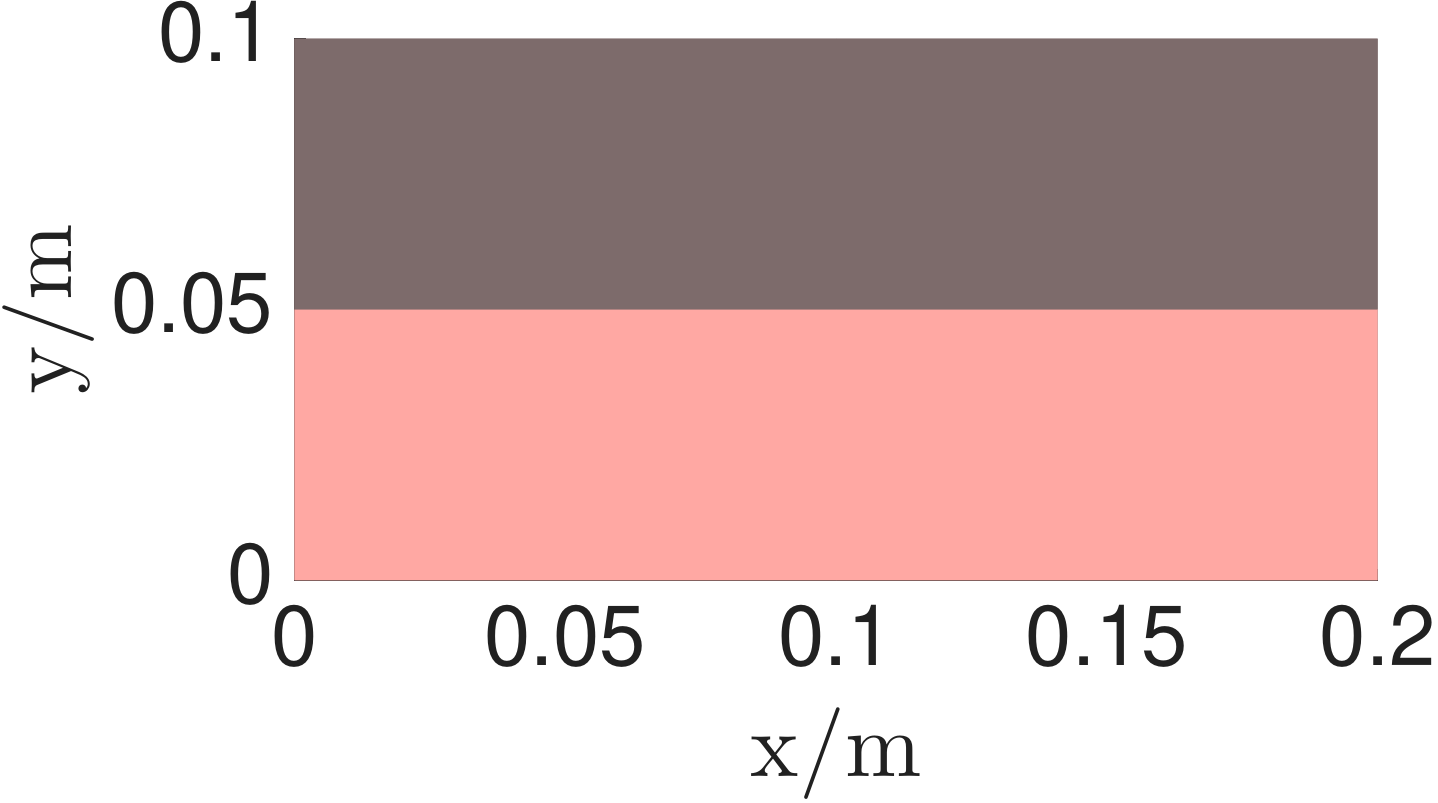} &
\includegraphics[width=0.065\linewidth]{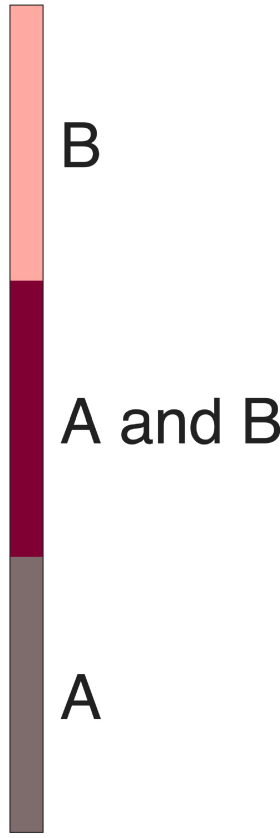} \\    
    {\text{b)}} & {\text{c)}} & {\text{d)}} \\   \includegraphics[width=0.31\linewidth]{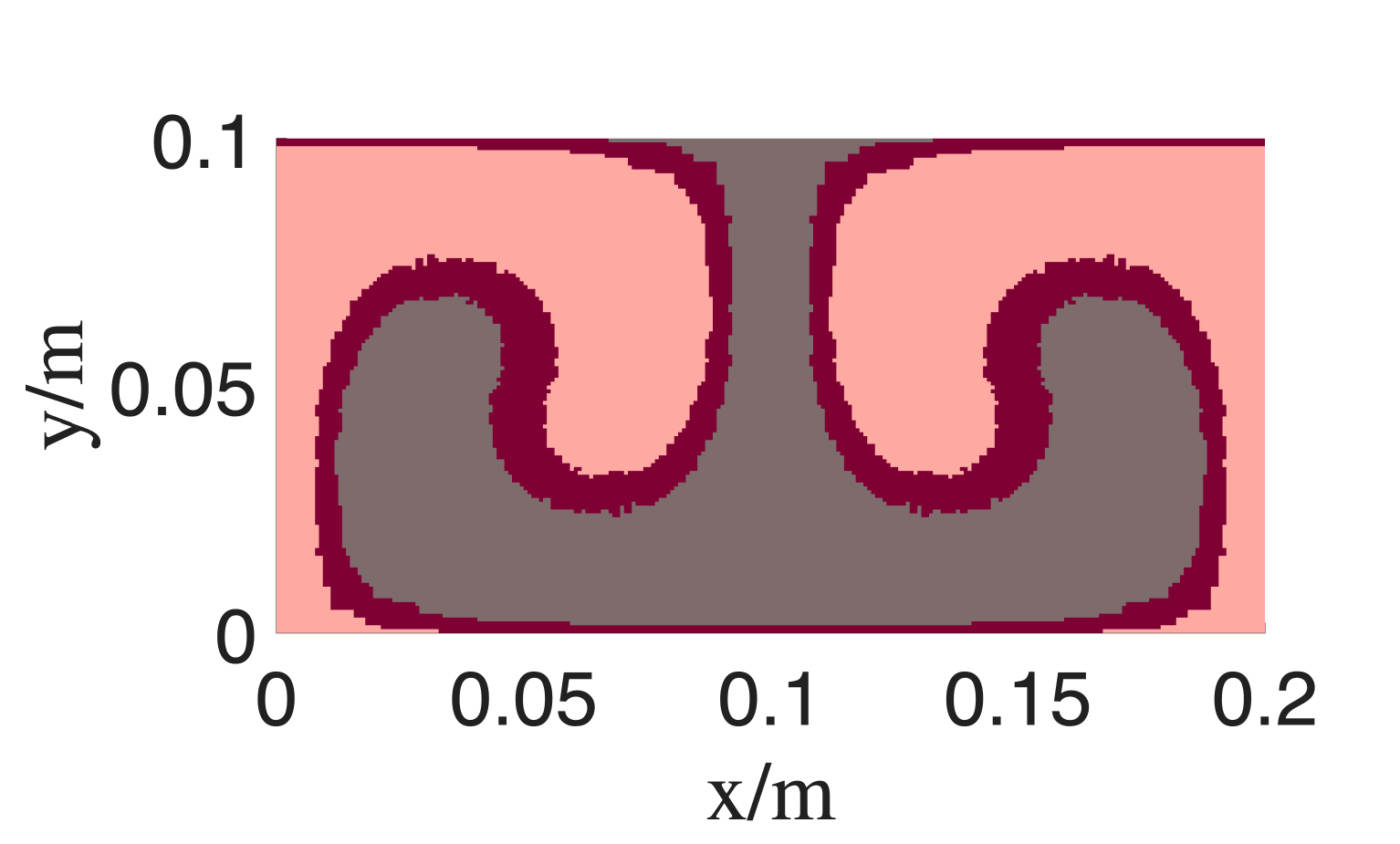} &    \includegraphics[width=0.31\linewidth]{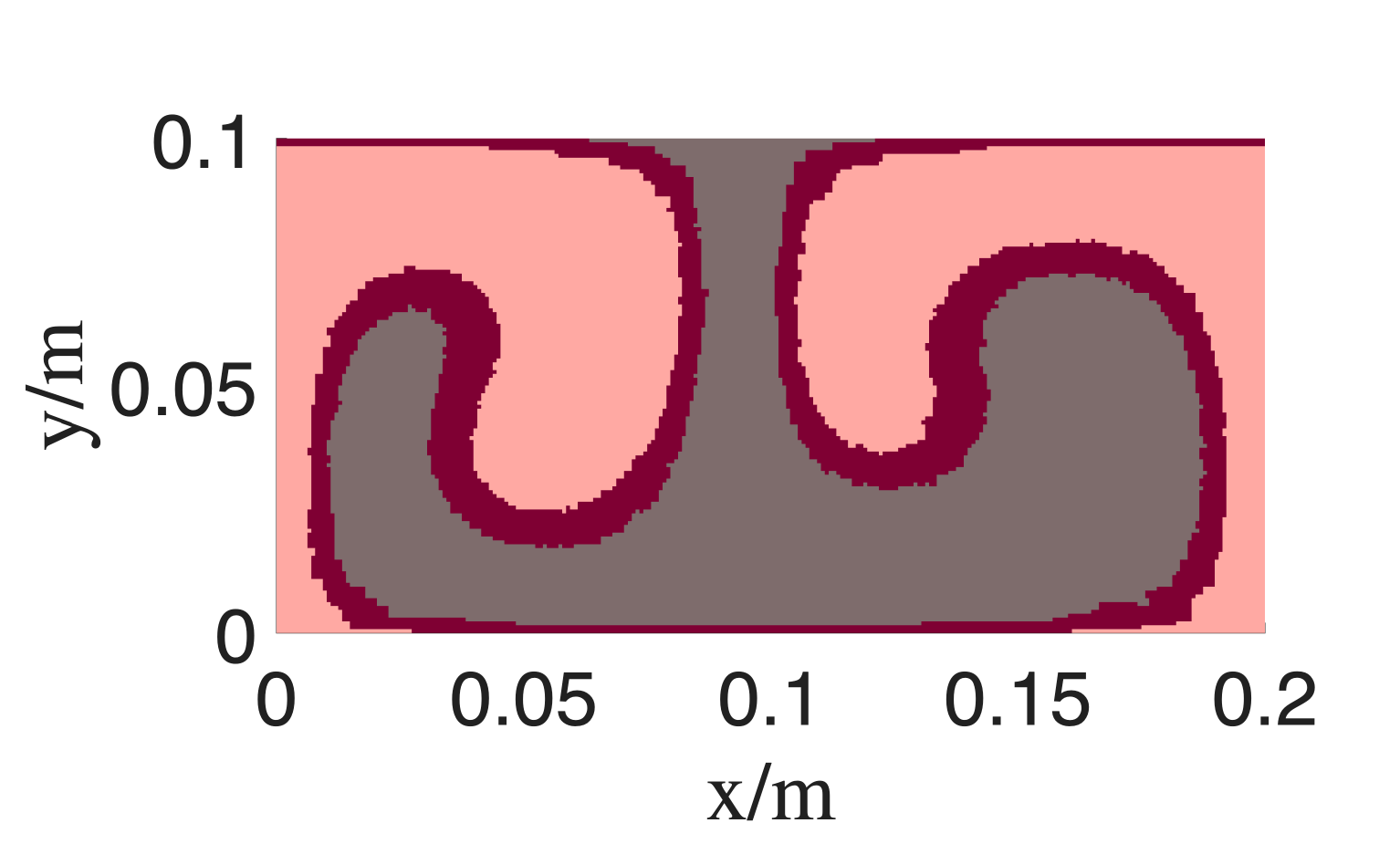} &    \includegraphics[width=0.31\linewidth]{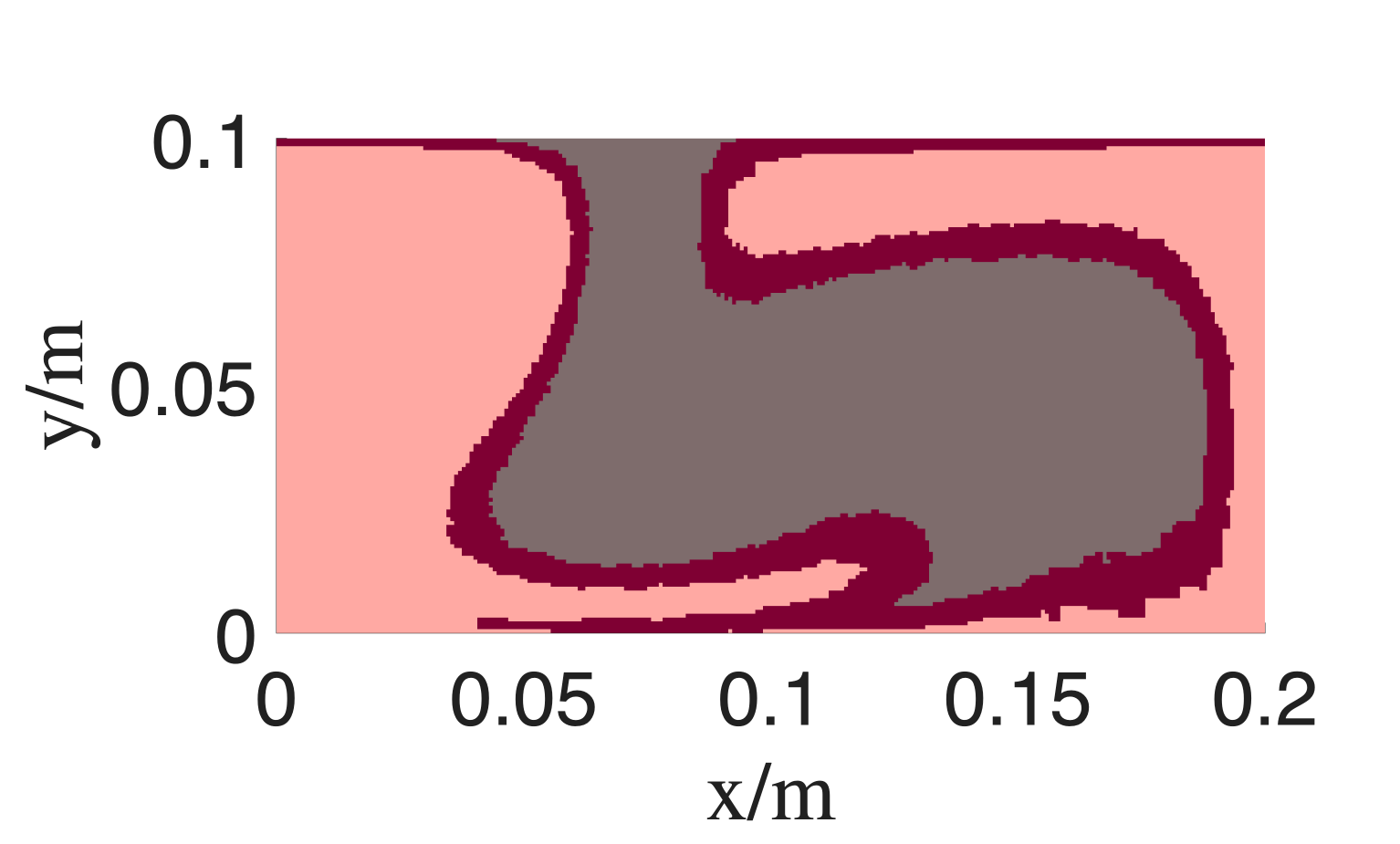} \\    
    {\text{e)}} & {\text{f)}} & {\text{g)}} \\   \includegraphics[width=0.31\linewidth]{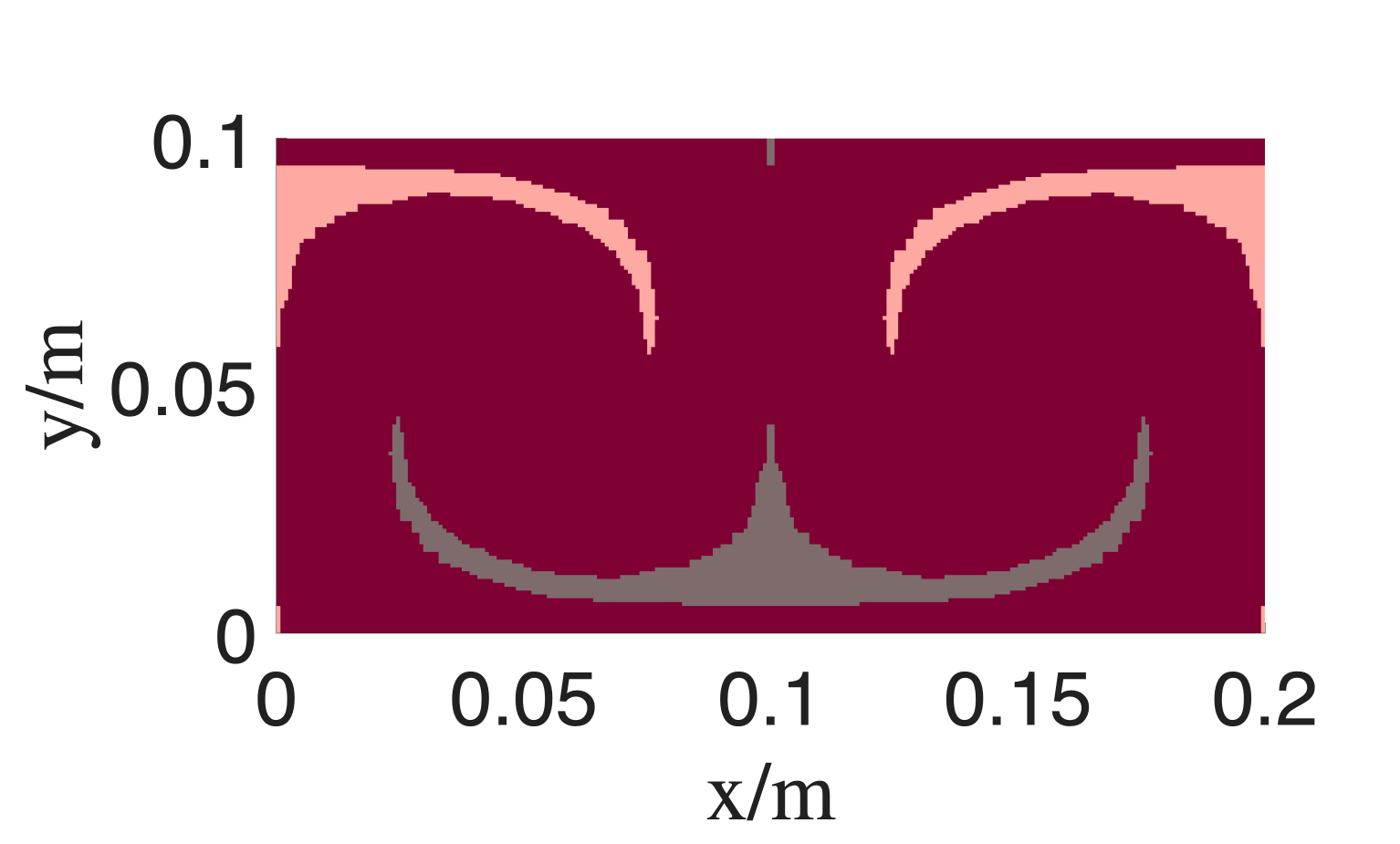} &   \includegraphics[width=0.31\linewidth]{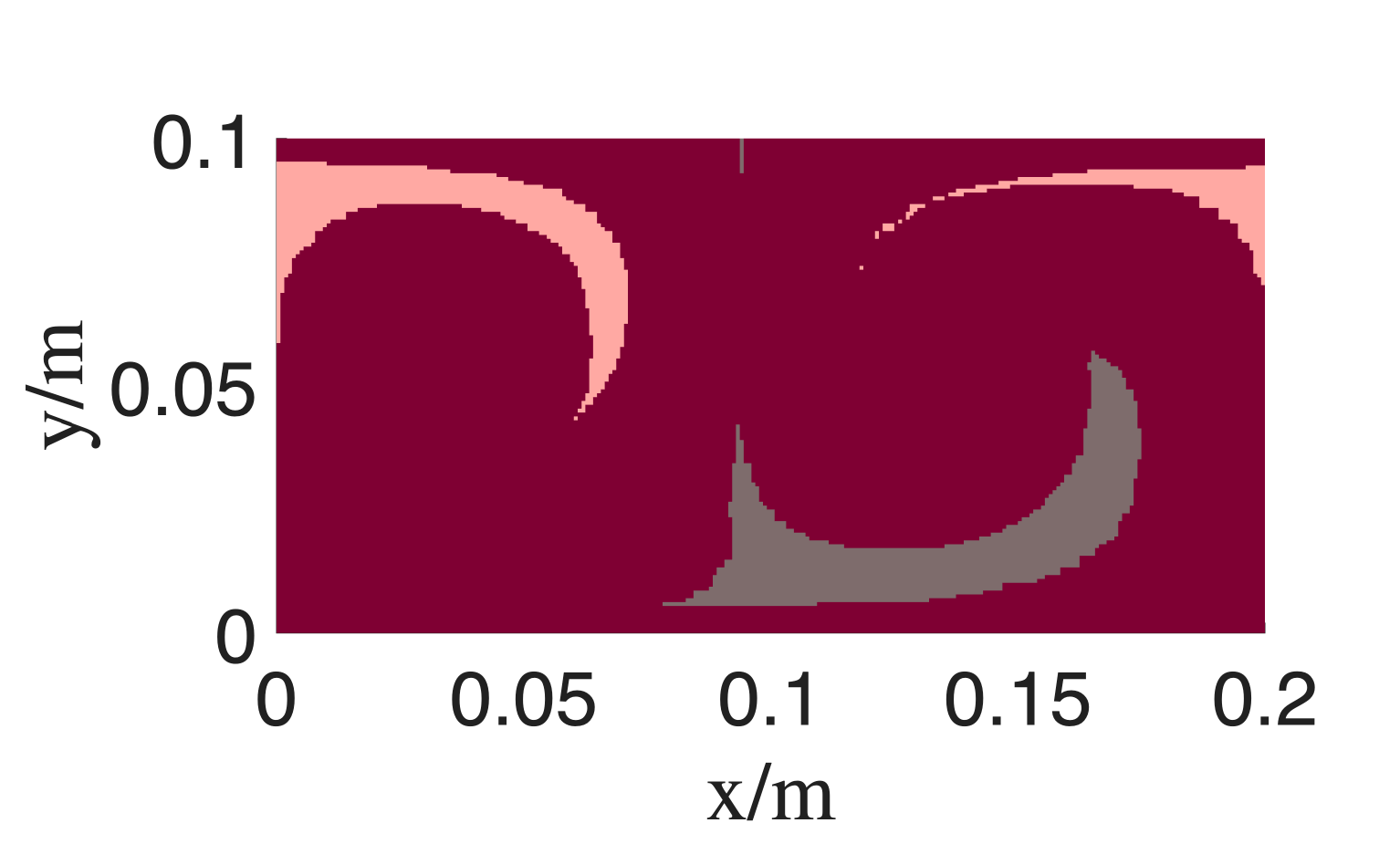} &  \includegraphics[width=0.31\linewidth]{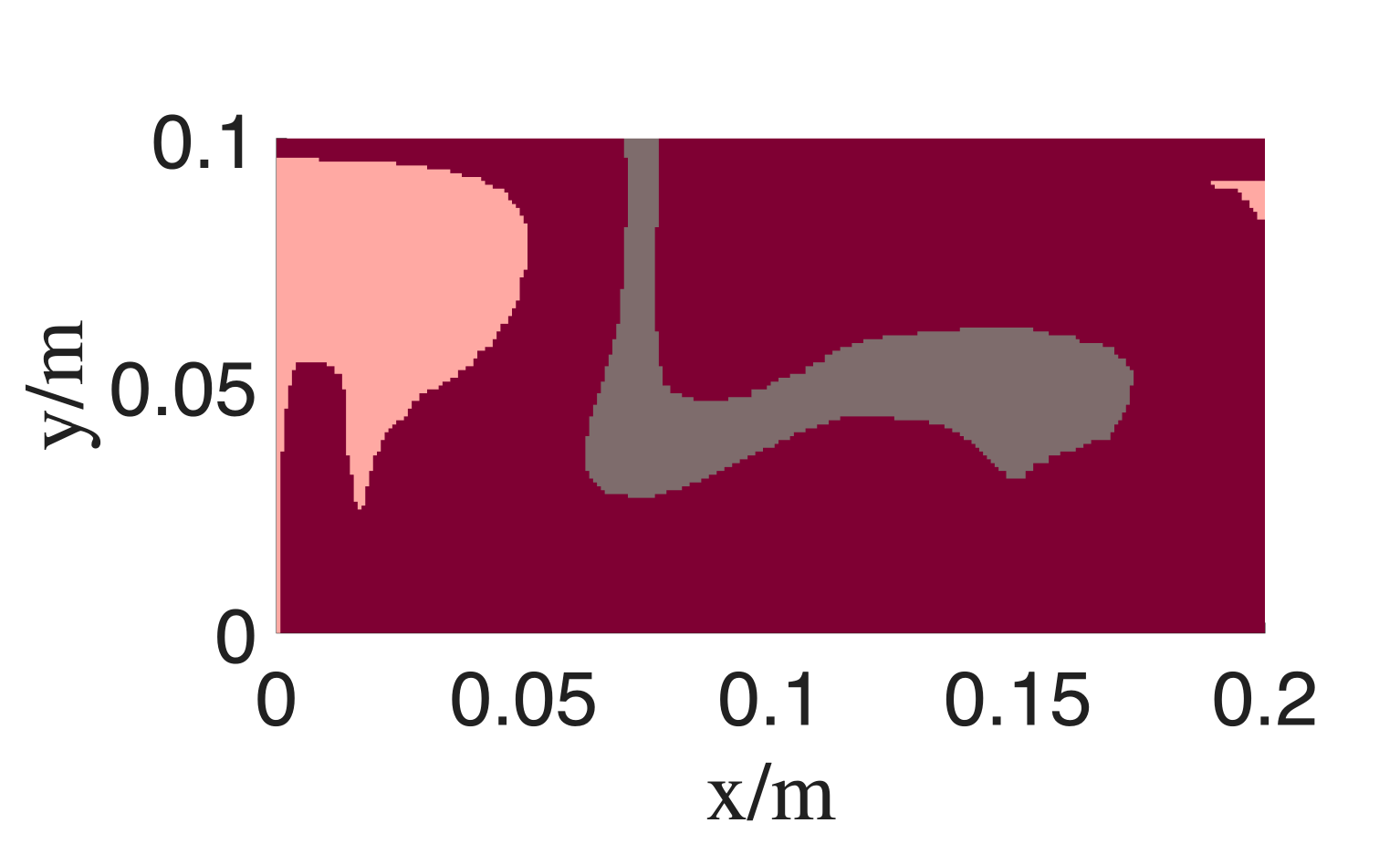} 
    \end{tabular}
    \end{center}
    \caption{Initial distribution (a) and product distribution at $t=1\unit{\s}$ after mixing (no reaction) for the DG without diffusion (b-d) and with diffusion (e-g) for $\delta=0$ (b,e), $\delta=0.1$ (c,f) and $\delta=0.5$ (d,g).}
    \label{fig:DG_mixing}
\end{figure}

We simulate the simple chemical reaction in the DG reactor model with an initial distribution of fluid A in the upper half of the reactor (see fig.~\ref{fig:DG_mixing}) and fluid B in the lower half, with each box $B_i$ containing $1\unit{\mol\per\l}$ of the substance present there.
For each time step $\tau = 0.1\unit{\s}$, the particles are transported and then react using the Euler scheme \eqref{eq:simplereact} with $h=0.01\unit{\s}$.

If we compare the reaction for the DG without and with diffusion, we see that the reaction with diffusion initially occurs faster (see fig.~\ref{fig:DG_total}).
For reaction rate constant $k_r=0.1\unit{\l\per\mol\per\s}$ and the maximum initial concentrations $c_{A_0}=c_{B_0}=1\unit{\mol\per\l}$, this implies the reaction time scale $\tau_{r}=\frac{1}{0.1\unit{\l\per\mol\per\s}\cdot 1\unit{\mol\per\l}}=10\unit{\s}$. For the flow time scales $\tau_{f_0}\approx 1.9\unit{\s}$ ($\delta=0$), $\tau_{f_{0.1}}\approx 1.9\unit{\s}$ ($\delta=0.1$) and $\tau_{f_{0.5}}\approx 2.0\unit{\s}$ ($\delta=0.5$), the Damköhler numbers are $Da_0 = \frac{\tau_{f_0}}{\tau_{r}}\approx \frac{1.9\unit{\s}}{10\unit{\s}}\approx 0.19$, similarly $Da_{0.1} =\approx 0.19$ and $Da_{0.5} = \approx  0.20$. For the DG reactor model, $Da_{0} \approx Da_{0.1} \approx Da_{0.5}\ll 1$ indicate that advection is faster than the chemical reaction.

At the beginning of the reaction process, the reaction front and the concentration distribution of C are nearly identical for the DG. Therefore, for $t=4\unit{\s}$, we only show the concentration distributions of C (see fig.~\ref{fig:DG_reaction_FTE}). These compare nicely to the corresponding concentration distributions from the numerical solution of the respective advection-diffusion-reaction partial differential equation (see fig.~\ref{fig:DG_reaction_ADR} in the appendix).
The reaction process is completed most quickly overall in the autonomous system with diffusion followed by the autonomous system without diffusion (see fig.~\ref{fig:DG_total}). In the time-dependent systems, the reaction is completed most quickly for $\delta=0.1$ with diffusion followed by $\delta=0.5$ with diffusion. If we consider the systems with and without diffusion separately, the reaction requires more time overall for a larger delta.
\begin{figure}
    \centering
    \includegraphics[width=0.7\linewidth]{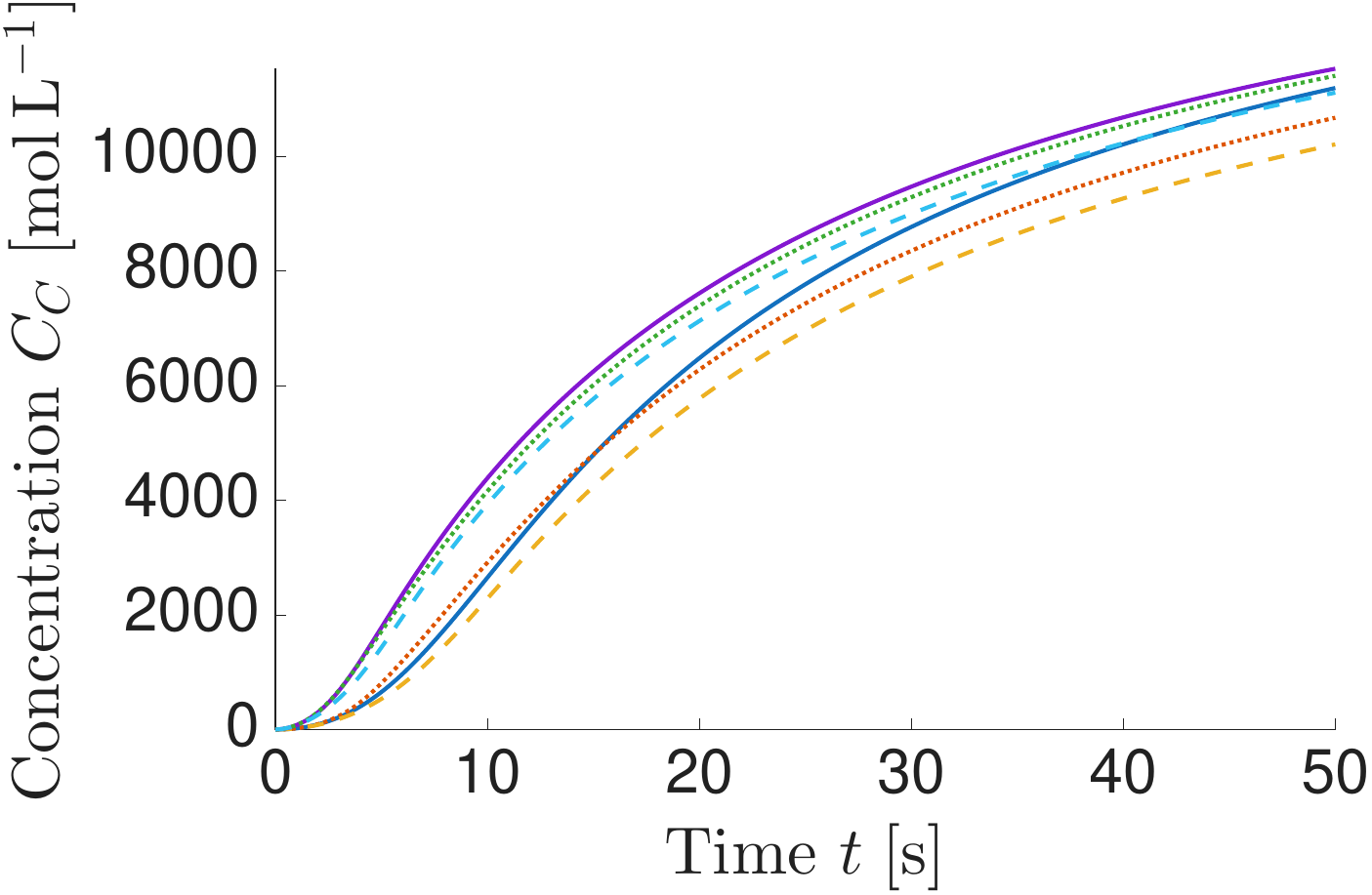}
    \includegraphics[width=0.7\linewidth]{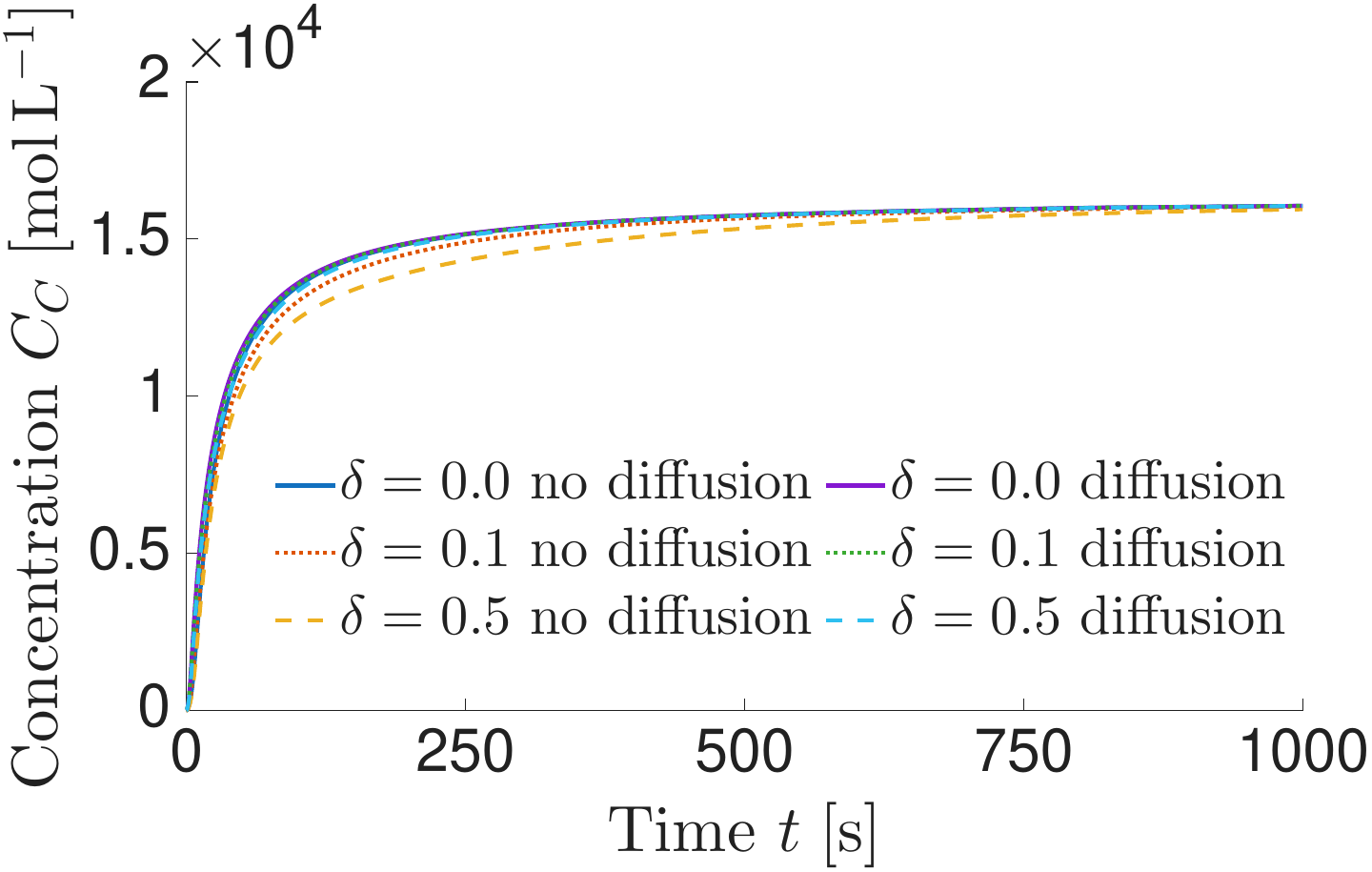}
    \caption{Total concentrations $C_C$ [\unit{\mol\per\l}] for the DG.}
    \label{fig:DG_total}
\end{figure}

Depending on $\delta$ and diffusion, the backward-time FTE differ for the DG, where higher values are observed in systems with diffusion (see fig.~\ref{fig:DG_reaction_FTE}). The higher the b-FTE, the more tracers collide. If these are the two reactants, A and B, they will react with each other. Of course, the initial distribution of the reactants affects where they collide and react. The question now is whether there is an increased reaction in boxes with high b-FTE.
\begin{figure*}
    \begin{center}
    \begin{tabular}{lll}
   {\text{a)}} & {\text{b)}} & {\text{c)}}   \\
\includegraphics[height=0.09\textheight]{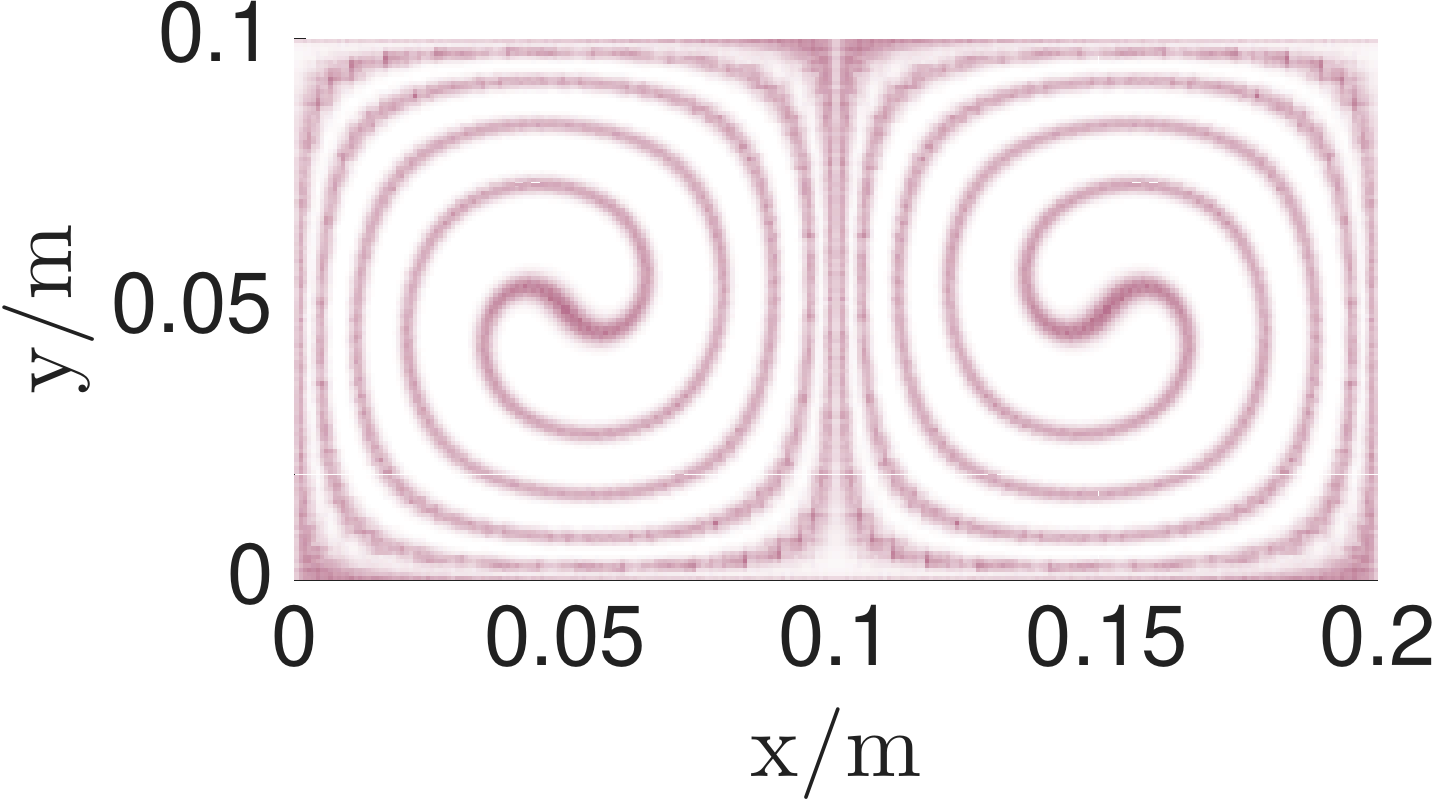} &    \includegraphics[height=0.09\textheight]{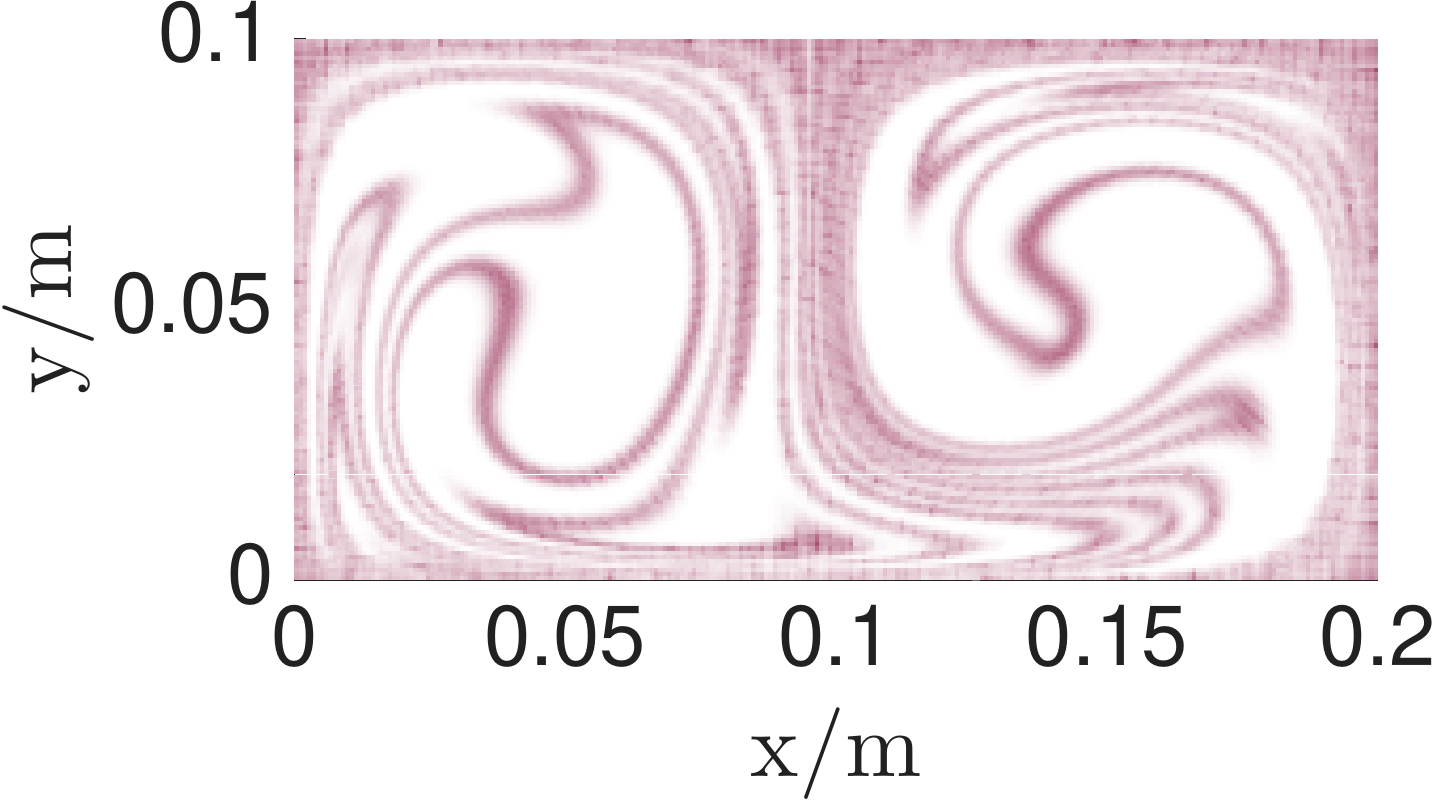} &    \includegraphics[height=0.09\textheight]{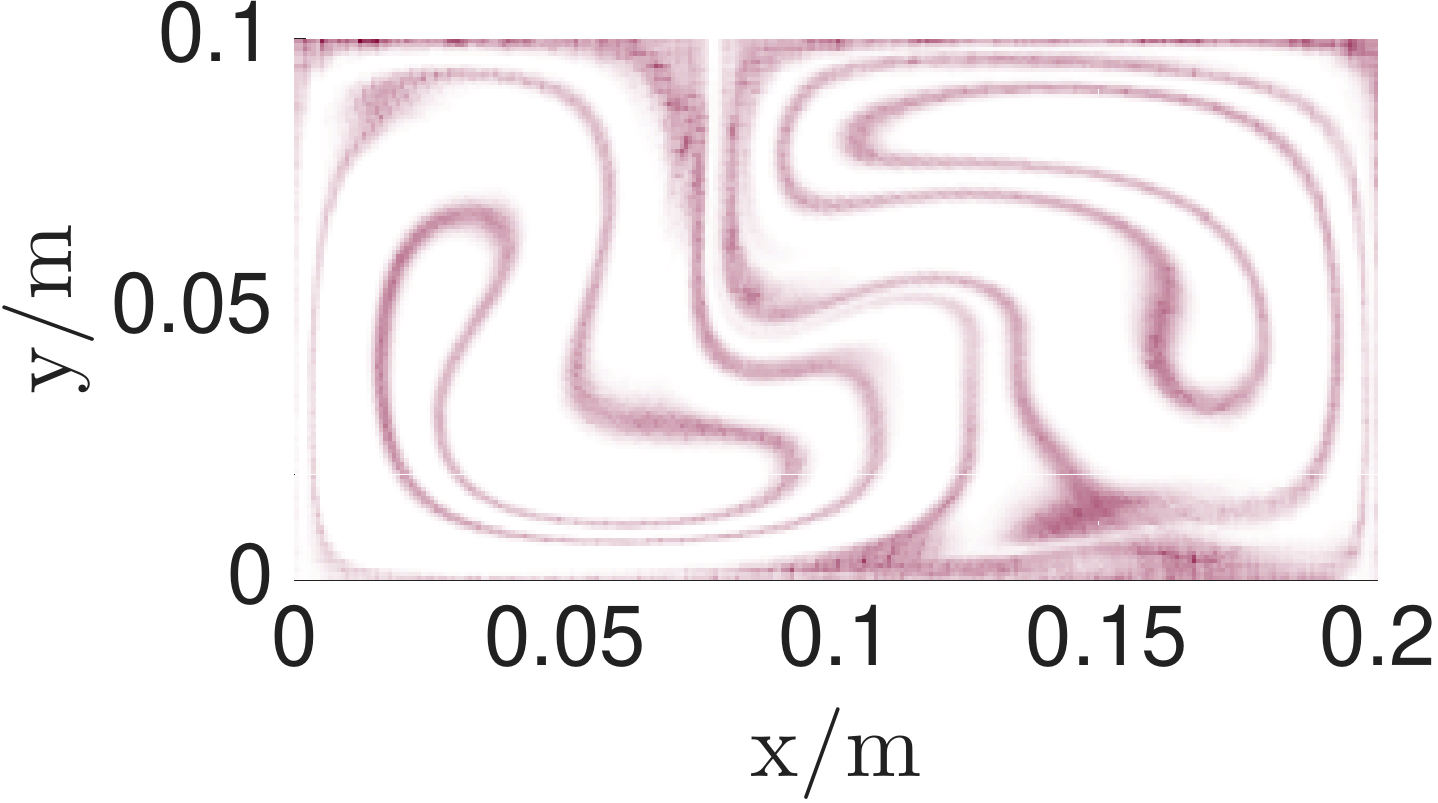}     \includegraphics[height=0.09\textheight]{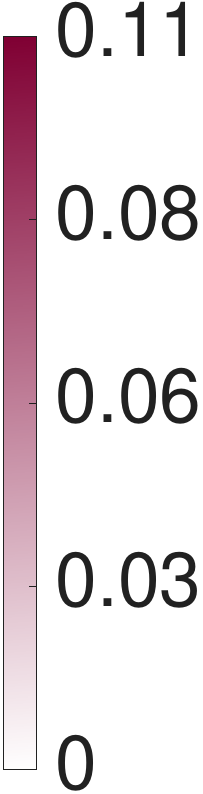} \\
    {\text{d)}} &
   {\text{e)}} &
    {\text{f)}}  \\
 \includegraphics[height=0.09\textheight]{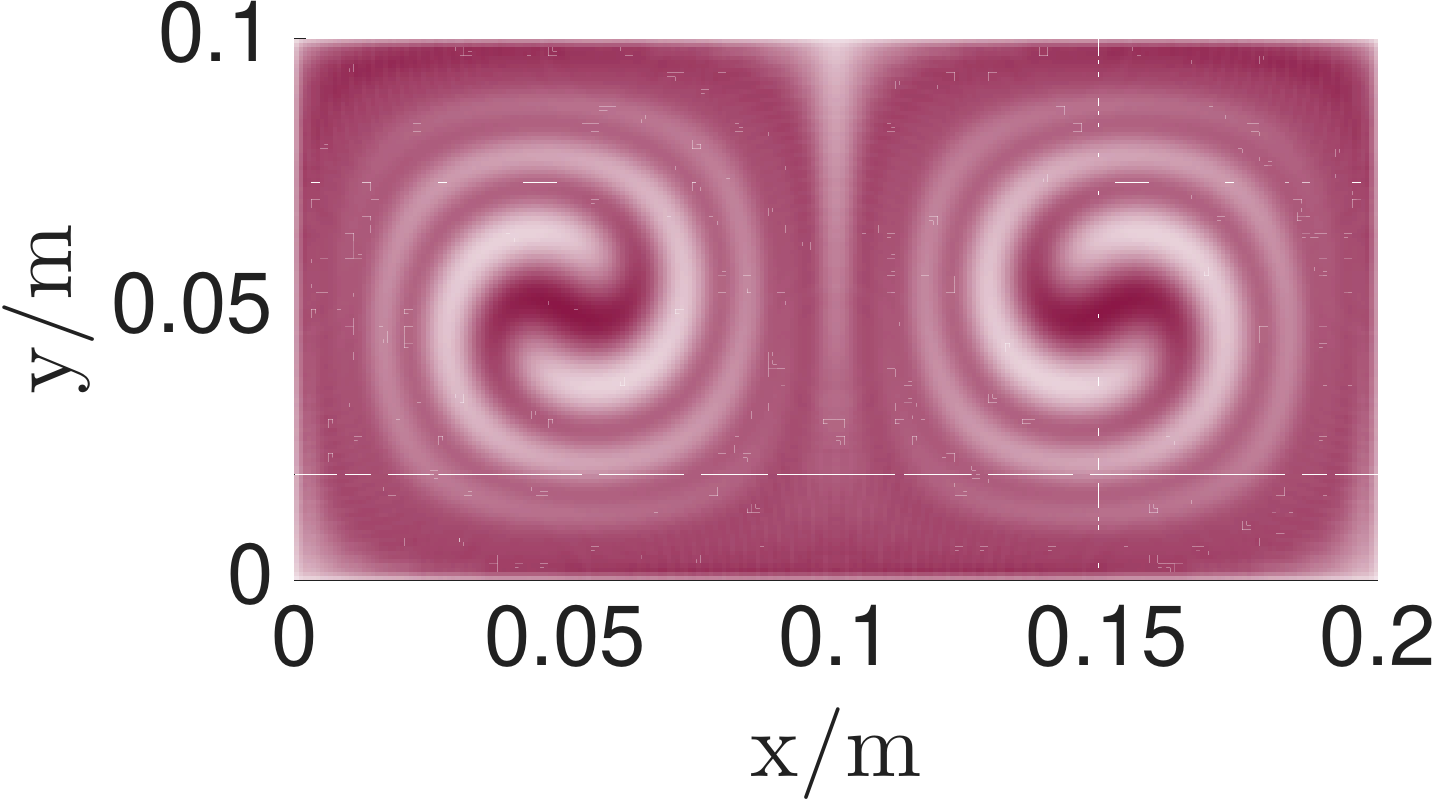} &
  \includegraphics[height=0.09\textheight]{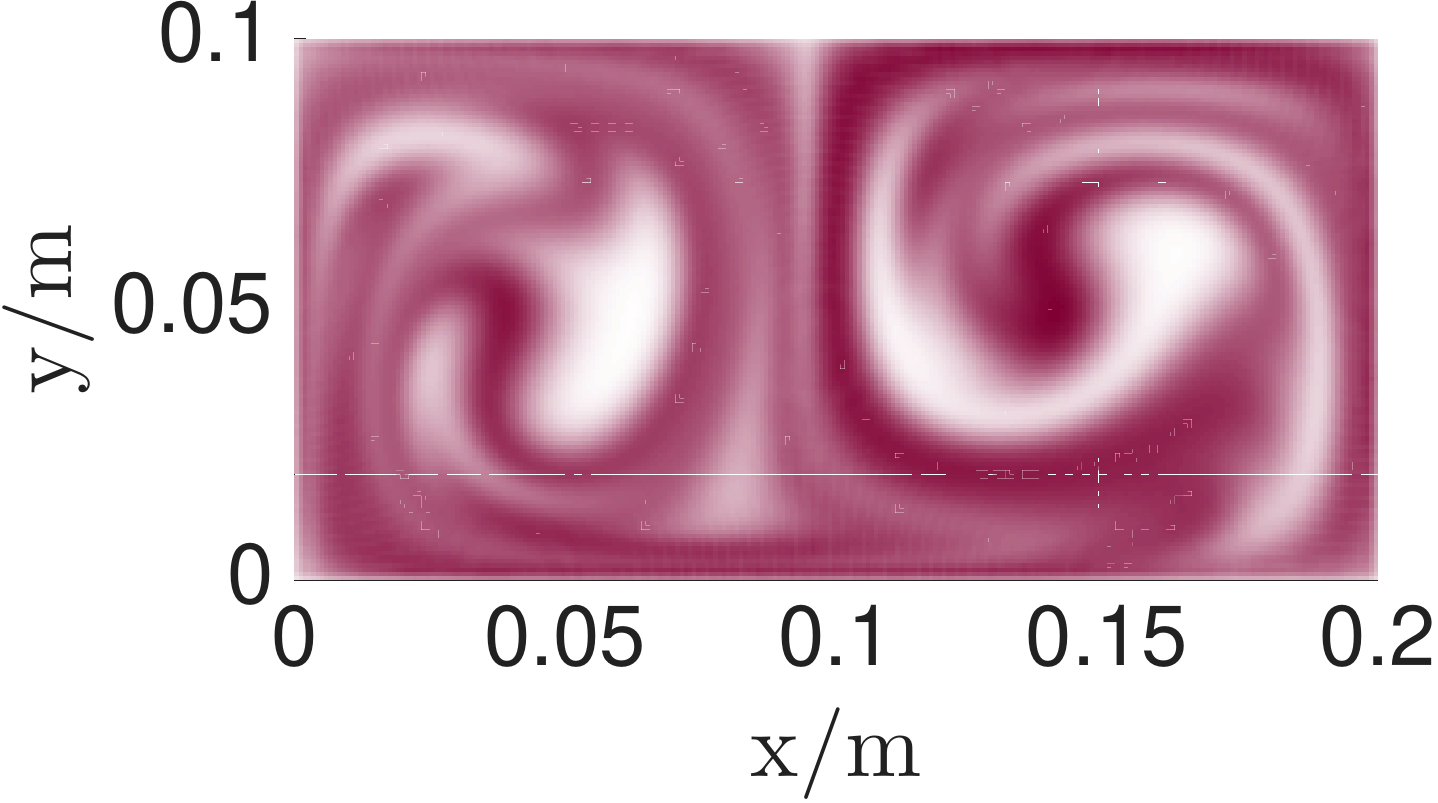} &    \includegraphics[height=0.09\textheight]{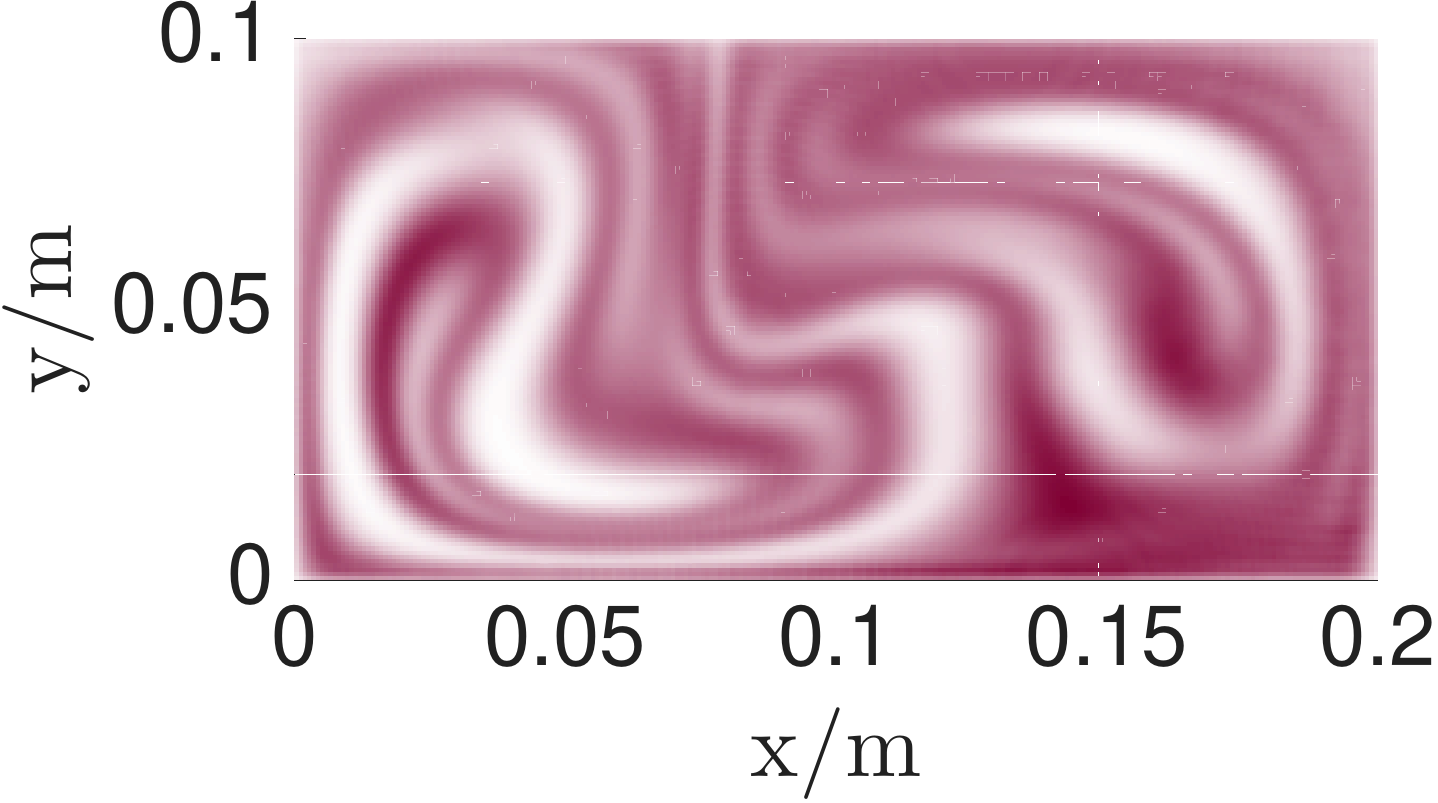}
  \includegraphics[height=0.09\textheight]{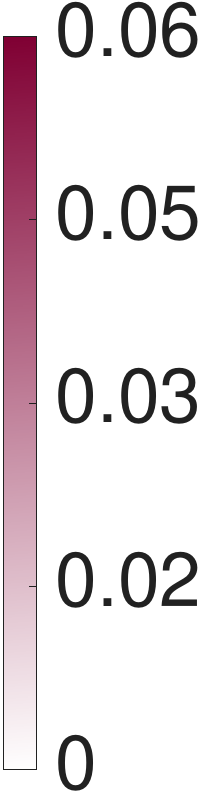} \\
    {\text{g)}}& 
    {\text{h)}} &
    {\text{i)}} \\
\includegraphics[height=0.09\textheight]{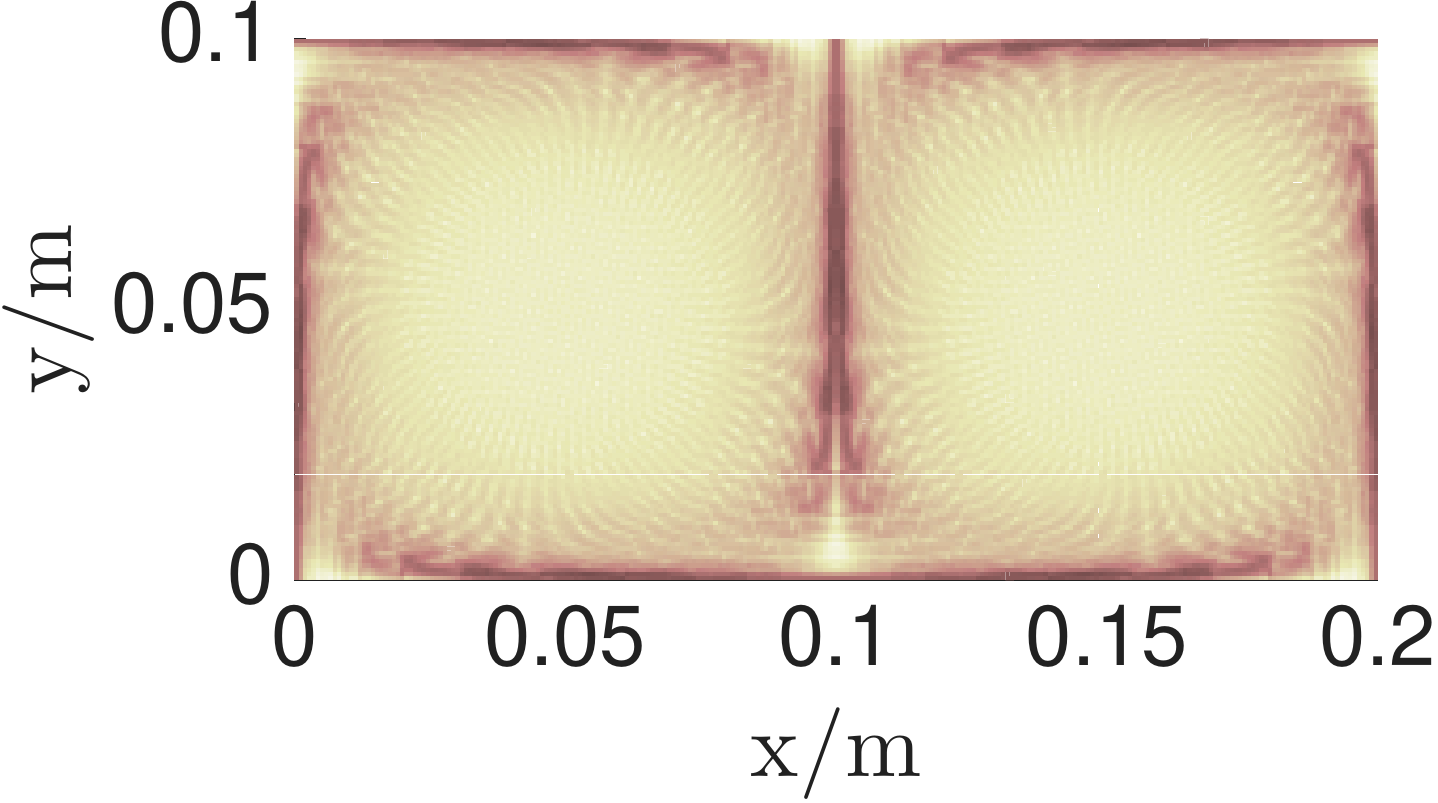} &
   \includegraphics[height=0.09\textheight]{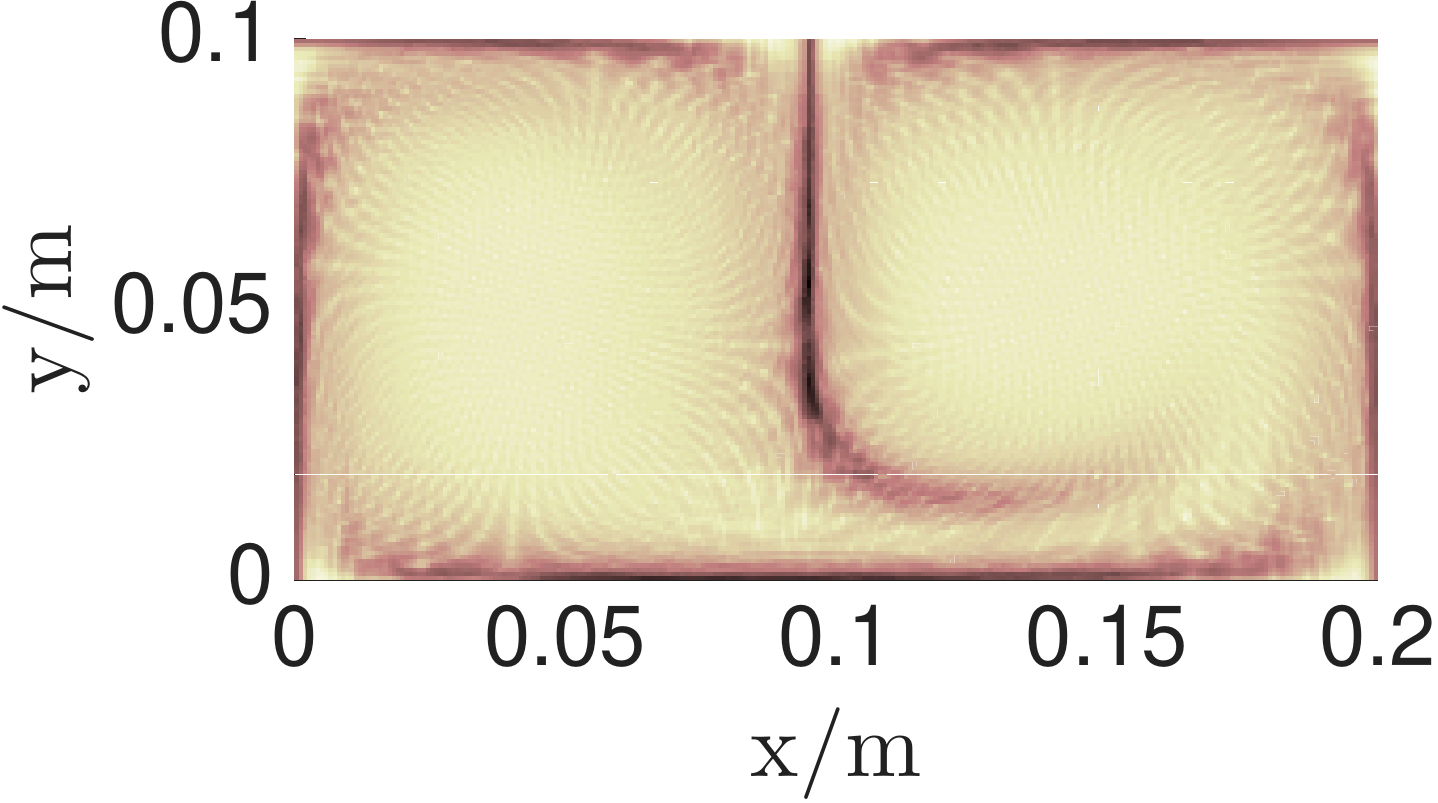} &
   \includegraphics[height=0.09\textheight]{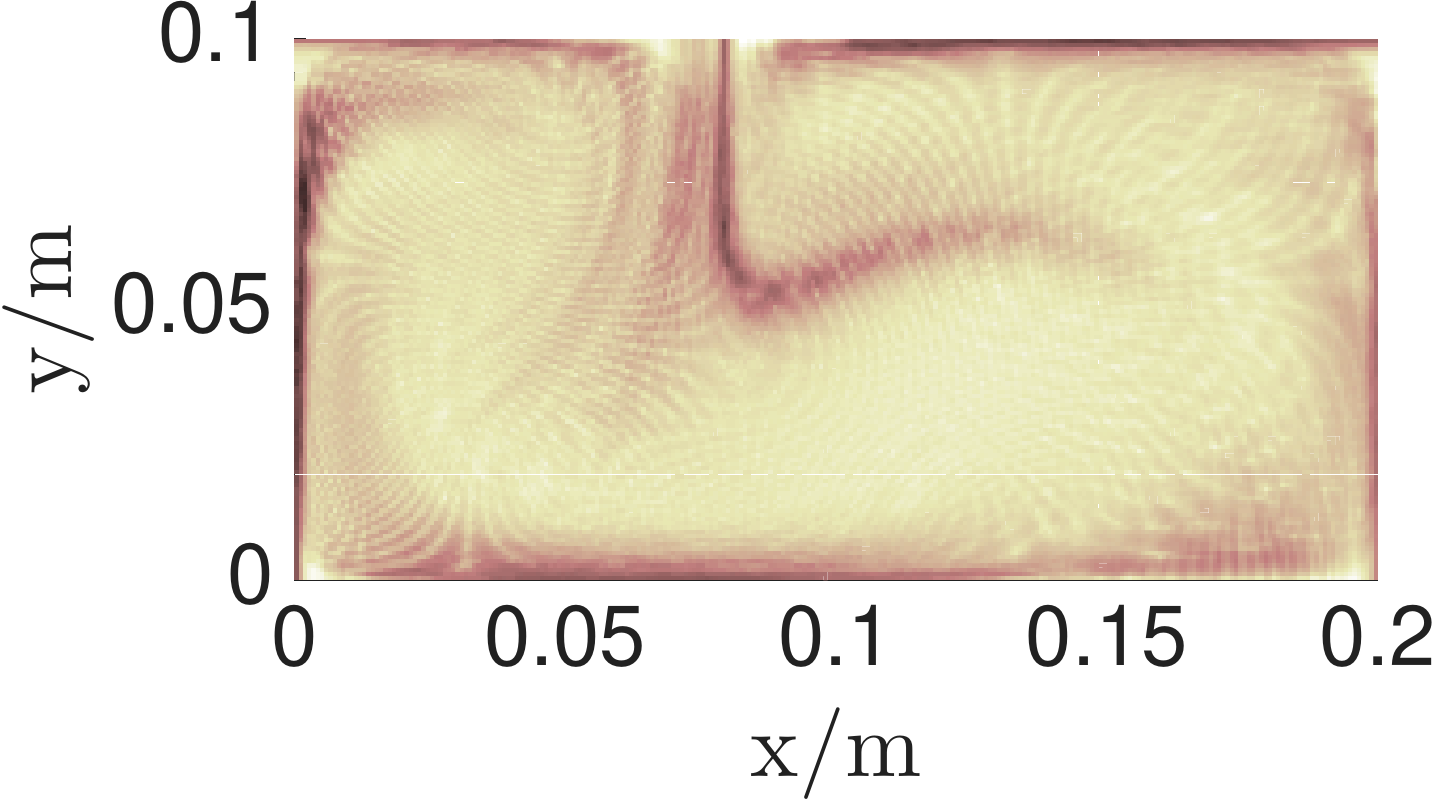}
   \includegraphics[height=0.09\textheight]{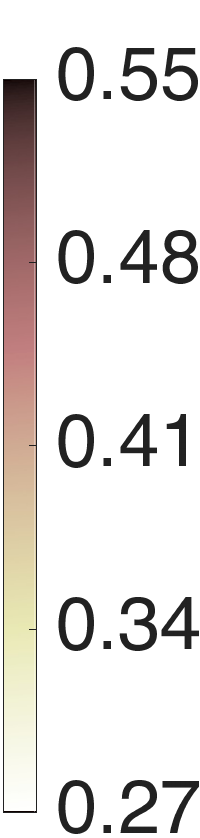} \\
    {\text{j)}} & {\text{k)}} & {\text{l)}} \\
  \includegraphics[height=0.09\textheight]{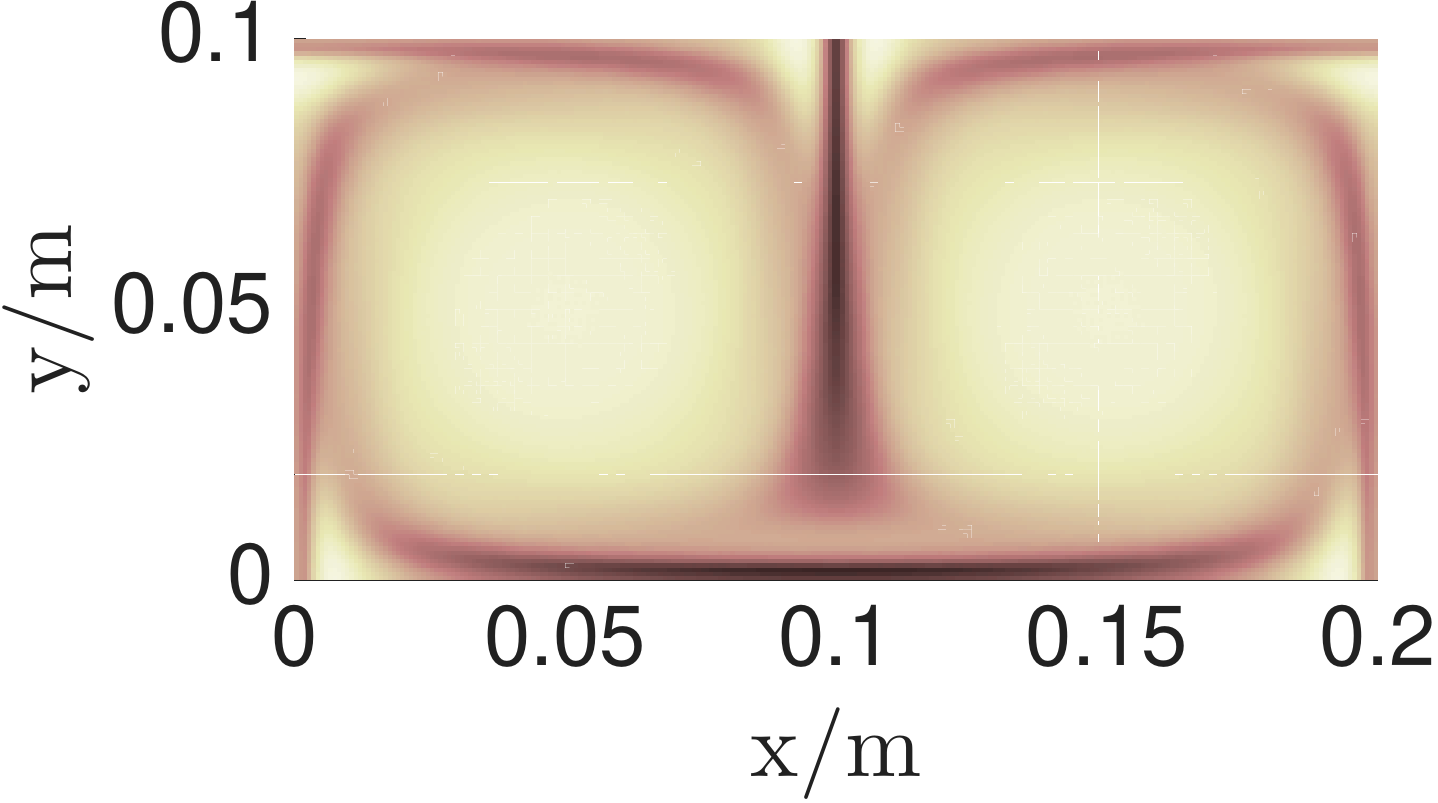}&  \includegraphics[height=0.09\textheight]{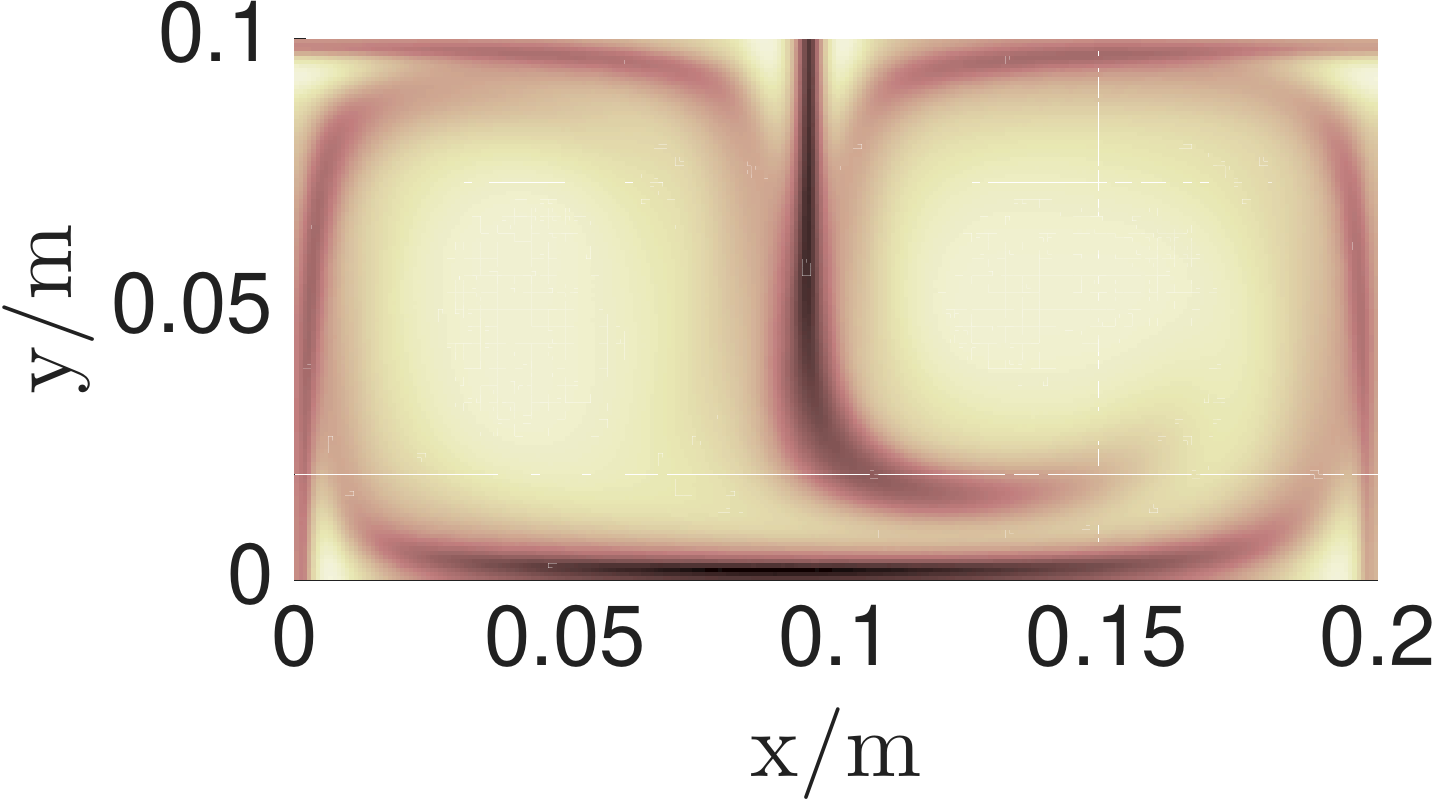} &   \includegraphics[height=0.09\textheight]{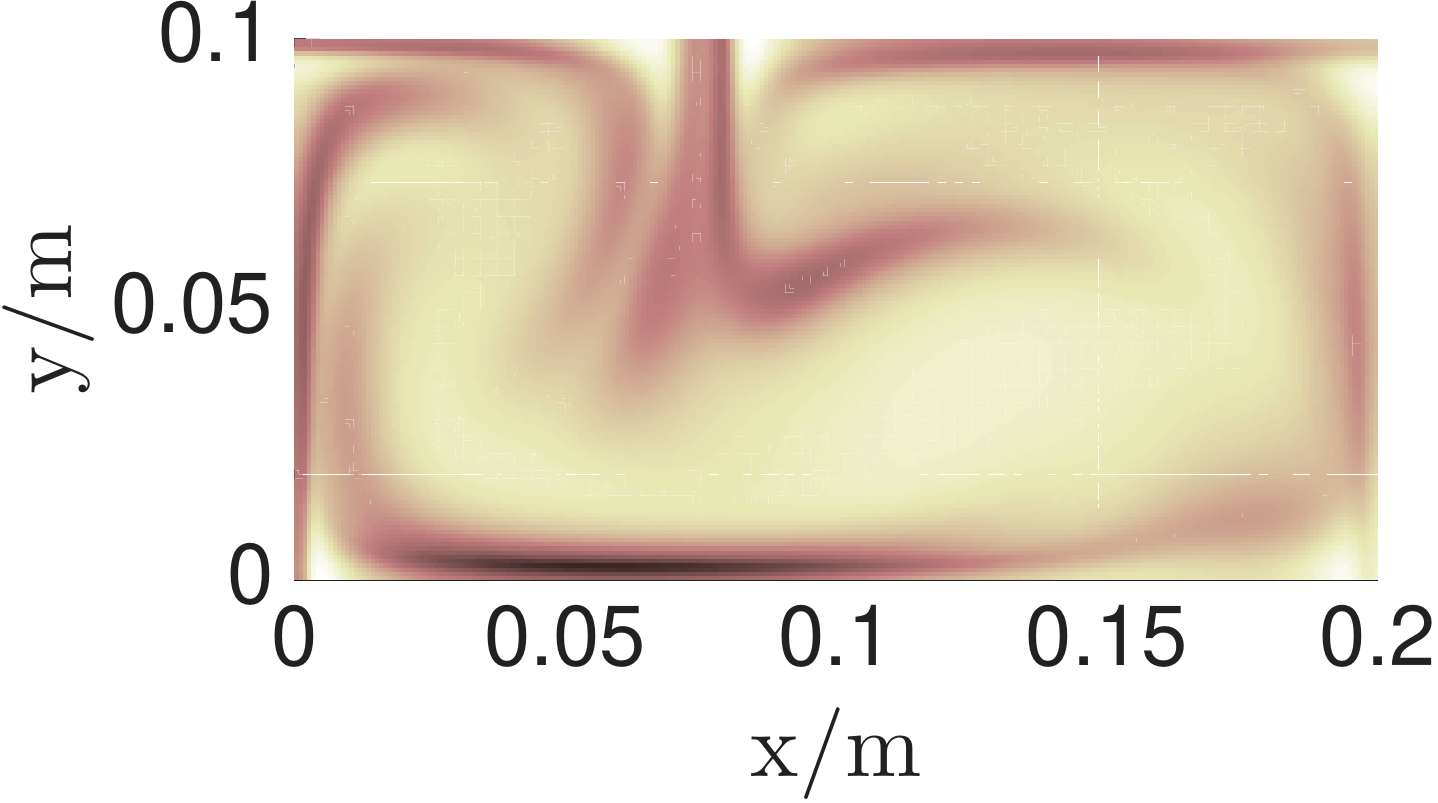}
  \includegraphics[height=0.09\textheight]{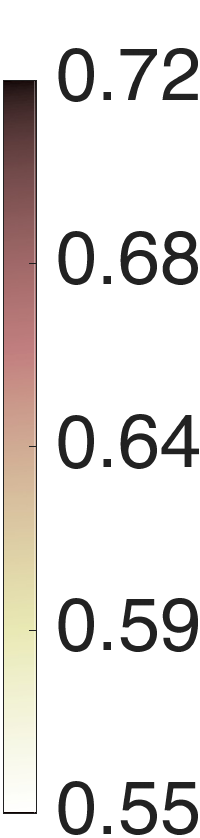} \\
    {\text{m)}} & {\text{n)}} & {\text{o)}} \\
\includegraphics[height=0.09\textheight]{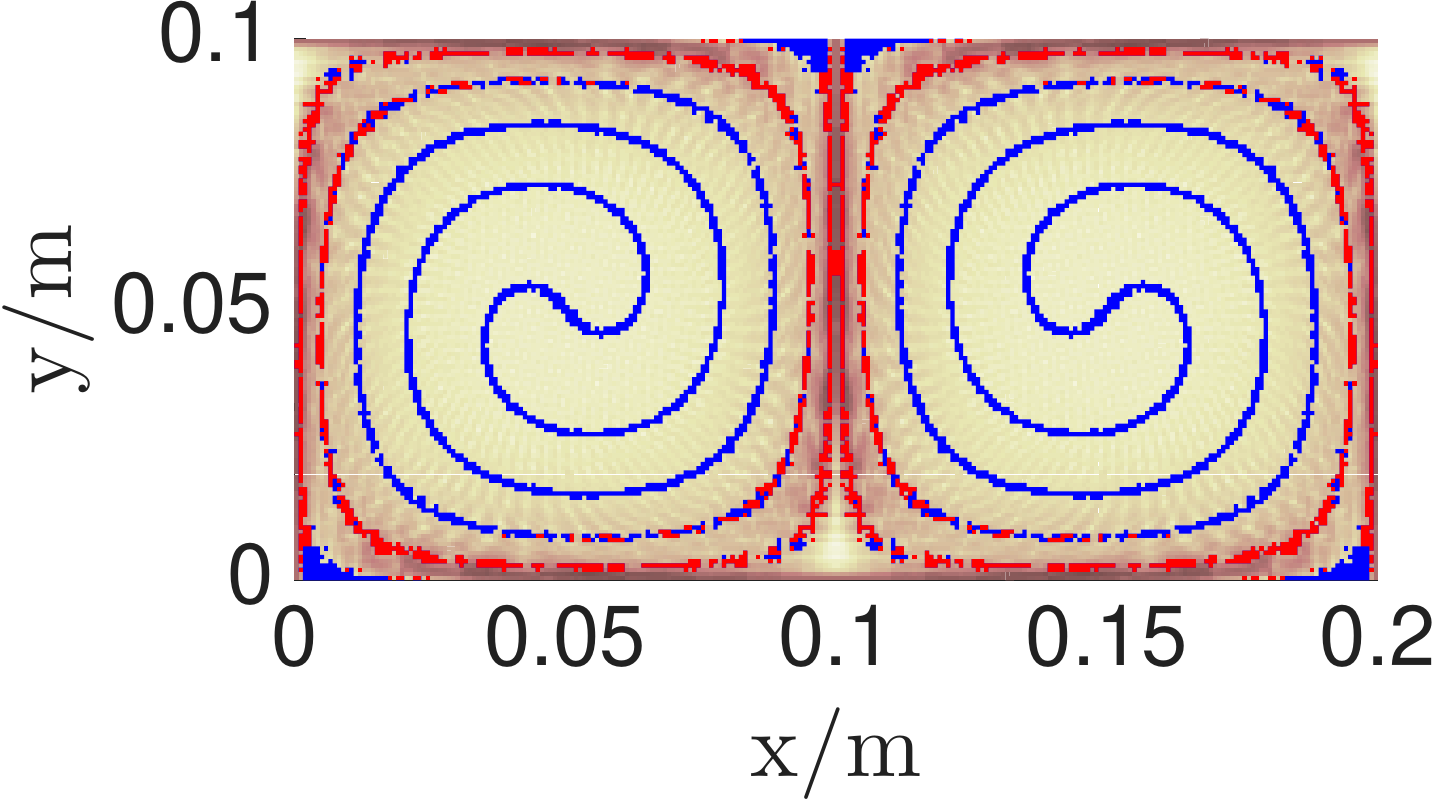} &     \includegraphics[height=0.09\textheight]{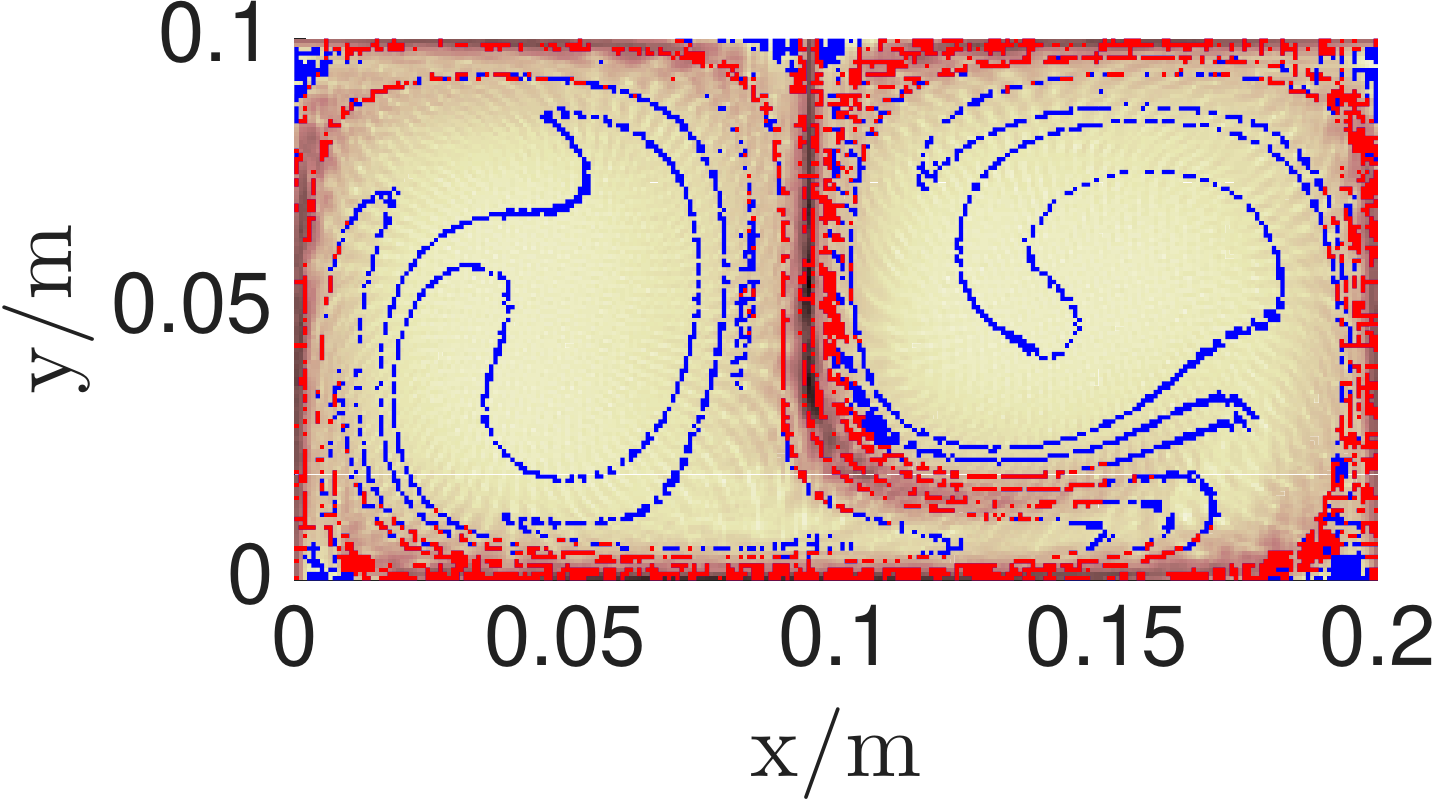} &
\includegraphics[height=0.09\textheight]{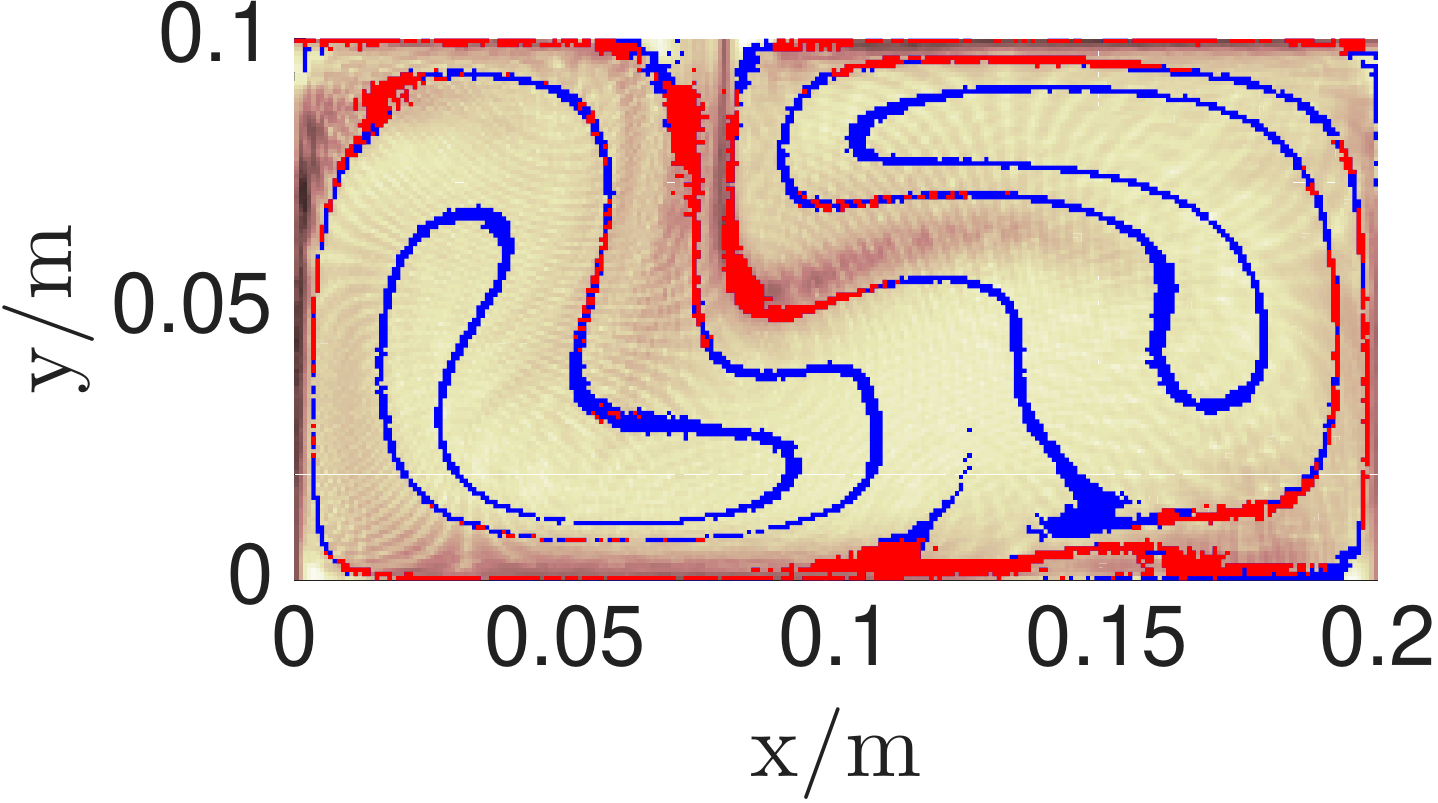}    \includegraphics[height=0.09\textheight]{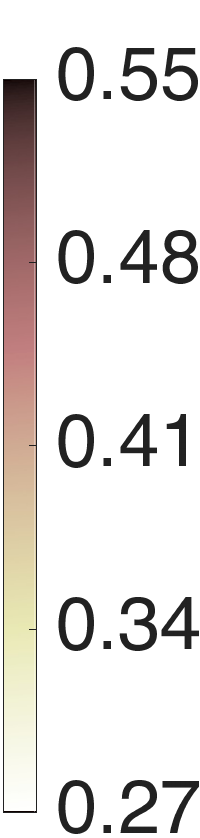} \\
    {\text{p)}} & {\text{q)}} & {\text{r)}} \\
  \includegraphics[height=0.09\textheight]{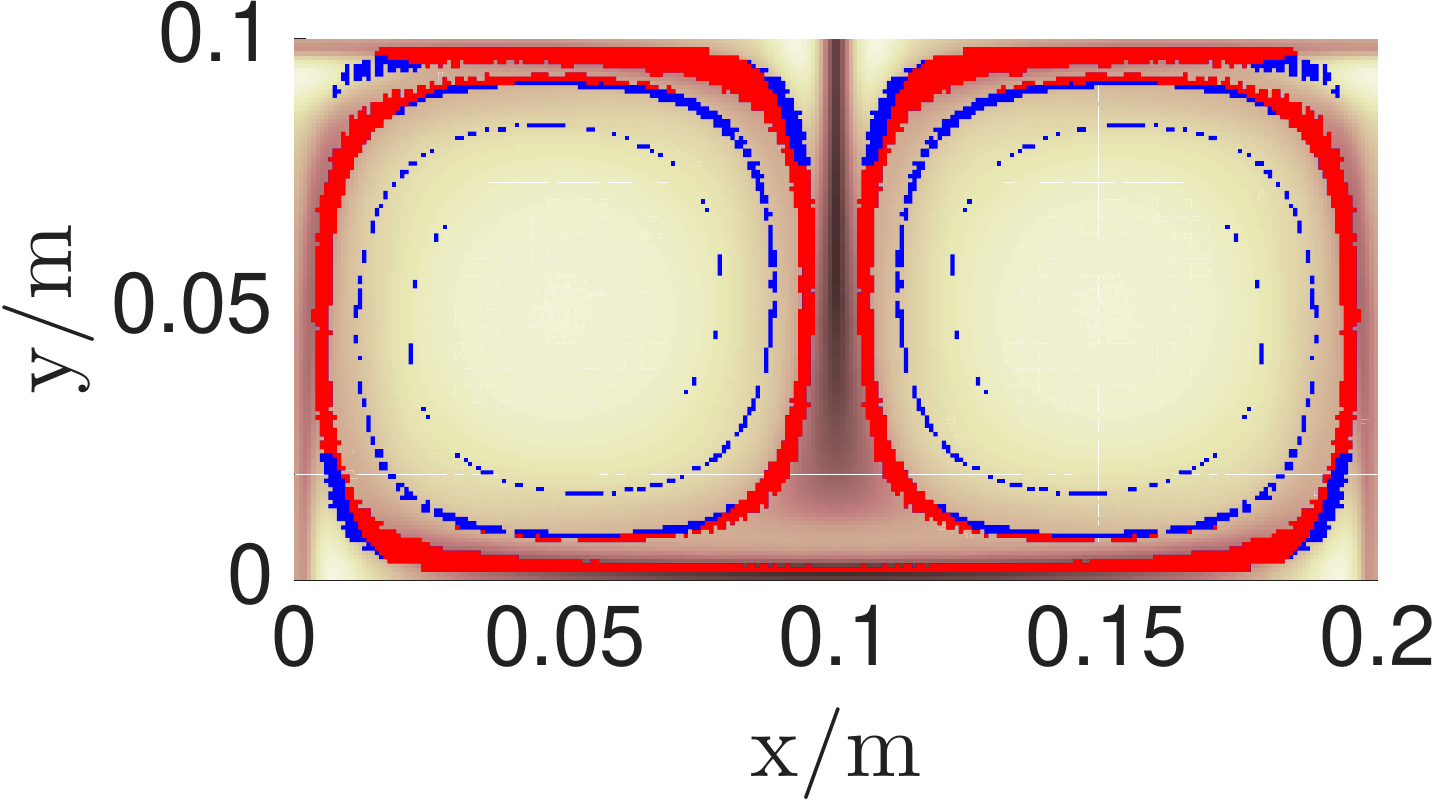} &
\includegraphics[height=0.09\textheight]{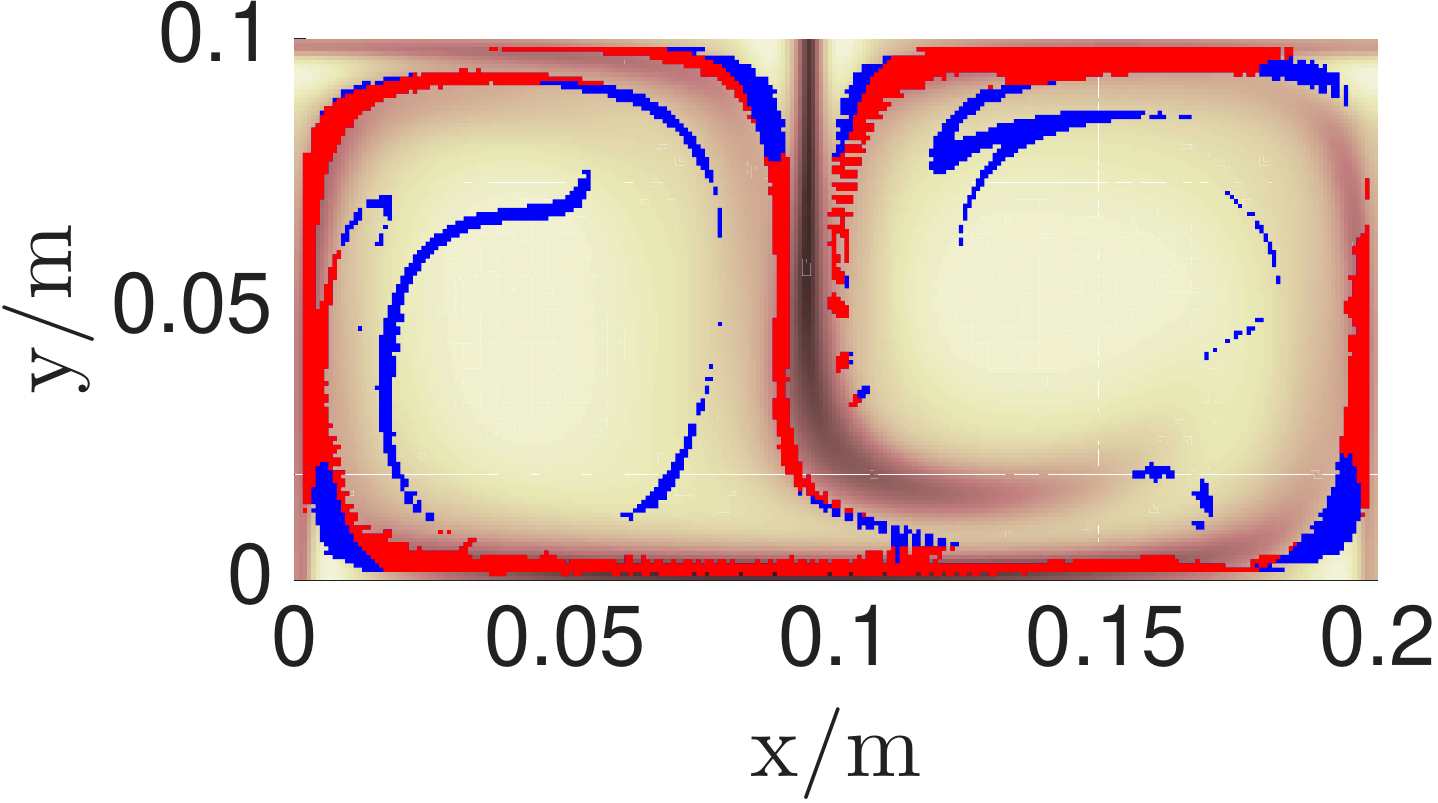} &
  \includegraphics[height=0.09\textheight]{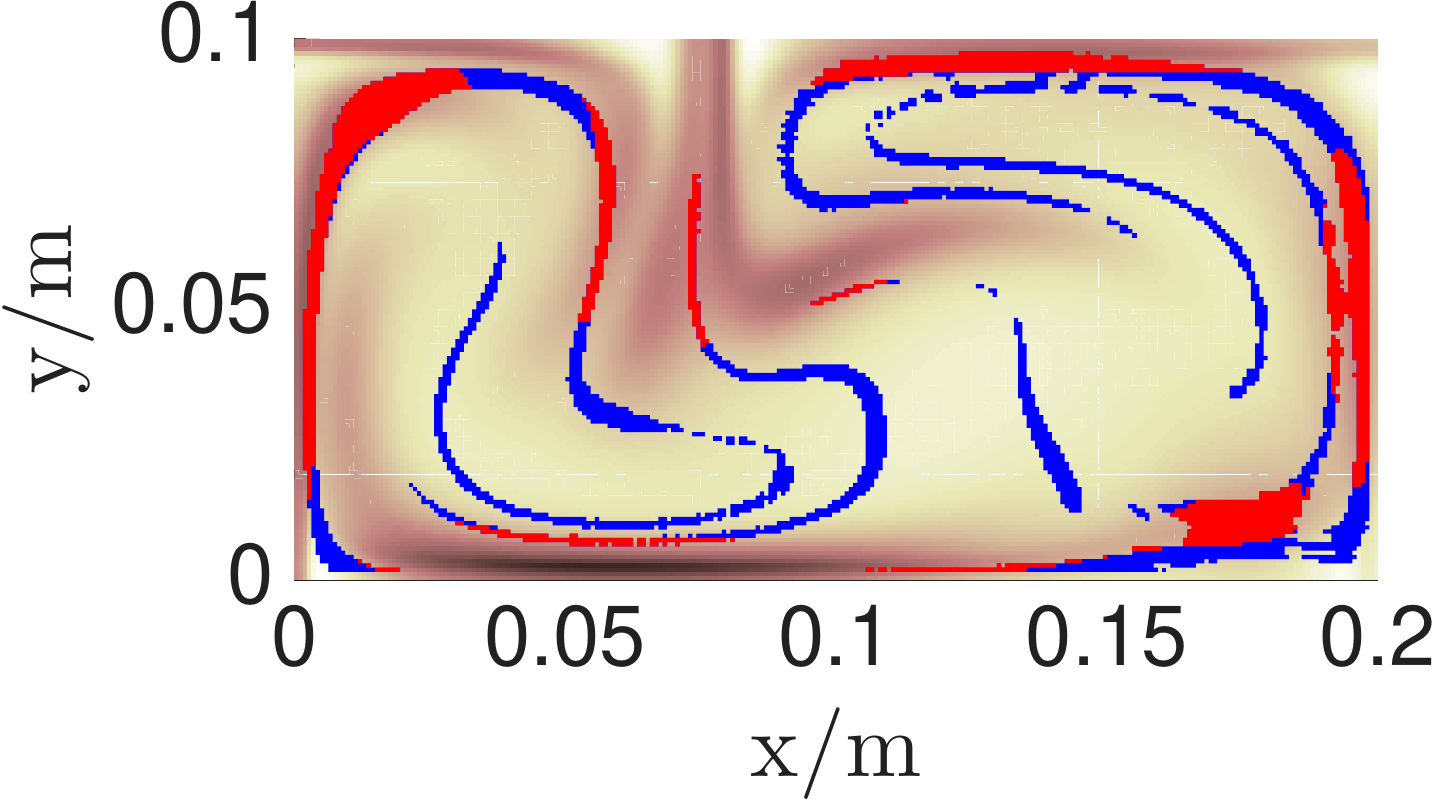}
  \includegraphics[height=0.09\textheight]{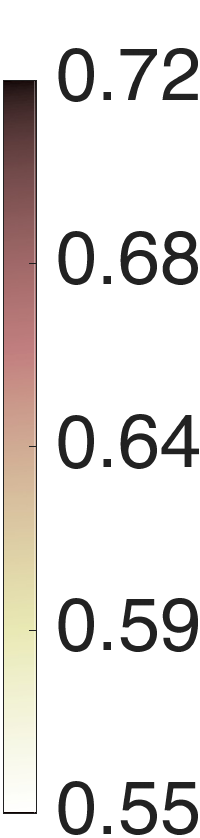}
  \end{tabular}
  \end{center}
    \caption{Concentrations $\bm{C}_C$ [\unit{\mol\per\l}] at $t=4\unit{\s}$ (a-f) and backward-time FTE at $t=1\unit{\s}$ (g-l) for the DG without diffusion (a-c,g-i,m-o) and with diffusion (d-f,j-l,p-r) for $\delta=0$  (column 1), $\delta=0.1$ (column 2) and $\delta=0.5$ (column 3). The highest reaction front (blue and red) is shown at $t=4\unit{\s}$, where the red boxes represent the matching of both high reactions and b-FTE (m-r).}
    \label{fig:DG_reaction_FTE}
\end{figure*}
To demonstrate that there are matches between high reactions and high b-FTE, we will consider the 5,000 highest reaction values and the 10,000 highest FTE in backward time values (see fig.~\ref{fig:DG_reaction_FTE}).
Without diffusion, the strongest reactions occur primarily in boxes with high backward FTE and outside or at the edge of the nearly invariant sets in the vortex center area.
At time $t=4\unit{\s}$, the concentration of C in the DG reactor model with diffusion is comparatively low around $x=0.1\unit{m}$. Although there are high b-FTE values in this area, there are also low reaction rates, consistent with the low concentration level.
The red boxes indicate the correspondence between high reaction and high b-FTE values for all systems. The agreement of high 
b-FTE values and high reactions emphasizes the importance of the FTE field in reactor design. However, depending on the initial distribution reactions also occur in regions with low b-FTE values.

\subsection{Mixing and reaction in an ABC flow reactor model}

We discretize the ABC flow reactor domain, which has an edge length of $0.2\pi\unit{\m}$, into $2^{18}$ boxes (see fig.~\ref{fig:ABC_mixing}). Each box $B_i$ is a cube with a side length of $d = 0.00982\unit{\m}$ and contains 125 test points on a uniform inner grid.
At time $t_0=0\unit{\s}$, the initial distribution is $c_{A}=1\unit{\mol\per\l}$ per box in the upper half of the reactor and {$c_{B}=1\unit{\mol\per\l}$ per box in the lower half (see fig.~\ref{fig:ABC_mixing}).
We simulate diffusion by approximating an $\epsilon$-ball for each test point with radius $\epsilon=0.025\unit{\m}$ by 125 grid points, so that $\epsilon \approx 5r$, with a radius $r= \frac{1}{2} d=0.00491\unit{\m}$ per box.
We calculate the transition matrix $\bm{P}$ (or $\bm{P}_{D}$) for the ABC flow for $\tau=1\unit{\s}$.

When we mix substance A for each time step $\tau=1\unit{\s}$ with substance B and compare the product distribution at time $t=2\unit{\s}$, again mixing is enhanced by diffusion (see fig.~\ref{fig:ABC_mixing}).
\begin{figure}
\begin{center}
\begin{tabular}{lll}
    {\text{a)}} & {\text{b)}} & {\text{c)}} \\
 \includegraphics[height=0.09\textheight, valign=t]{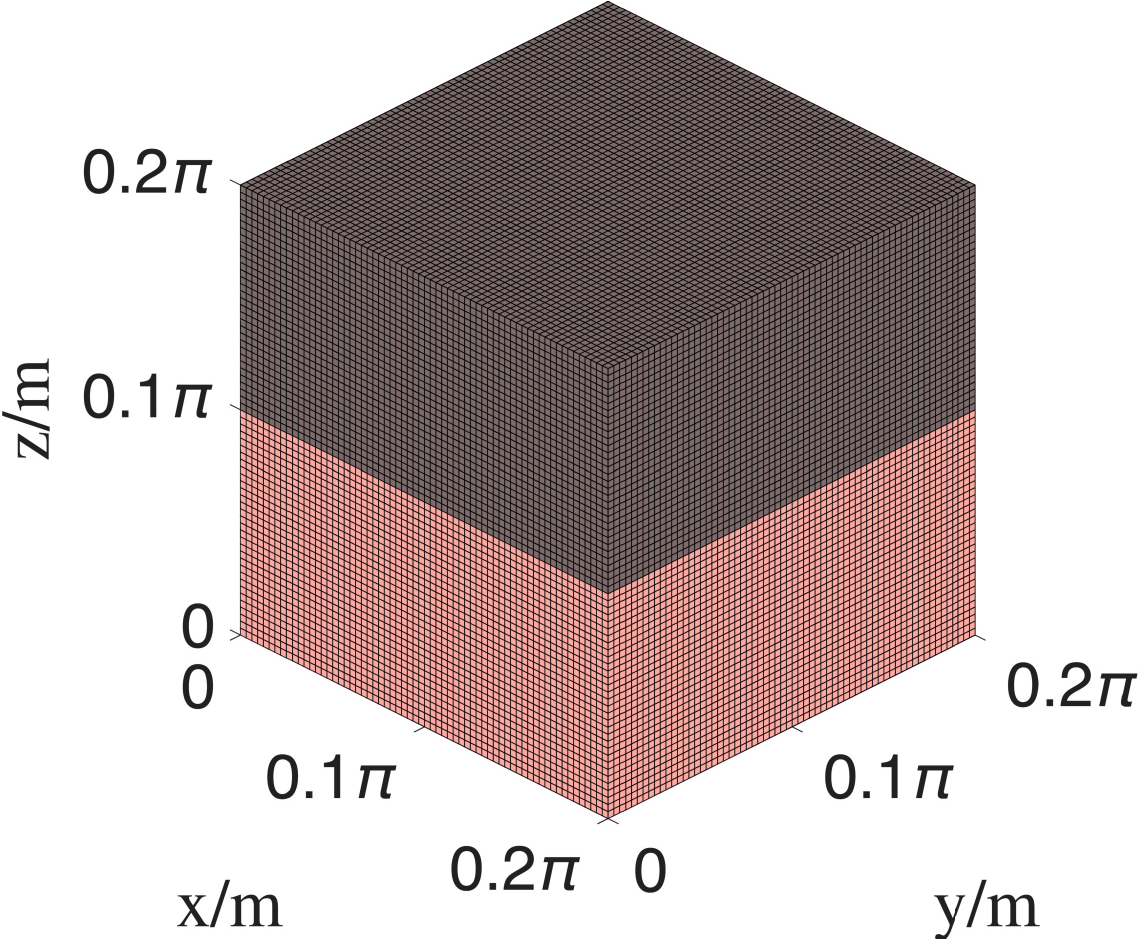} &
  \includegraphics[height=0.09\textheight, valign=t]{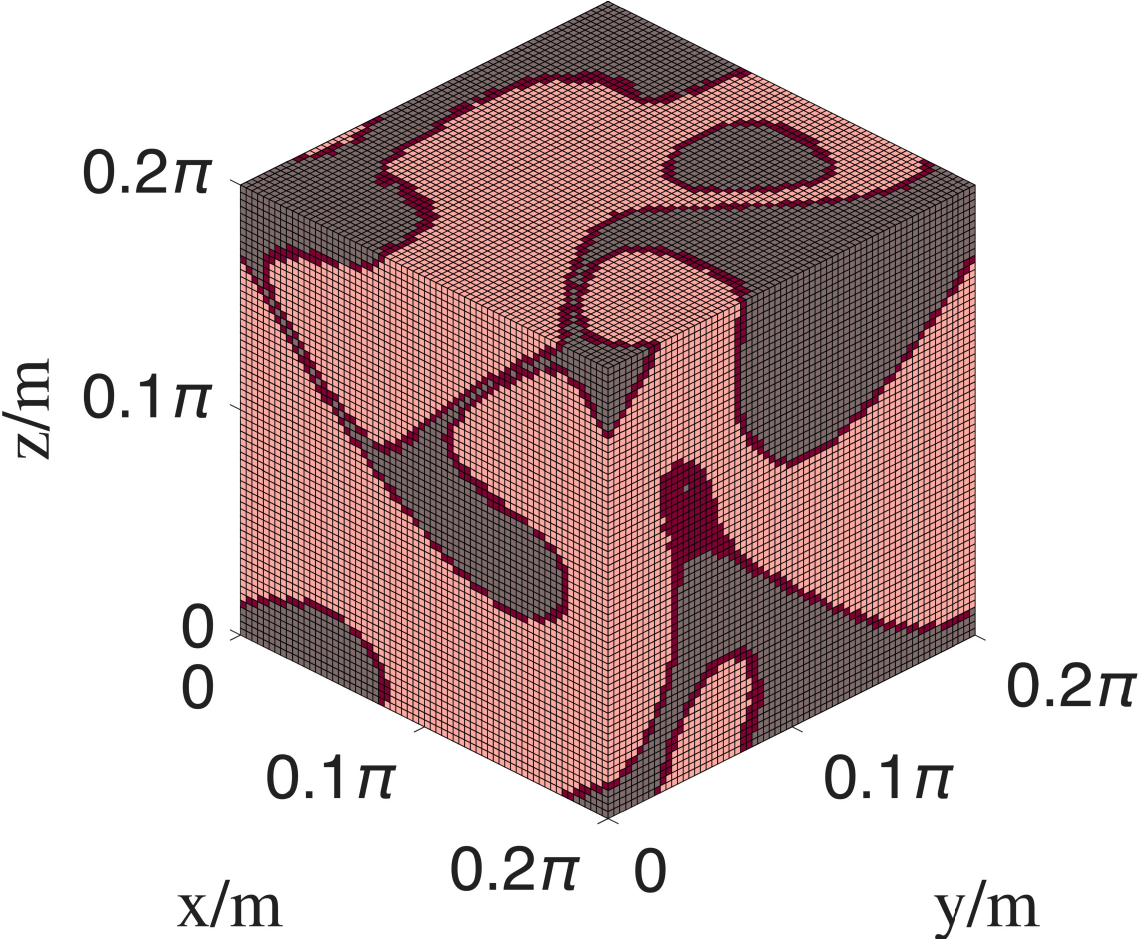} &   \includegraphics[height=0.09\textheight, valign=t]{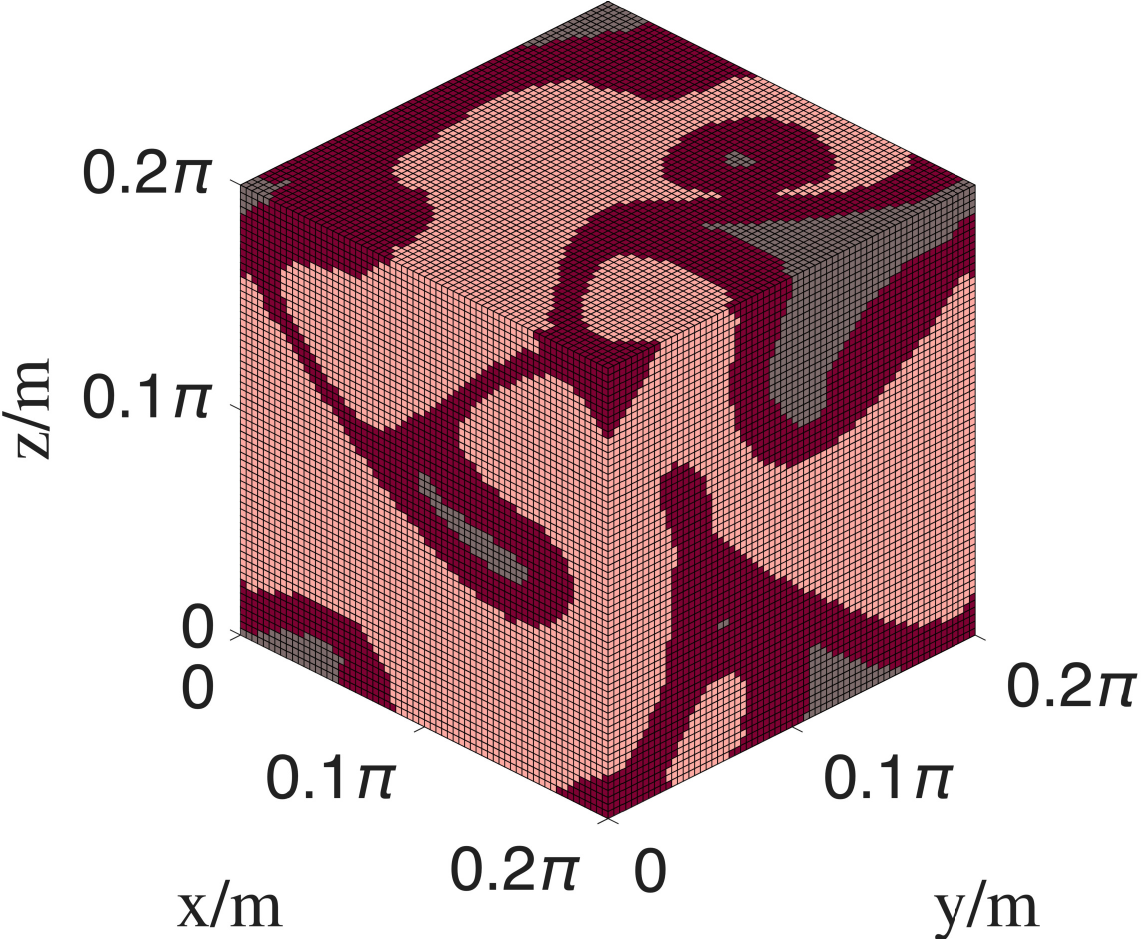}   \includegraphics[height=0.08\textheight, valign=t]{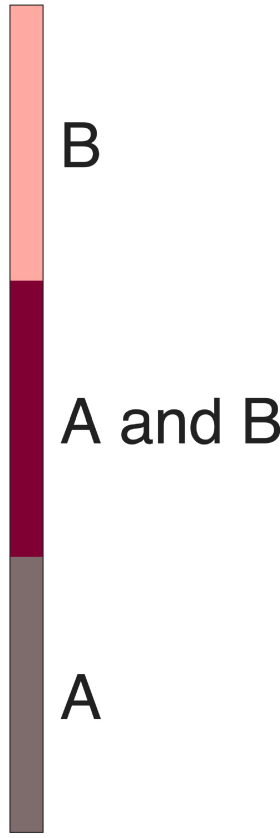}
  \end{tabular}
  \end{center}
    \caption{Initial distribution (a) and product distribution after mixing (no reaction) for the ABC flow without diffusion (b) and with diffusion (c) at $t=2\unit{\s}$.}
    \label{fig:ABC_mixing}
\end{figure}

We simulate the simple chemical reaction in the ABC flow reactor model with an initial distribution of fluid A in the upper half of the reactor and fluid B in the lower half, with each box $B_i$ containing $1\unit{\mol\per\l}$ of the substance present there (see fig.~\ref{fig:ABC_mixing}).
For each time step $\tau = 1.0\unit{\s}$, the particles are transported and then react using the Euler discretization \eqref{eq:simplereact} with time-stepping $h=0.01\unit{\s}$.

If we compare the reaction without and  with diffusion, we see that the reaction with diffusion initially occurs faster and is completed most quickly (see fig.~\ref{fig:ABC_total}).
For reaction rate constant $k_r=0.1\unit{\l\per\mol\per\s}$ and the maximum initial concentration $c_{A_0}=c_{B_0}=1\unit{\mol\per\l}$, this implies the reaction time scale $\tau_{r}=\frac{1}{0.1\unit{\l\per\mol\per\s}\cdot 1\unit{\mol\per\l}}=10\unit{\s}$. For the flow time scale $\tau_{f}\approx 2.7\unit{\s}$, the Damköhler number is thus $Da \approx 0.27$, indicating that advection is faster than the chemical reaction. Because the ABC flow is three-dimensional and chaotic, the reaction front between A and B becomes severely stretched and folded. $Da \approx 0.27$ indicates that the chemical reaction occurs during this complex deformation, resulting in a broad, dispersed reaction zone (see fig.~\ref{fig:ABC_reaction_FTE}). Because the Damköhler number is only slightly larger for the ABC flow reactor model than for the DG reactor model, both systems exhibit a similar increase in total concentrations $C_C$ [\unit{\mol\per\l}] (see fig.~\ref{fig:ABC_total}).
\begin{figure}
    \centering
    \includegraphics[width=0.7\linewidth]{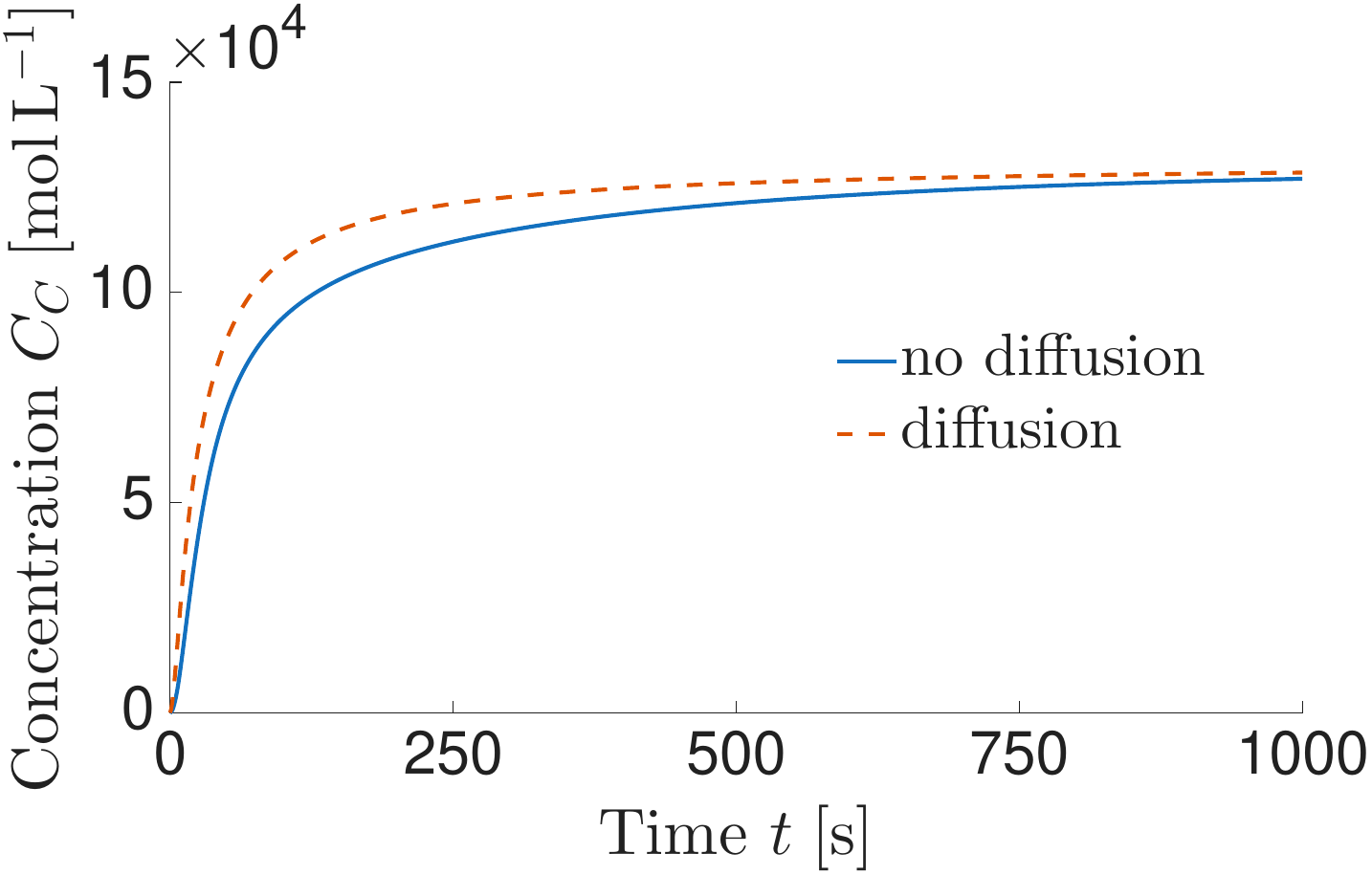}
    \caption{Total concentrations $C_C$ [\unit{\mol\per\l}] for the  ABC flow.}
    \label{fig:ABC_total}
\end{figure}

The distribution of different concentrations (see fig.~\ref{fig:ABC_reaction_FTE}) suggests that higher concentrations are present in certain boxes $B_i$ with high b-FTE values, i.e., in the ridges of the b-FTE field.
At the beginning of the reaction process, the reaction front and the concentration distribution of C are nearly identical. Therefore,  we only show the concentration distribution.

The backward time FTE fields highlight attracting coherent structures (see fig.~\ref{fig:ABC_reaction_FTE}).
The b-FTE fields differ for the ABC flow depending on whether diffusion is applied or not, where higher values are observed for the diffusive case.
Although the range of $FTE_b(\cdot,2)$ is greater compared to the $FTE_b(\cdot,5)$ as shown in figure 11 in reference \onlinecite{FROYLAND20121612}, the same coherent structures are identified.
The attractive coherent structures in the b-FTE fields correspond to the hyperbolic repelling LCS in the ABC flow, as shown in figure 4 in references \onlinecite{10.1063/5.0044151} and figure 6 in \onlinecite{Mowlavi_2022}.
\begin{figure}
    \centering
        \begin{tabular}{cccc}
        \makebox[0.39\linewidth][l]{\text{a)}} & &
        \makebox[0.39\linewidth][l]{\text{b)}} \\
        \includegraphics[height=0.12\textheight]{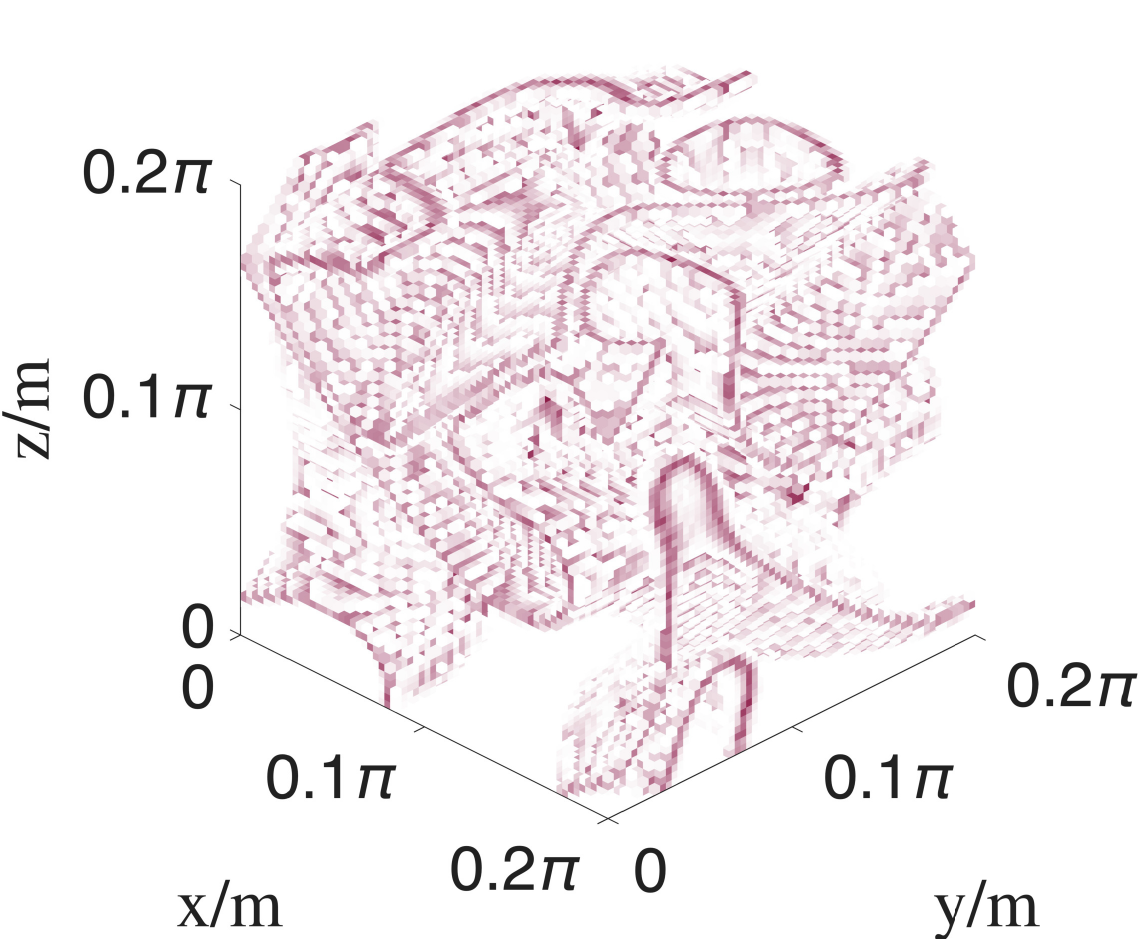} & &
        \includegraphics[height=0.12\textheight]{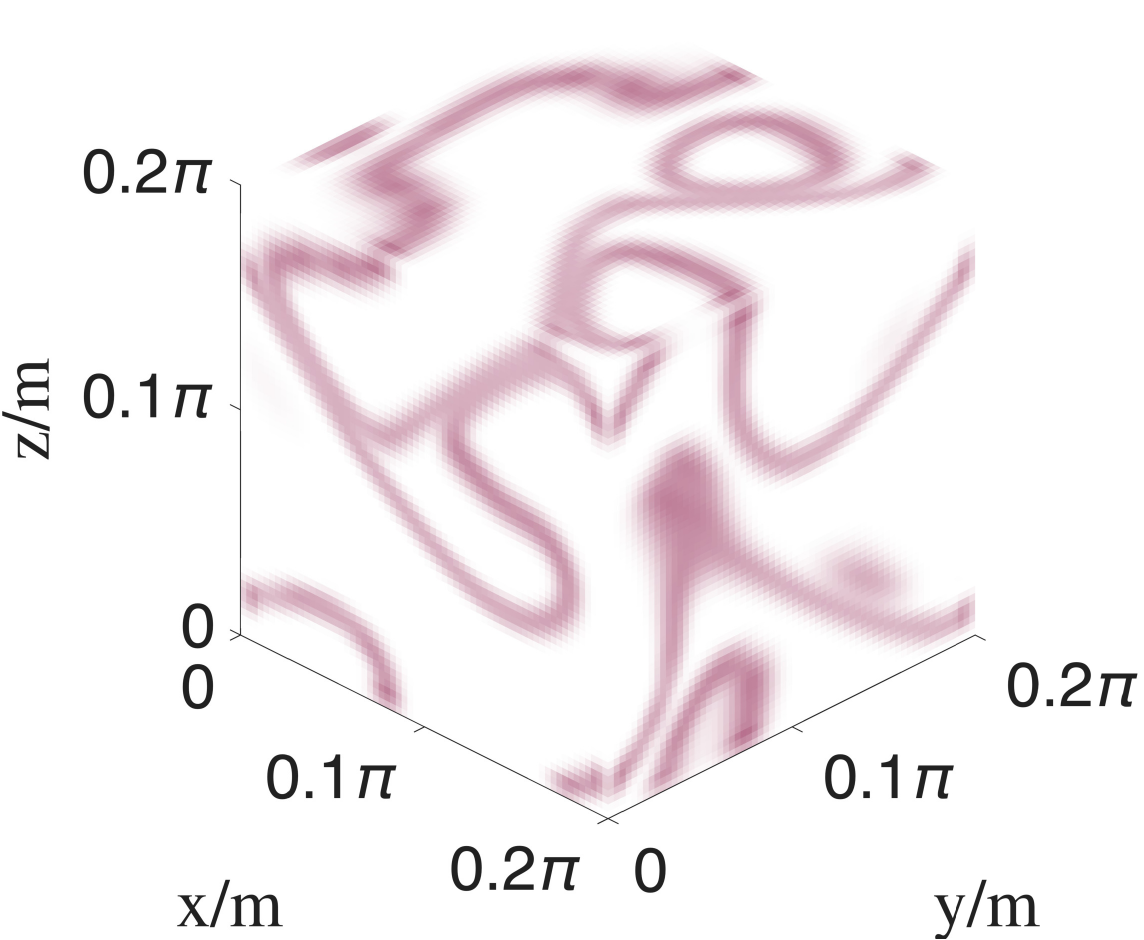} &
        \includegraphics[height=0.12\textheight]{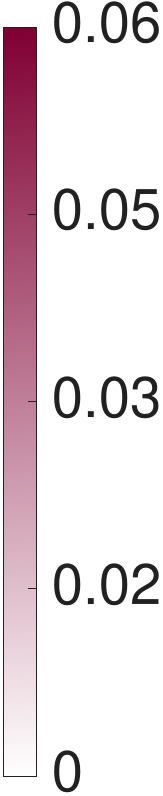} \\
        \makebox[0.39\linewidth][l]{\text{c)}} & &
        \makebox[0.39\linewidth][l]{\text{d)}} \\
        \includegraphics[height=0.12\textheight]{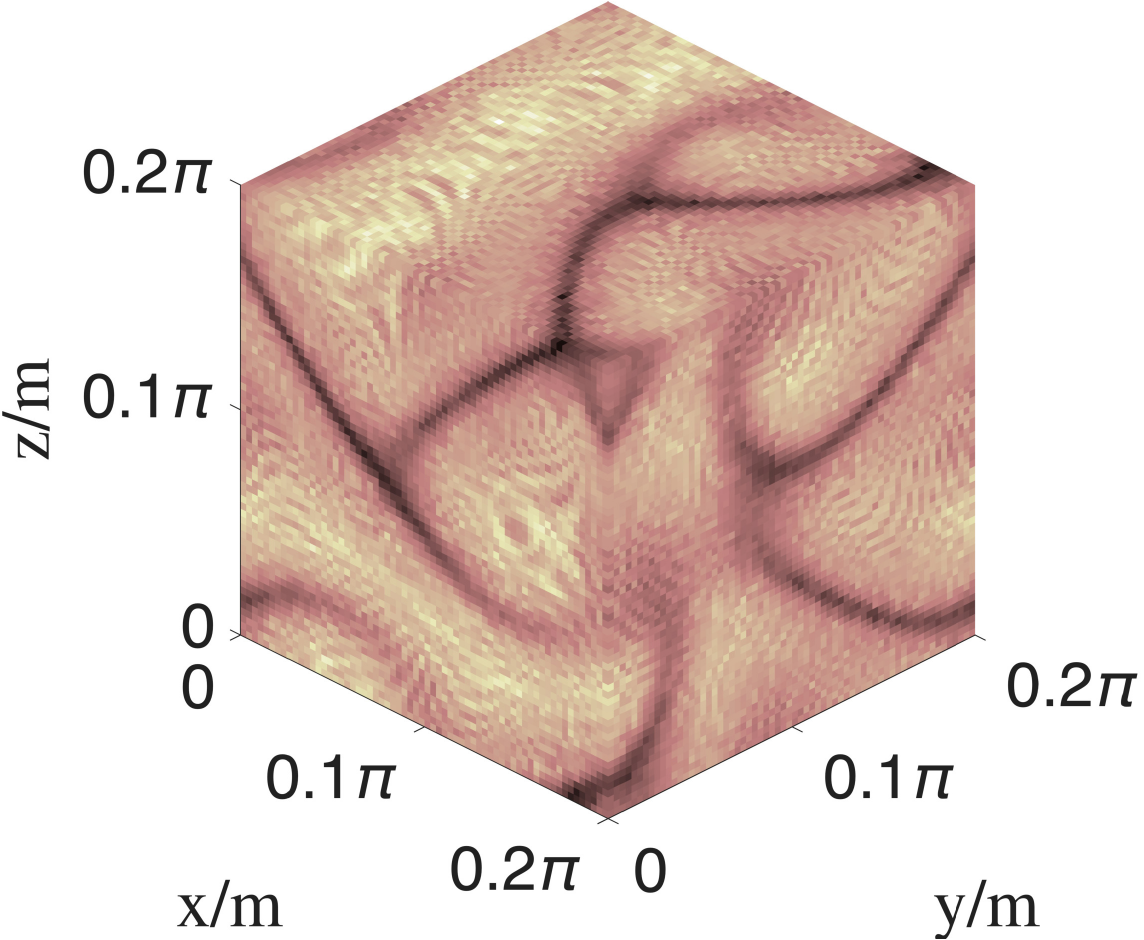} &
        \includegraphics[height=0.12\textheight]{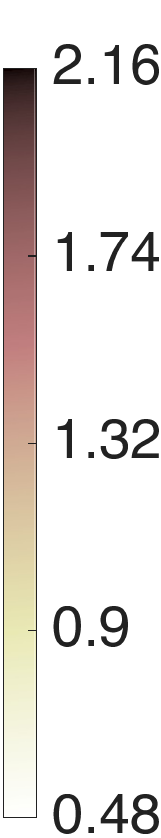} &
        \includegraphics[height=0.12\textheight]{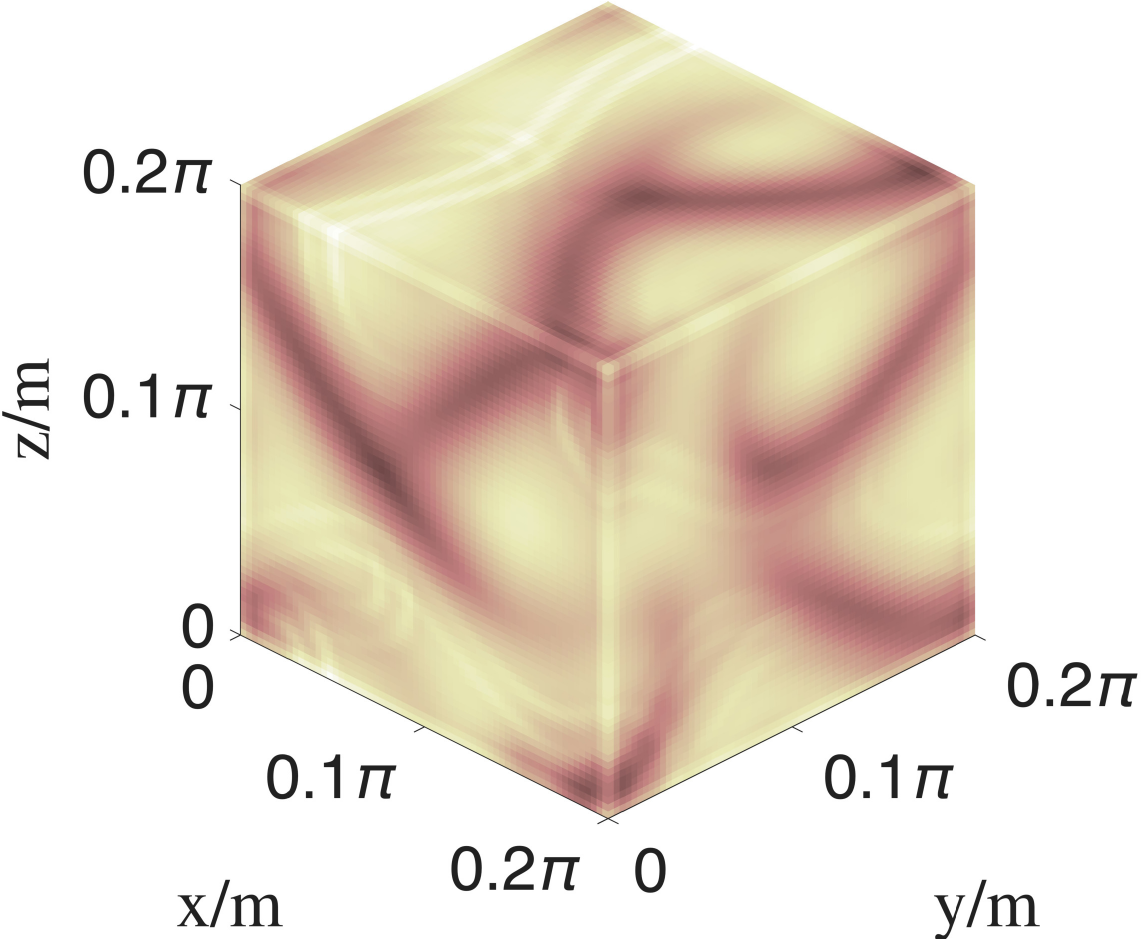} &
        \includegraphics[height=0.12\textheight]{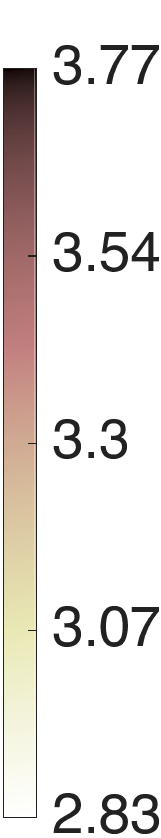} \\
        \makebox[0.39\linewidth][l]{\text{e)}} & &
        \makebox[0.39\linewidth][l]{\text{f)}} \\
        \includegraphics[height=0.12\textheight]{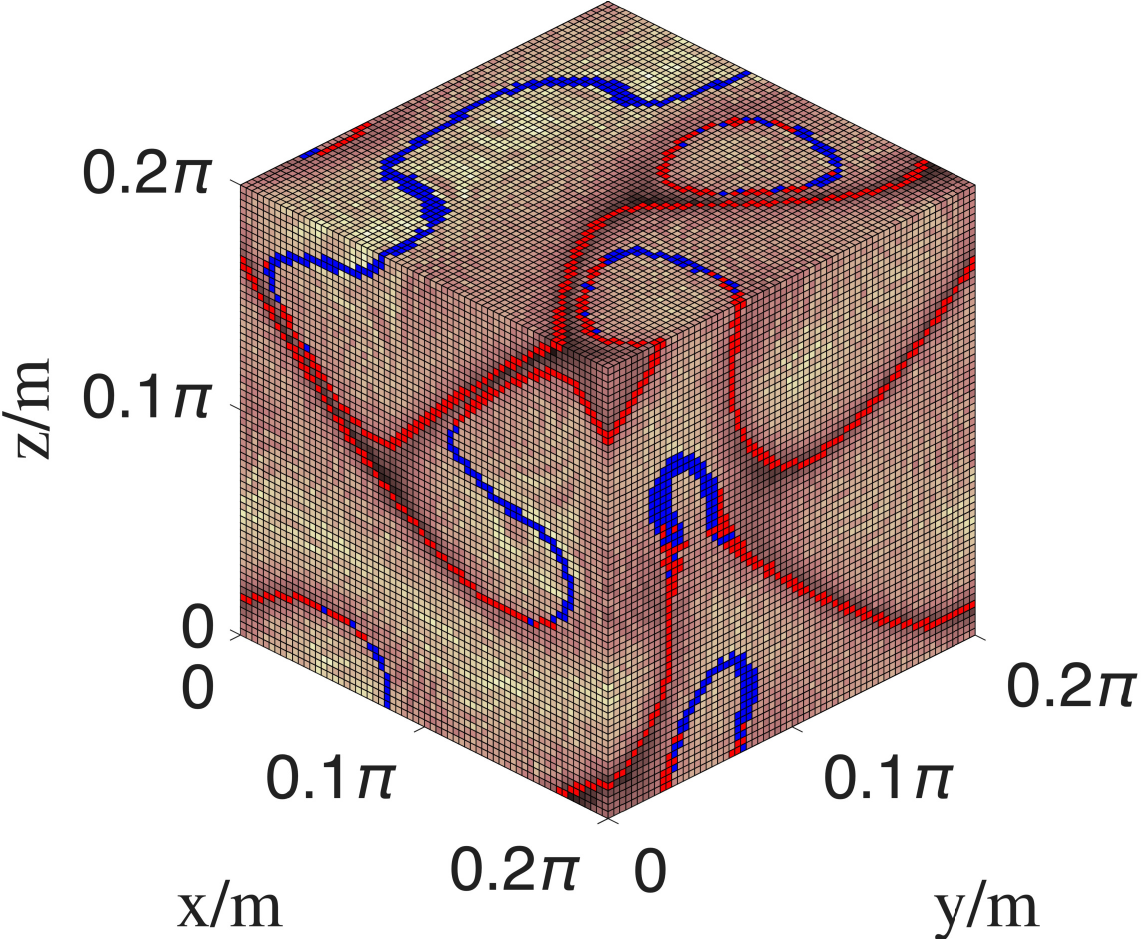} &
        \includegraphics[height=0.12\textheight]{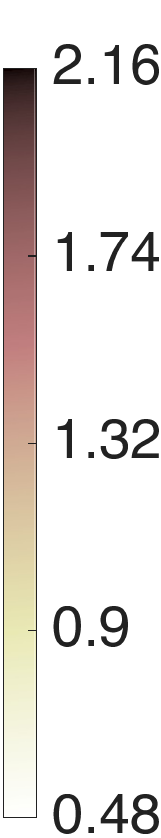} &
        \includegraphics[height=0.12\textheight]{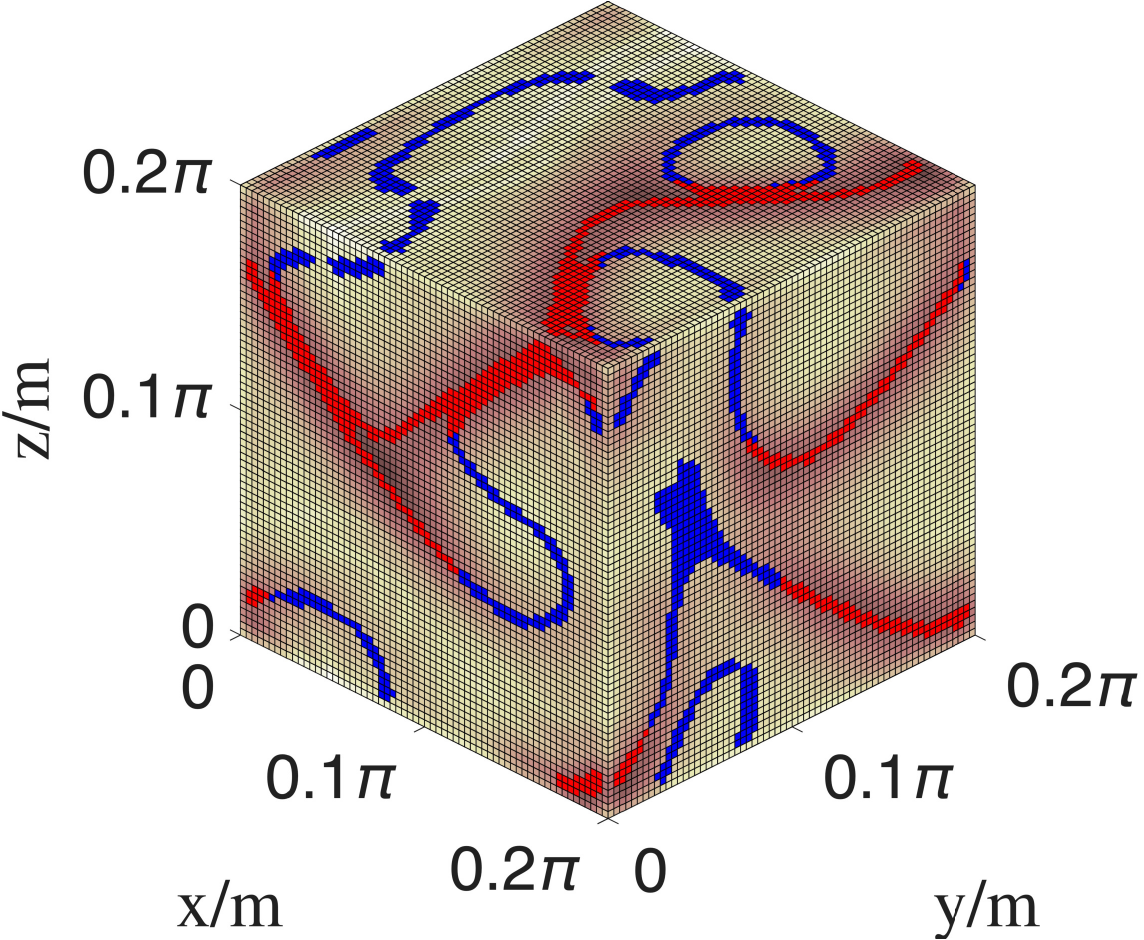} &
        \includegraphics[height=0.12\textheight]{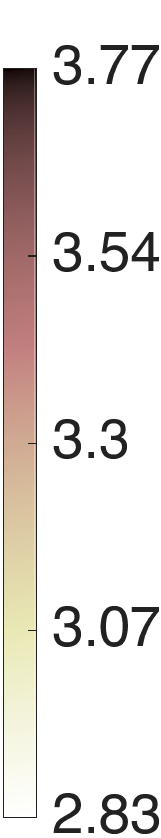}
        \end{tabular}
        \caption{Concentrations $\bm{C}_{C}$ [\unit{\mol\per\l}]  at t=2\unit{\s}$,$(a-b) and $FTE_b(\cdot,2)$ for the ABC flow without (a,c,e) and with diffusion(b,d,f). The highest reaction front (blue and red) is shown, where red boxes represent the matching of both high reactions and b-FTE.}
        \label{fig:ABC_reaction_FTE}
\end{figure}
To demonstrate that there are matches between high reactions and high backward time FTE, we will consider the 75,000 highest b-FTE values and we differentiate between the 25,000 highest reaction values without diffusion and the 75,000 highest reaction values with diffusion.
We found both the highest b-FTE and the highest reactions near the center of the diffusive ABC flow reactor model, which is not visible from the outside.
Therefore, we consider 75,000 instead of 25,000 highest reaction values for the diffusive system to make the correlation visible from the outside of the ABC flow reactor model (see fig.~\ref{fig:ABC_reaction_FTE}).
The red boxes indicate the correspondence between high reaction and high b-FTE values for both systems. The agreement of high b-FTE values and high reactions emphasizes the importance of the b-FTE field in reactor design. However, depending on the initial distribution, reactions may also occur in regions with low backward FTE, which corresponds to elliptic regions with longer residence time.

\subsection{Mixing and reaction in a STR model}

We model the STR by discretizing a cuboid with height $l$, width $2r$, and depth $2r$ into $2^{16}$ boxes. Each box $B_i$ is a cuboid with side lengths $d_x=\qty{0.004}{\m}$, $d_y=\qty{0.0038}{\m}$ and $d_z=\qty{0.004}{\m}$. We then remove boxes that do not contain trajectories and obtain a STR with 49,983 boxes whose union approximate the desired cylindrical shape (see fig.~\ref{fig:STR_mixing}). Initially, the median is 77 test points in each box $B_i$ from which trajectories start.
In the closed STR model, the initial distribution at time $t_0$ is $c_{A}=10\unit{\mol\per\l}$ per box in a sphere with a radius of $\qty{0.02}{\m}$, located in the center of the upper half of the reactor, and $c_B=10\unit{\mol\per\l}$ per box in a sphere with a radius of $\qty{0.02}{\m}$, located in the center of the lower half (see fig.~\ref{fig:STR_mixing}).

We simulate diffusion by approximating an $\epsilon$-ball for each test point with radius $\epsilon=\qty{0.01}{\m}$ by 125  grid points, so that $\epsilon \approx 5r$, with a radius $r \approx \qty{0.002}{\m}$ ($r_x=\qty{0.002}{\m}$, $r_y=\qty{0.0019}{\m}$ and $r_z=\qty{0.002}{\m}$) per box.
We calculate the transition matrix $\bm{P}$ (or $\bm{P}_{D}$) for one rotation, i.e.\ $\tau=\frac{\qty{60}{\s}}{252}\approx\qty{0.2381}{\second}$.

When we mix substance A for each time step $\tau=\qty{0.2381}{\s}$ (one rotation) with substance B and compare the distribution of A and B at time $t=t_0+\qty{0.95238}{s}$, we again observe the mixing-enhancing effect of diffusion (see fig.~\ref{fig:STR_mixing}).
The boxes containing A and B after mixing correlate with those with high b-FTE values, particularly in the diffusion-free scenario (see fig.~\ref{fig:STR_FTE}).
\begin{figure}
    \centering
    \makebox[0.3\linewidth][l]{\text{a)}}
    \makebox[0.3\linewidth][l]{\text{b)}}
    \makebox[0.3\linewidth][l]{\text{c)}} \\
    \includegraphics[height=0.1\textheight, valign=t]{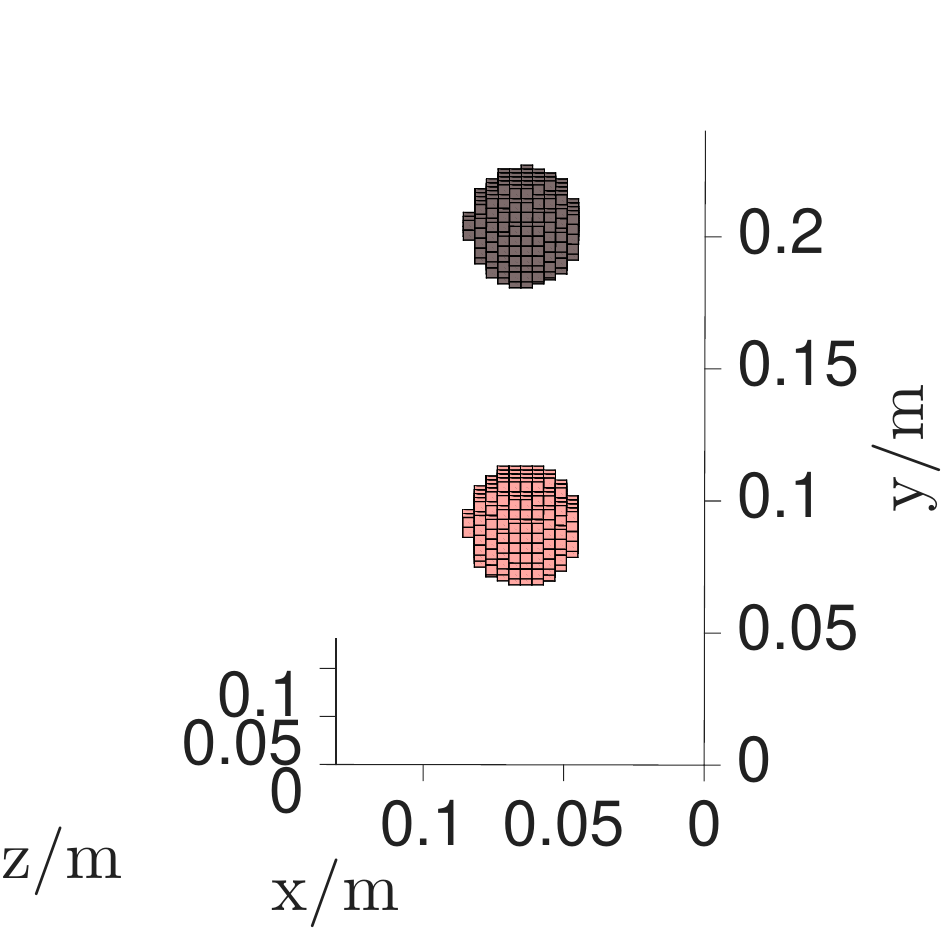}
    \includegraphics[height=0.1\textheight, valign=t]{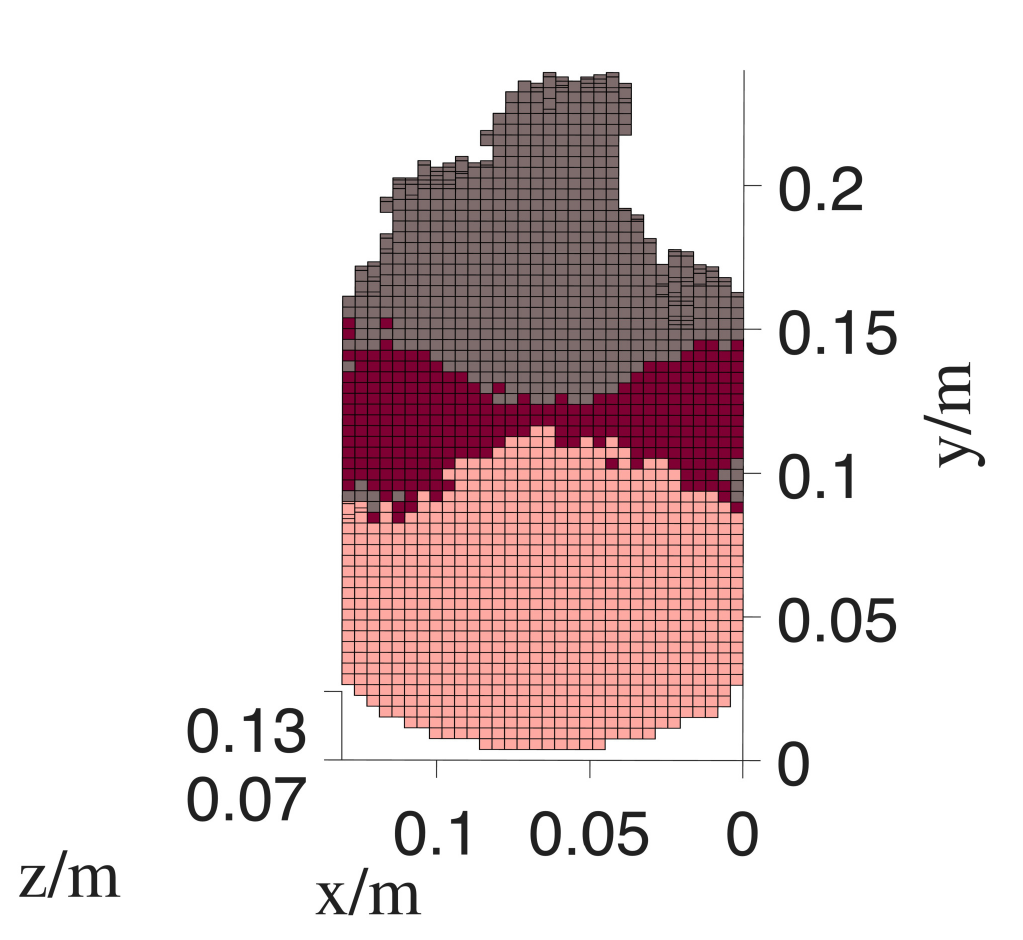}
    \includegraphics[height=0.1\textheight, valign=t]{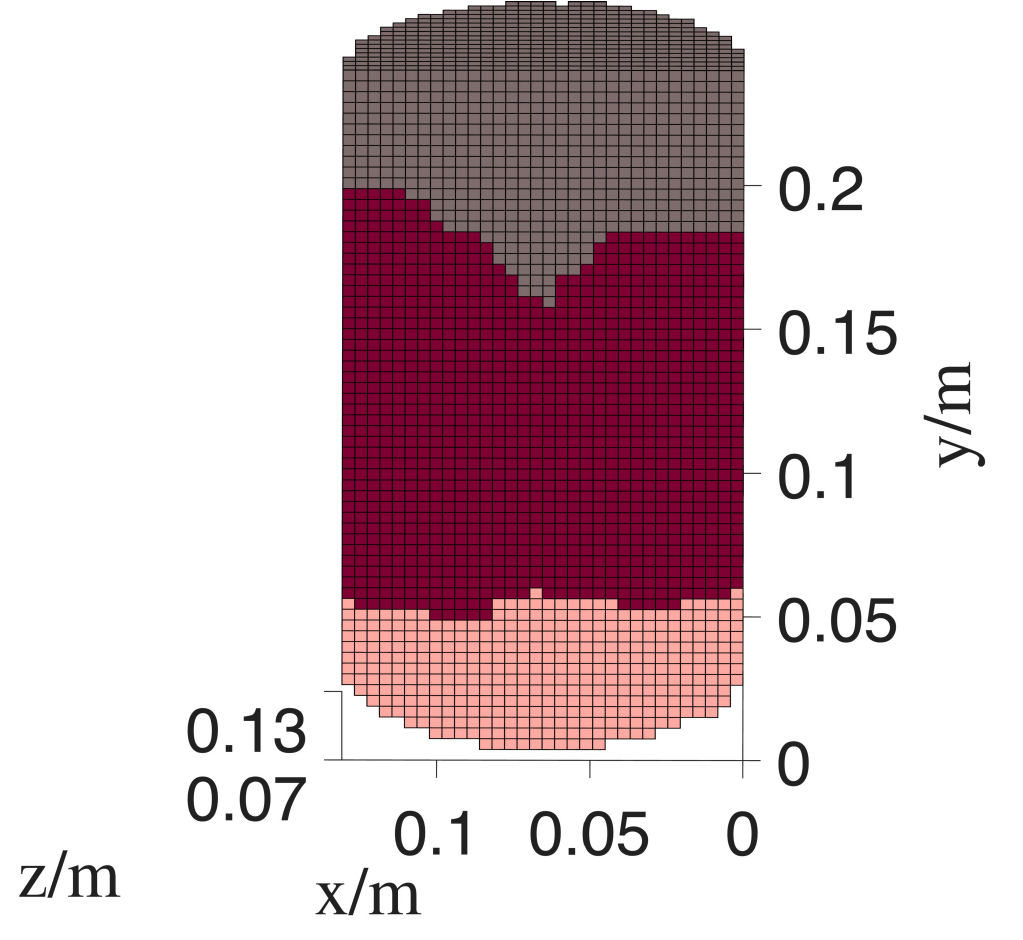}
    \includegraphics[height=0.09\textheight, valign=t]{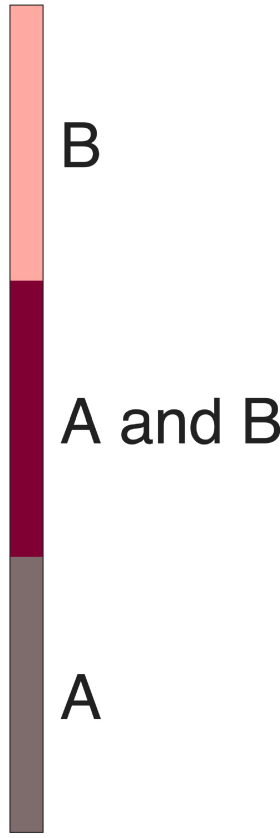} 
    \caption{Initial distribution (a) and longitudinal section of reactant distribution after mixing (no reaction) in the STR without diffusion (b) and with diffusion (c) at $t=t_0+\qty{0.95238}{\s}$ (four rotations).}
    \label{fig:STR_mixing}
\end{figure}

We simulate the simple chemical reaction in the STR model with the same initial distribution of fluids A and B (see fig.~\ref{fig:STR_mixing}).
For each time step $\tau = \qty{0.2381}{\s}$ (one rotation), the particles are transported and then react using the Euler scheme \eqref{eq:simplereact} with step size $h=\frac{1}{5}\tau$.

Due to the initial spatial separation of the reactants, no reaction can occur initially. In this case, the reaction is mixture-limited. For this reason, we use an average initial reactant concentration of $0.1068\unit{\mol\per\l}$ per box in the STR model to calculate the Damköhler number.
For reaction rate constant $k_r=0.1\unit{\l\per\mol\per\s}$, this implies the reaction time scale $\tau_{r}=\frac{1}{0.1\unit{\l\per\mol\per\s}\cdot 0.1068\unit{\mol\per\l}}\approx 93.6\unit{\s}$. For the flow time scale $\tau_{f}\approx 3.9\unit{\s}$, the Damköhler number is $Da \approx 0.04$, indicating again that advection is much faster than the chemical reaction. Compared to the DG and the ABC flow reactor models, transport dominates the reaction even more strongly in the STR model. This is the reason for the increase in product concentration that can still be observed even after $1,000\unit{\s}$ (see fig.~\ref{fig:STR_total}).

The first reactions and the formation of product C also occur in boxes with high b-FTE values (see fig.~\ref{fig:STR_FTE}).
Comparing the reaction in our STR model without and with diffusion shows that the absence of diffusion leads to a marginally faster reaction (see fig.~\ref{fig:STR_total}). However, at the beginning of the reaction process up to $t = t_0 + \qty{19.2857}{s}$ (81 rotations), higher desired concentrations are generated in the system with diffusion, as in the models considered previously.
\begin{figure}
    \centering
    \includegraphics[width=0.7\linewidth]{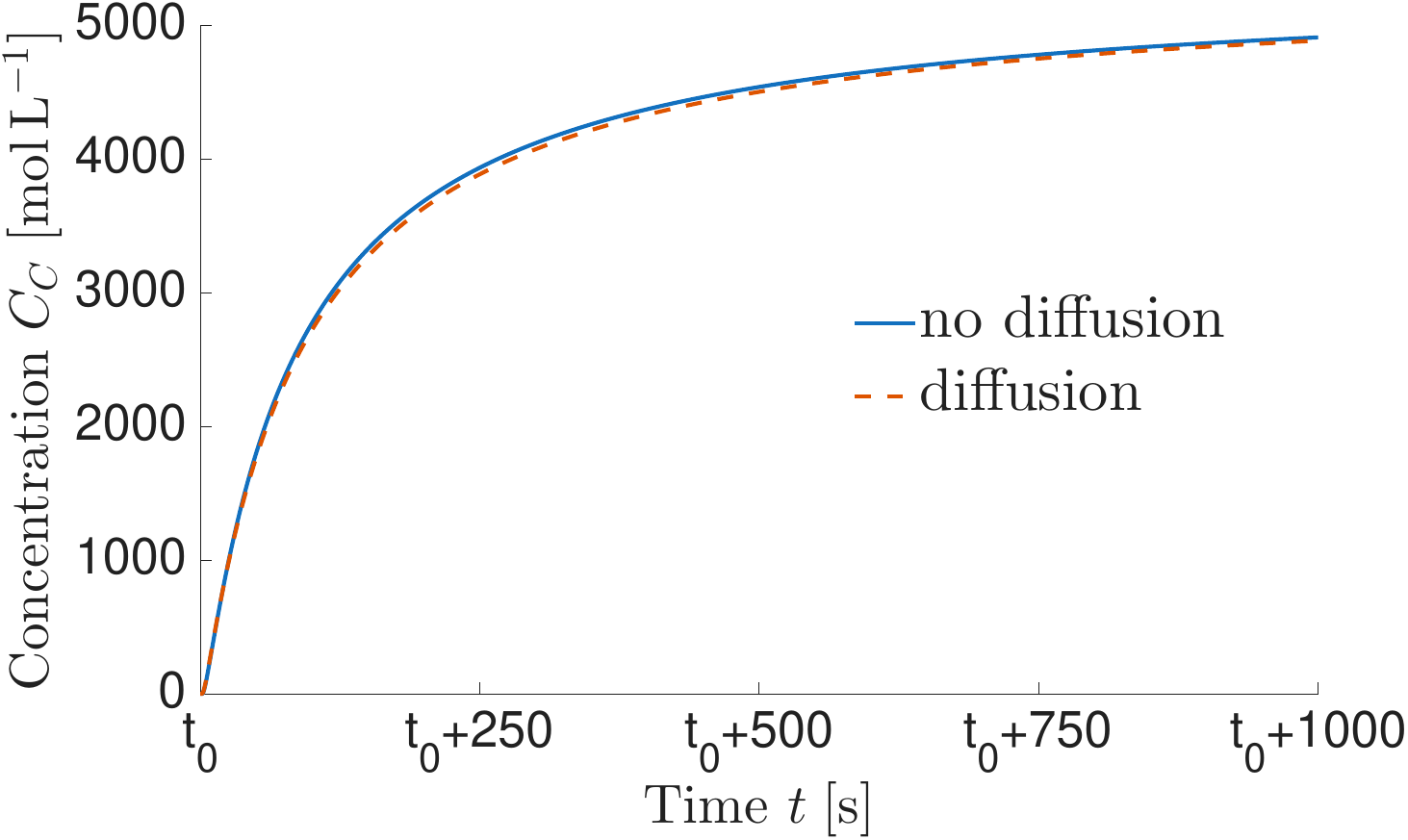}
    \caption{Total concentrations $C_C$ [\unit{\mol\per\l}] for the STR.}
    \label{fig:STR_total}
\end{figure}

Depending on diffusion, the b-FTE fields differ for the STR model, where higher values are observed with diffusion (see fig.~\ref{fig:STR_FTE}).
The b-FTE without diffusion at time $t=t_0+\qty{0.2381}{\s}$ (one rotation) can be compared with the b-FTLE at time $t=t_0+\qty{0.16}{\s}$ shown in figure 7 in reference \onlinecite{WEILAND2023100448}, which exhibits the same structures.

To demonstrate that there are matches between high reactions and high b-FTE values, we will consider the 2,000 highest reaction values and the 5,000 highest b-FTE values.
The red boxes indicate the correspondence between high reaction and FTE values.
Because the two systems have nearly identical matching boxes, diffusion in the STR model has little impact. This is confirmed by the nearly identical total concentrations $C_C$ in the STR (see fig.~\ref{fig:STR_total}). This is probably due to the unresolved small scales in the Lattice Boltzmann simulation that already have a similar effect on the dynamics as additional diffusion.
\begin{figure}
    \centering
    \begin{tabular}{cccc}
        \makebox[0.39\linewidth][l]{\text{a)}} & &
        \makebox[0.39\linewidth][l]{\text{b)}} \\
    \includegraphics[height=0.12\textheight]{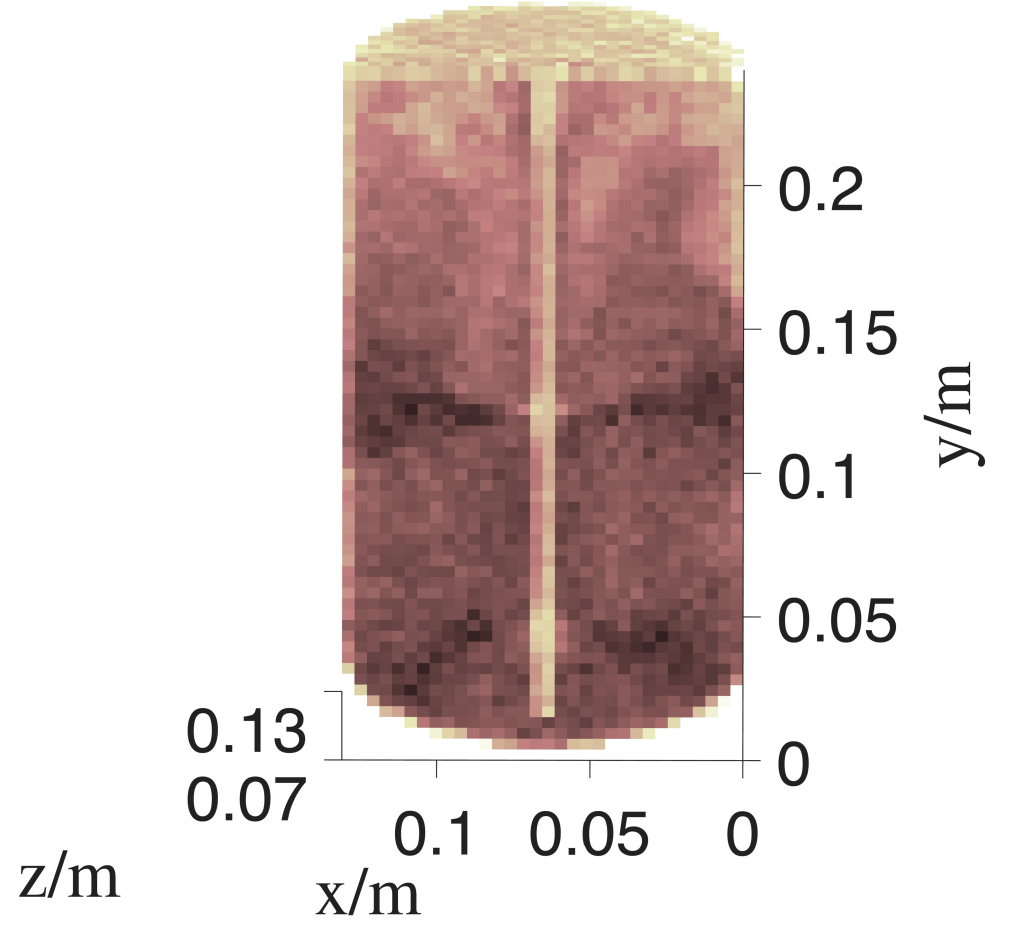} &
    \multicolumn{1}{l}{\includegraphics[height=0.12\textheight]{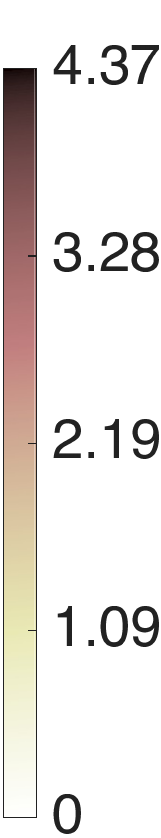}} &
    \includegraphics[height=0.12\textheight]{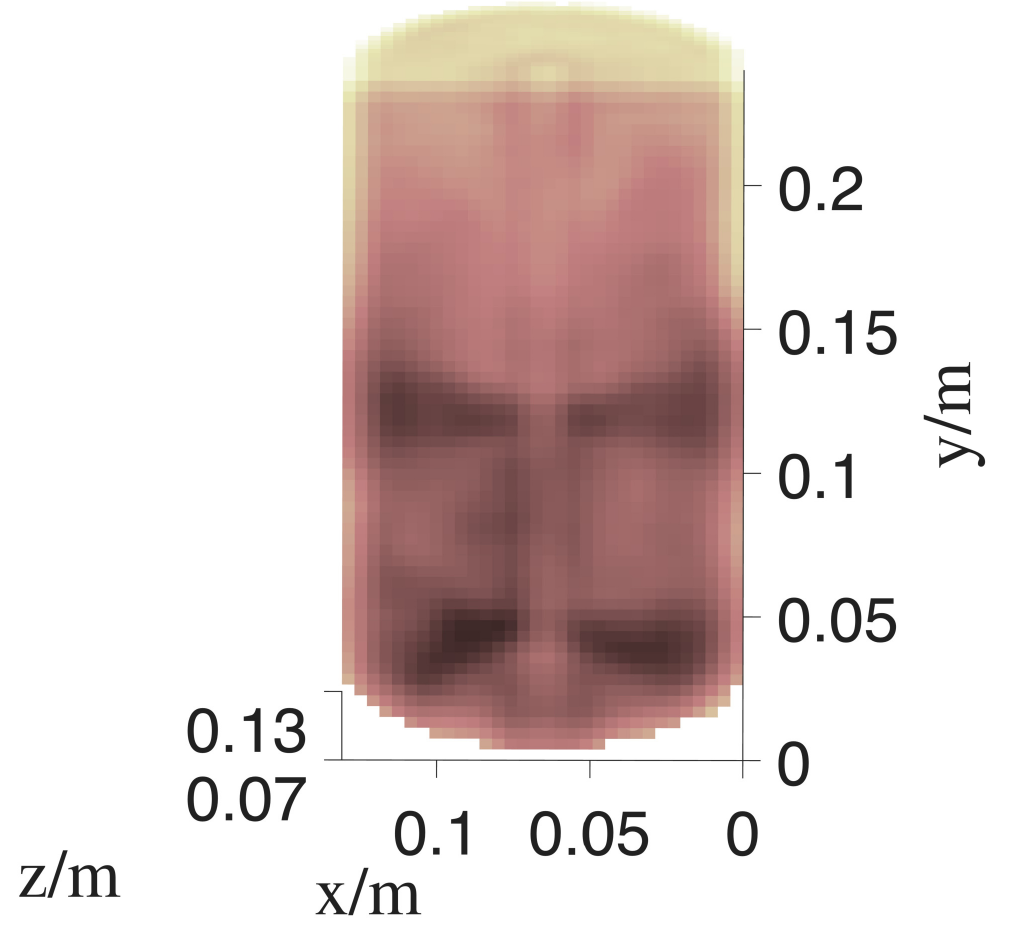} &
    \multicolumn{1}{l}{\includegraphics[height=0.12\textheight]{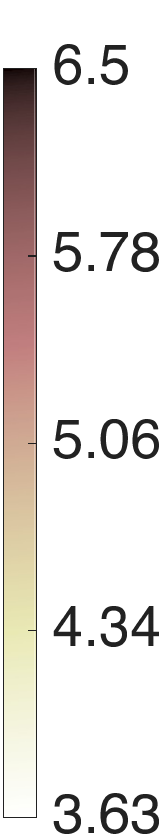}} \\
        \makebox[0.39\linewidth][l]{\text{c)}} & &
        \makebox[0.39\linewidth][l]{\text{d)}} \\
    \includegraphics[height=0.12\textheight]{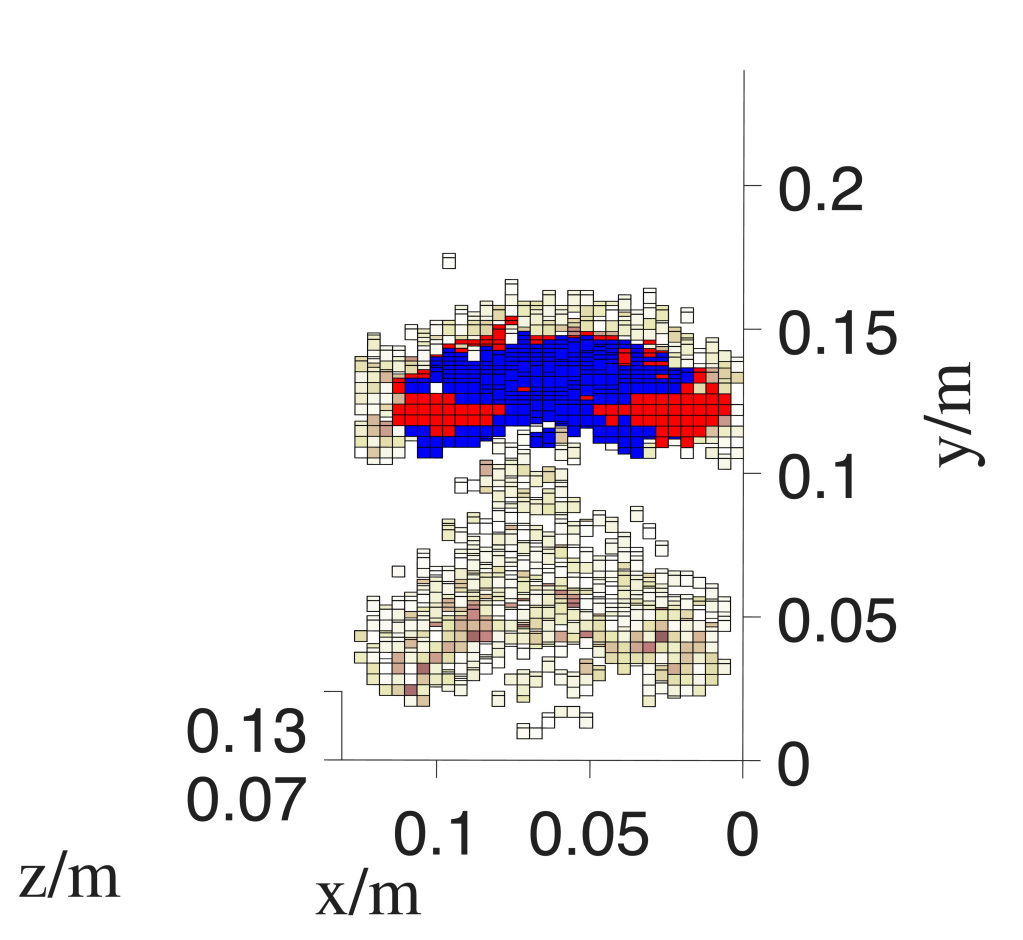} &
    \multicolumn{1}{l}{\includegraphics[height=0.12\textheight]{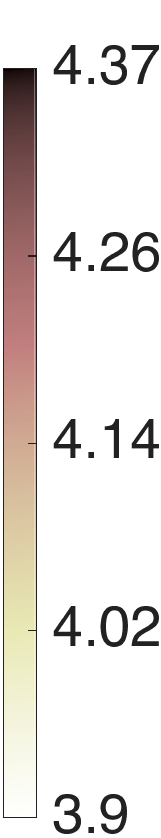}} &
    \includegraphics[height=0.12\textheight]{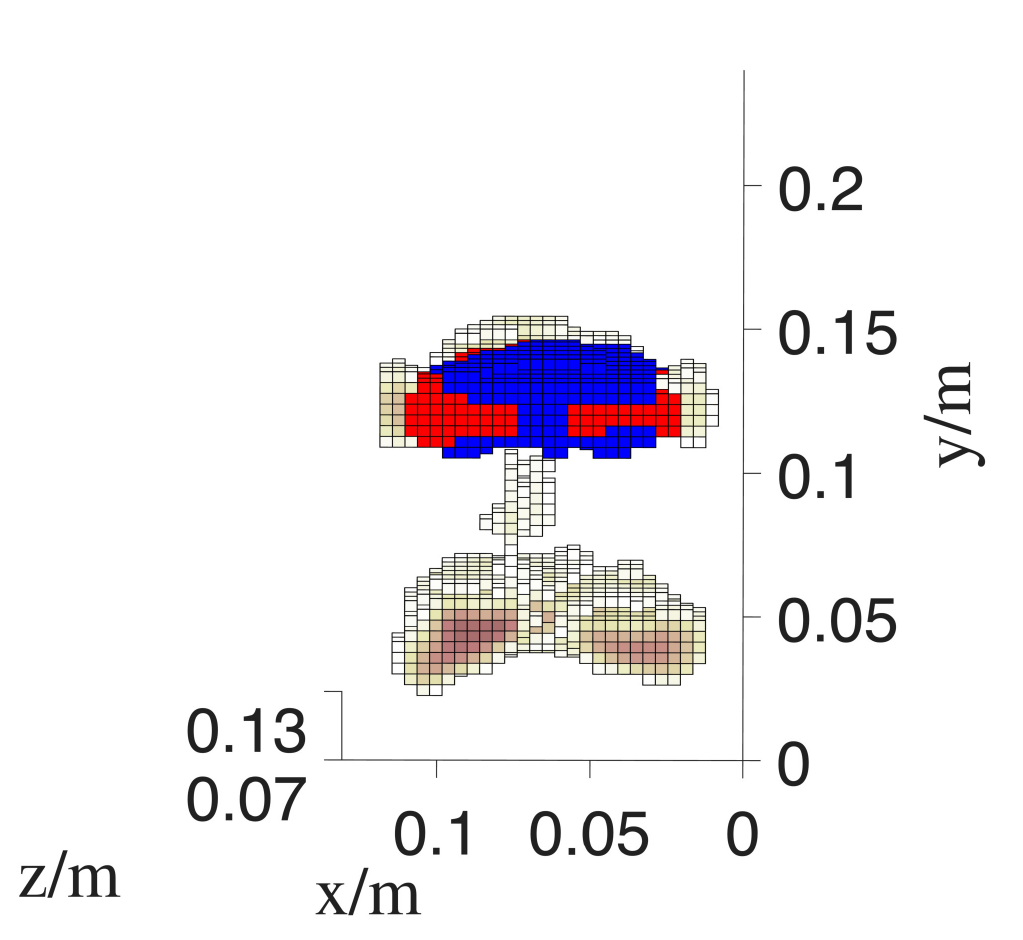} &
    \multicolumn{1}{l}{\includegraphics[height=0.12\textheight]{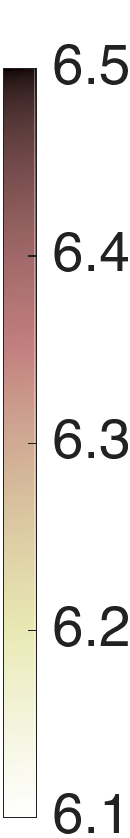}} \\
        \makebox[0.39\linewidth][l]{\text{e)}} & &
        \makebox[0.39\linewidth][l]{\text{f)}} \\
    \includegraphics[height=0.12\textheight]{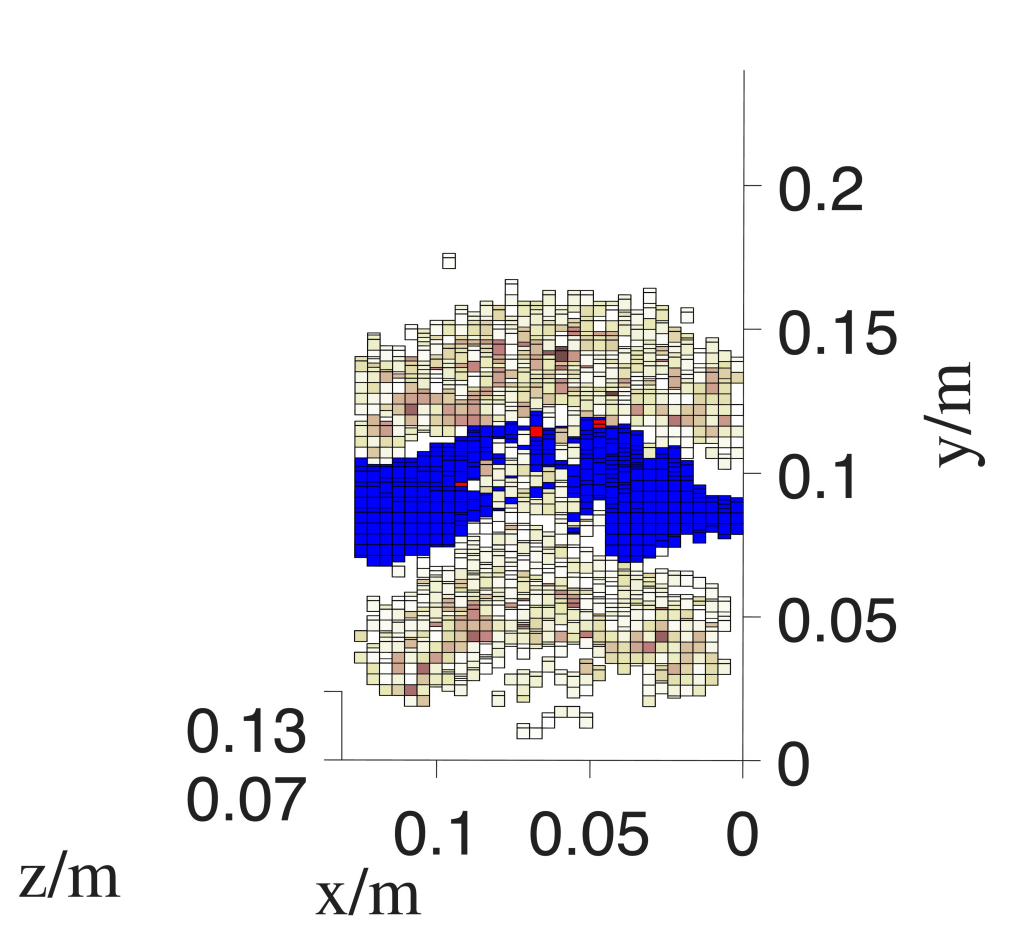} &
    \multicolumn{1}{l}{\includegraphics[height=0.12\textheight]{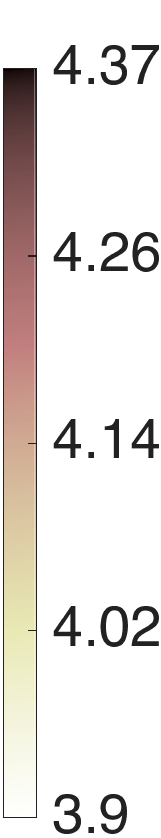}} &
    \includegraphics[height=0.12\textheight]{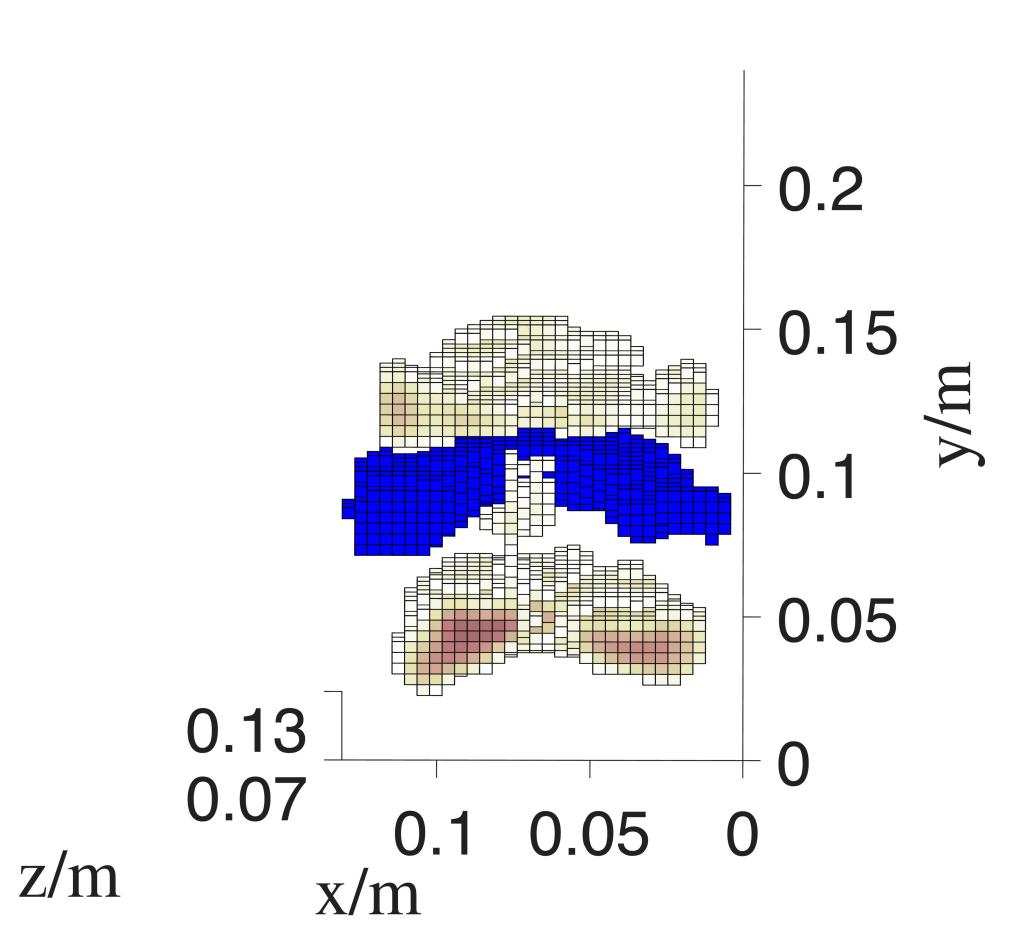} &
    \multicolumn{1}{l}{\includegraphics[height=0.12\textheight]{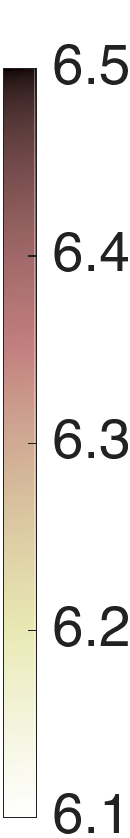}}
    \end{tabular}
    \caption{Longitudinal section of $FTE_b(\cdot,1)$ (one rotation) for the STR without diffusion (a,c,e) and with diffusion (b,d,f).
    The highest reaction front (blue and red) is shown at $t=t_0+\qty{0.95238}{\s}$ (four rotations)(c-d) and $t=t_0+\qty{2.3810}{\s}$ (ten rotations)(e-f), where red boxes represent the matching of both high reactions and b-FTE.}
    \label{fig:STR_FTE}
\end{figure}
The distribution of different concentrations and reaction levels suggests that higher concentrations are present in certain boxes $B_i$ due to entropy and stronger reactions occurring there (see fig.~\ref{fig:STR_FTE} and \ref{fig:STR_reaction_5}).
\begin{figure}
    \centering
    \makebox[0.39\linewidth][l]{\text{a)}}
    \makebox[0.39\linewidth][l]{\text{b)}} \\
    \includegraphics[height=0.12\textheight]{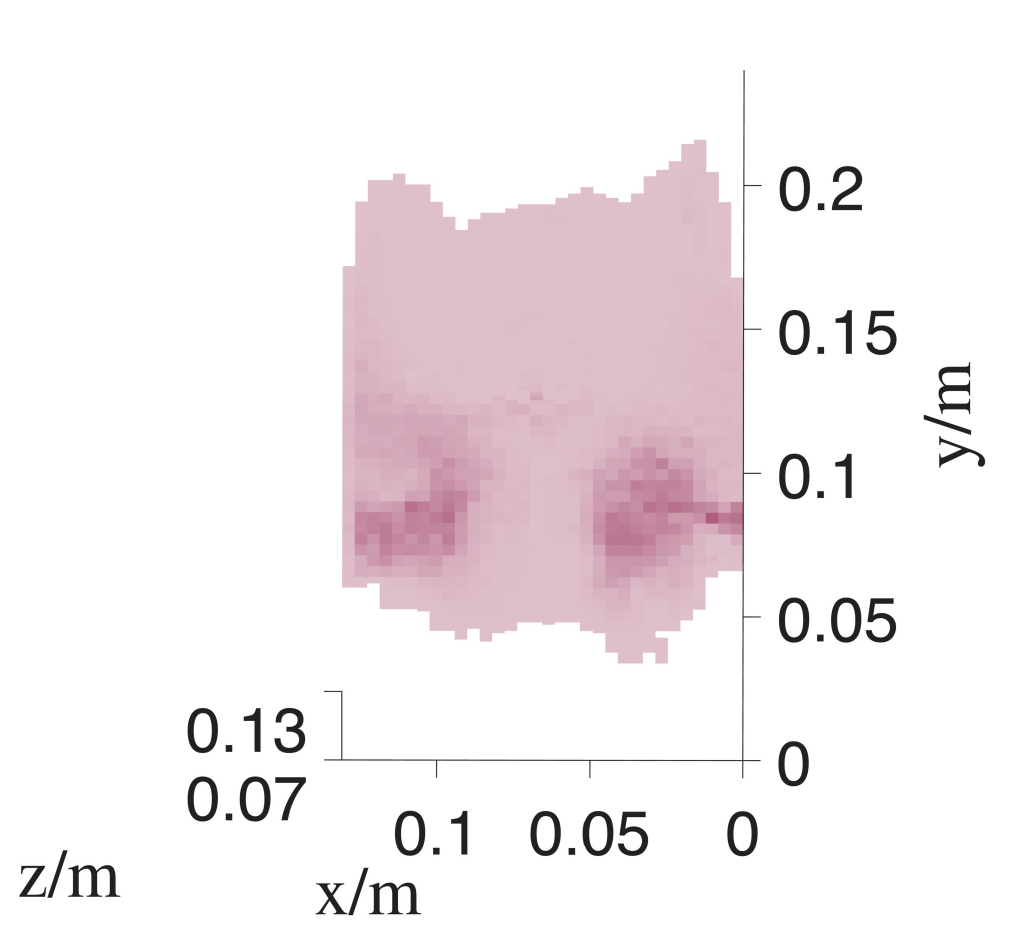}
    \includegraphics[height=0.12\textheight]{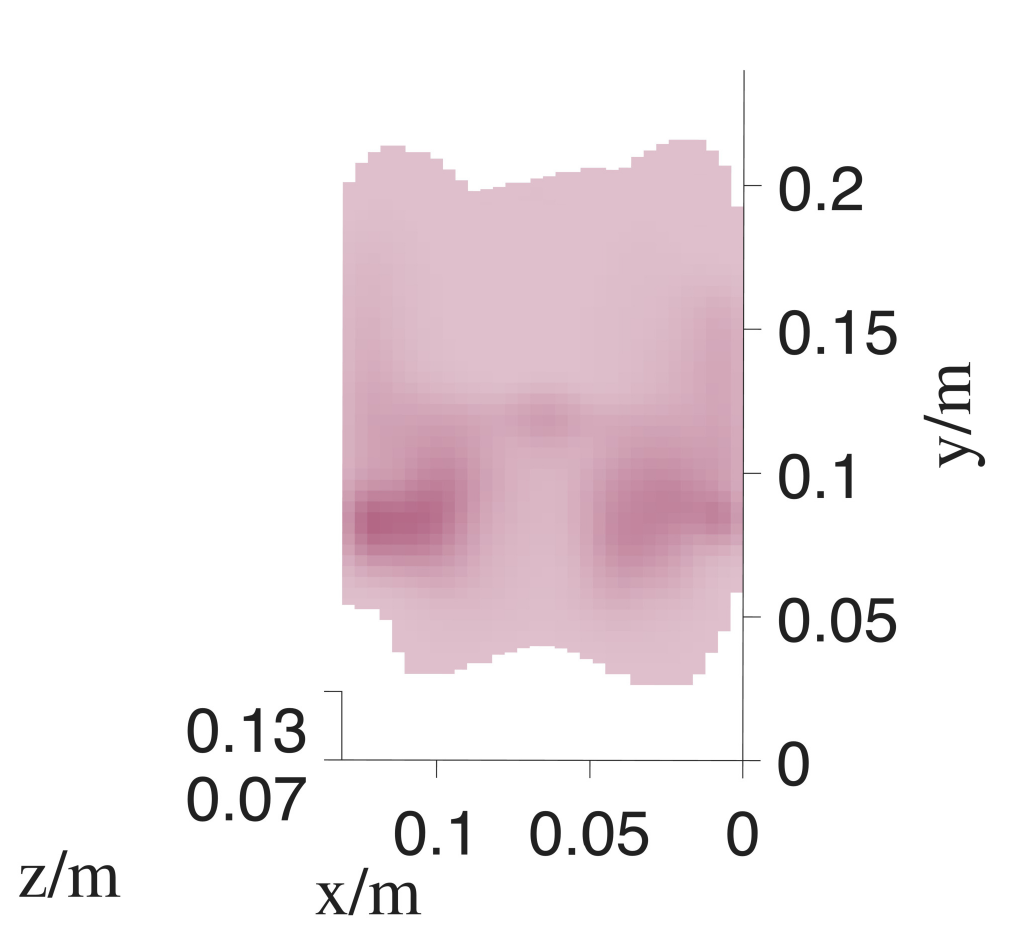}   \includegraphics[height=0.12\textheight]{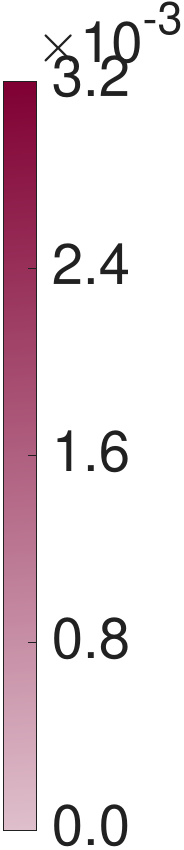}\\
    \makebox[0.39\linewidth][l]{\text{c)}}
    \makebox[0.39\linewidth][l]{\text{d)}} \\  \includegraphics[height=0.12\textheight]{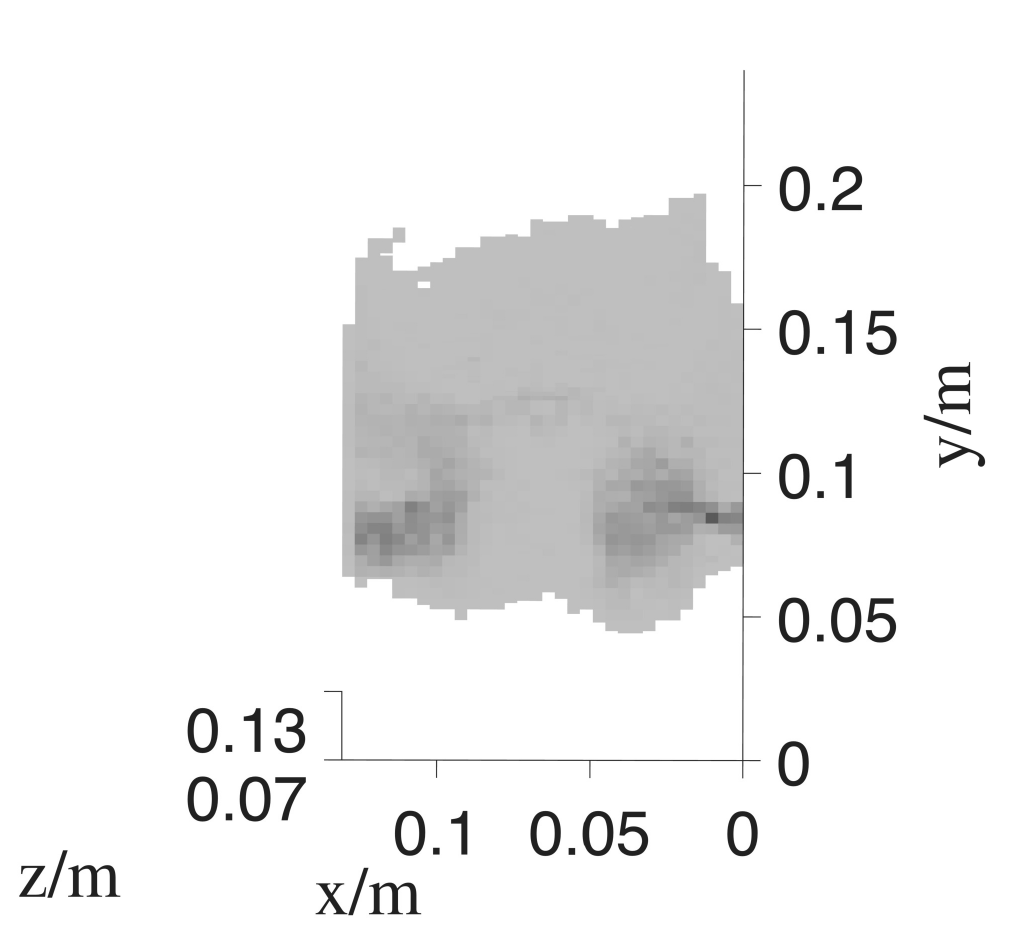}
\includegraphics[height=0.12\textheight]{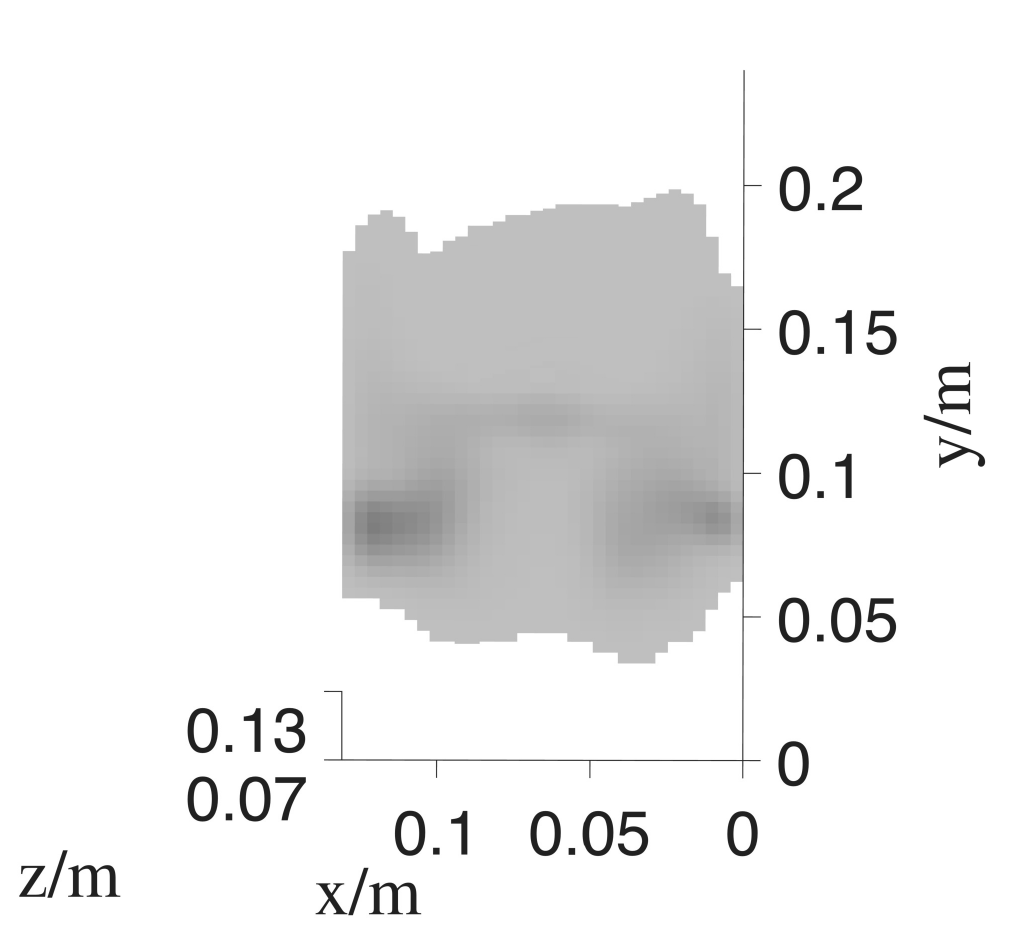}  \includegraphics[height=0.12\textheight]{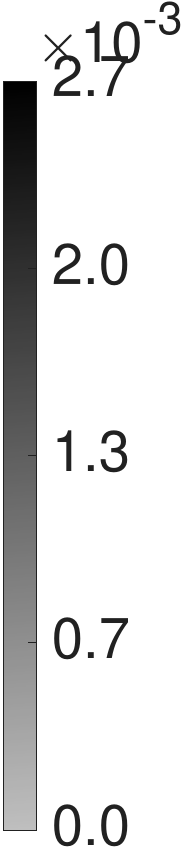}
    \caption{Longitudinal section of concentrations $C_C$ [\unit{\mol\per\l}] (red) and reaction front [\unit{\mol\per\l}] (gray) without diffusion (a,c) and with diffusion (b,d) for the STR at $t=t_0+\qty{2.3810}{\s}$ (ten rotations).}
    \label{fig:STR_reaction_5}
\end{figure}
At the beginning of a reaction process, the strongest reactions initially occur in regions with high b-FTE values (see fig.~\ref{fig:STR_FTE}). Subsequently, most of the strongest reactions occur between these regions (see fig.~\ref{fig:STR_FTE} and \ref{fig:STR_reaction_5}).
At $t=t_0+\qty{2.3810}{\s}$, both reactants react in the central boxes and the reaction front continues to spread through the middle of the reactor (see fig.~\ref{fig:STR_reaction_5}).
After A and B react in the center of the reactor to form C, C could act at a $y$-height of approx. \qty{0.1}{\m}, i.e. in the upper region with high b-FTE values, as a barrier, preventing further high reactions between A and B (see fig.~\ref{fig:STR_FTE} and \ref{fig:STR_reaction_5}).
Weiland et al.~\cite{WEILAND2023100448} identified five coherent regions, three in the upper, one in the lower area of the STR and one at a $y$-height of approximately $\qty{0.15}{\m}$. Due to the initial distribution, after mixing, A is located in one or more upper coherent regions and B is located in the lower coherent region. Consequently, the strongest reactions can only occur between these coherent regions. This strongest reaction area is precisely located between the highest b-FTE regions (see fig.~\ref{fig:STR_FTE} and \ref{fig:STR_reaction_5}).
In the long term, reactions occur throughout the reactor, but they are minor because the reaction process is almost complete.

\subsection{Competitive consecutive reaction in a CSTR model}

The open STR model represents a CSTR with input and output dynamics. It contains only a neutral liquid at time $t_0$. Reactant A is introduced into the input area at a concentration of $c_{A_0}=10\unit{\mol\per\l}$ per box and a flow rate of $v_0=\num{1e-4}\unit{\l\per\s}$, where each box has a volume of $V=d_x d_y d_z=\num{6.08e-5}\unit{\l}$ (see fig.~\ref{fig:STR_VDV}). All substances in the output area are released at the same rate. To ensure that the input and output area contain the same number of boxes (here 1,000), these boxes are determined using the Euclidean distance function with the desired number of boxes, so that the input boxes are close to $(max(x), max(y), max(z))$ and the output boxes are close to $(0, 0, 0)$ (see fig.~\ref{fig:STR_VDV}).

We simulate diffusion by approximating an $\epsilon$-ball for each test point with radius $\epsilon=\qty{0.01}{\m}$ by 125  grid points, so that $\epsilon \approx 5r$, with a radius $r \approx \qty{0.002}{\m}$ ($r_x=\qty{0.002}{\m}$, $r_y=\qty{0.0019}{\m}$ and $r_z=\qty{0.002}{\m}$) per box.
We calculate the transition matrix $\bm{P}$ (or $\bm{P}_{D}$) for one rotation, i.e. $\tau=\frac{\qty{60}{\s}}{252}\approx\qty{0.2381}{\second}$.
For each time step $\tau=\qty{0.2381}{\s}$ (one rotation), the particles are transported and then react based on the Euler scheme \eqref{eq:vandevusse} with time step $h=\frac{1}{5}\tau$.

Since the CSTR contains no chemical species at the start of the reaction process, with concentrations subsequently increasing, we use the concentration $c_{A_0}=10\unit{\mol\per\l}$ to calculate the Damköhler number $Da_3$.
For the reaction rate constants $k_1=\frac{1}{72}\unit{\per\s}$ and $k_3=\frac{1}{360}\unit{\l\per\mol\per\s}$, this implies the Damköhler numbers
$Da_1=k_1\cdot\tau_f=\frac{1}{72}\unit{\per\s}\cdot3.9\unit{\s}\approx0.05$ for the first-order main reaction and $Da_3=k_3\cdot c_{A_0}\cdot\tau_f=\frac{1}{360}\unit{\l\per\mol\per\s}\cdot10\unit{\mol\per\l}\cdot3.9\unit{s}\approx0.11$ for the second-order side reaction. Because $Da_1\ll1$ and $Da_3<1$, transport in the CSTR model is much faster than the chemical reactions. The concentration of A increases very rapidly in the reactor, and the fluid leaves the reactor before a significant portion of A can react to B or D (see fig.~\ref{fig:STR_VDV_output}). Furthermore, the output concentrations of B and D reflect $Da_1<Da_3$, since transport dominates the reaction to B more than the reaction to D.

If we compare the competitive consecutive reaction in our CSTR model  without and with diffusion, we get higher output concentrations $C_B$ with diffusion as expected (see fig.~\ref{fig:STR_VDV_output}).
\begin{figure}
    \centering
    \includegraphics[width=0.7\linewidth]{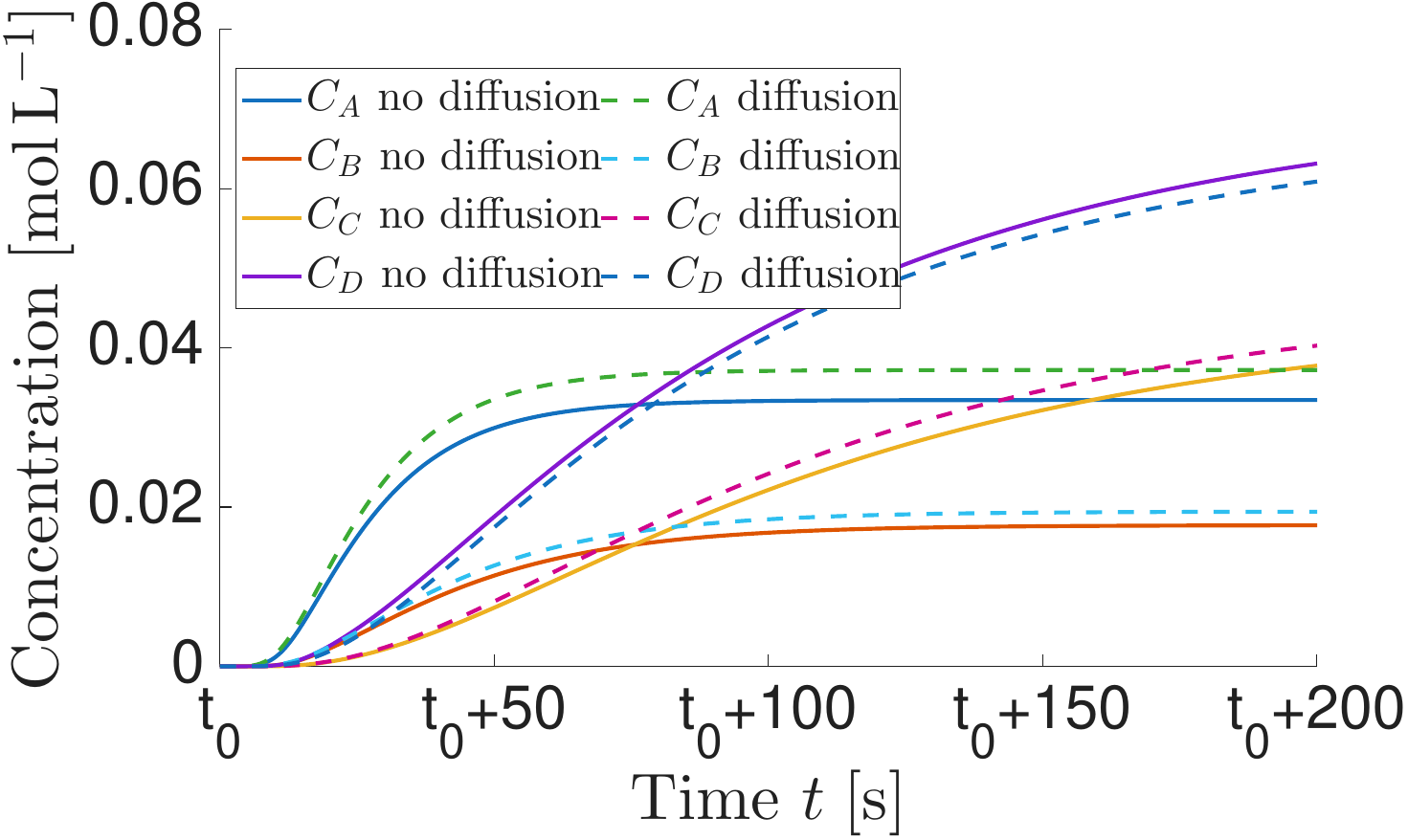}
    \caption{Output concentrations $C_B$ [\unit{\mol\per\l}] for the CSTR model.}
    \label{fig:STR_VDV_output}
\end{figure}
Because B is initially located in a coherent region in the upper half of the reactor, its distribution in the CSTR model is slow, and only very low concentrations reach the output area (see fig.~\ref{fig:STR_VDV}).
\begin{figure}
    \centering
    \makebox[0.3\linewidth][l]{\text{a)}}
    \makebox[0.3\linewidth][l]{\text{b)}}
    \makebox[0.3\linewidth][l]{\text{c)}} \\
\includegraphics[height=0.10\textheight]{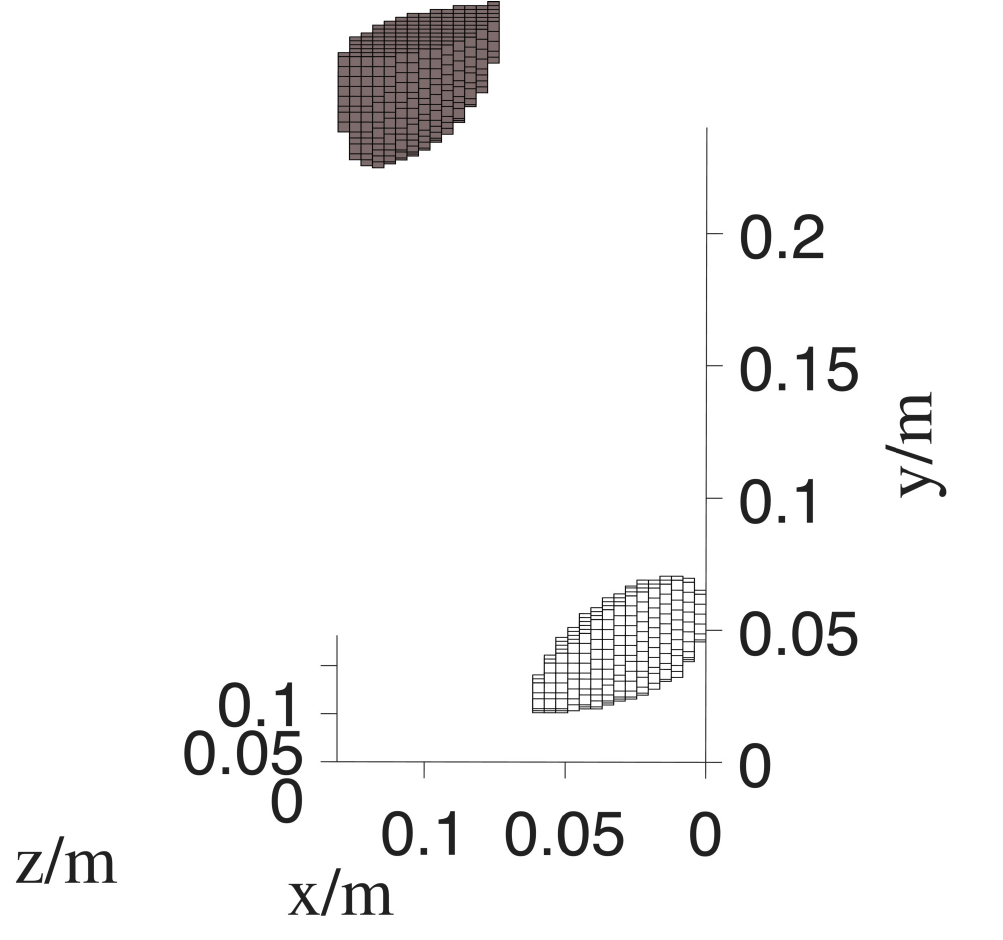}
\includegraphics[height=0.10\textheight]{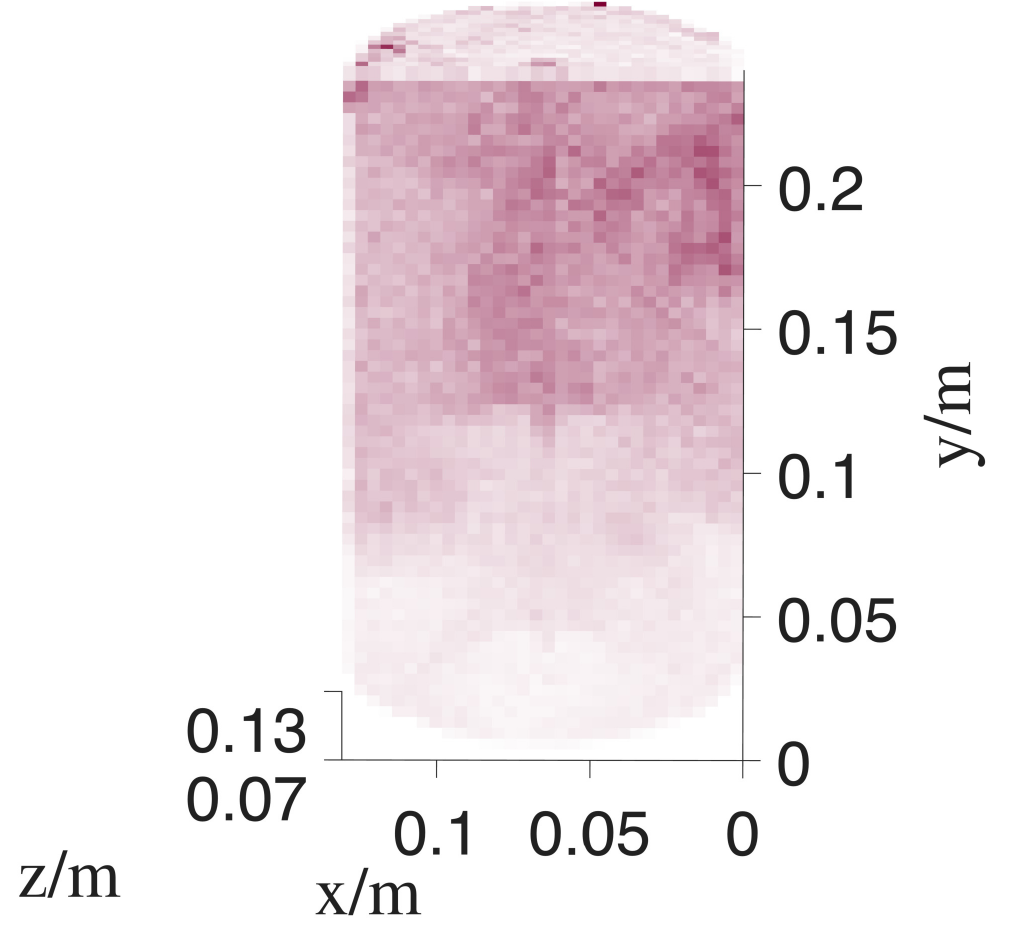}
\includegraphics[height=0.10\textheight]{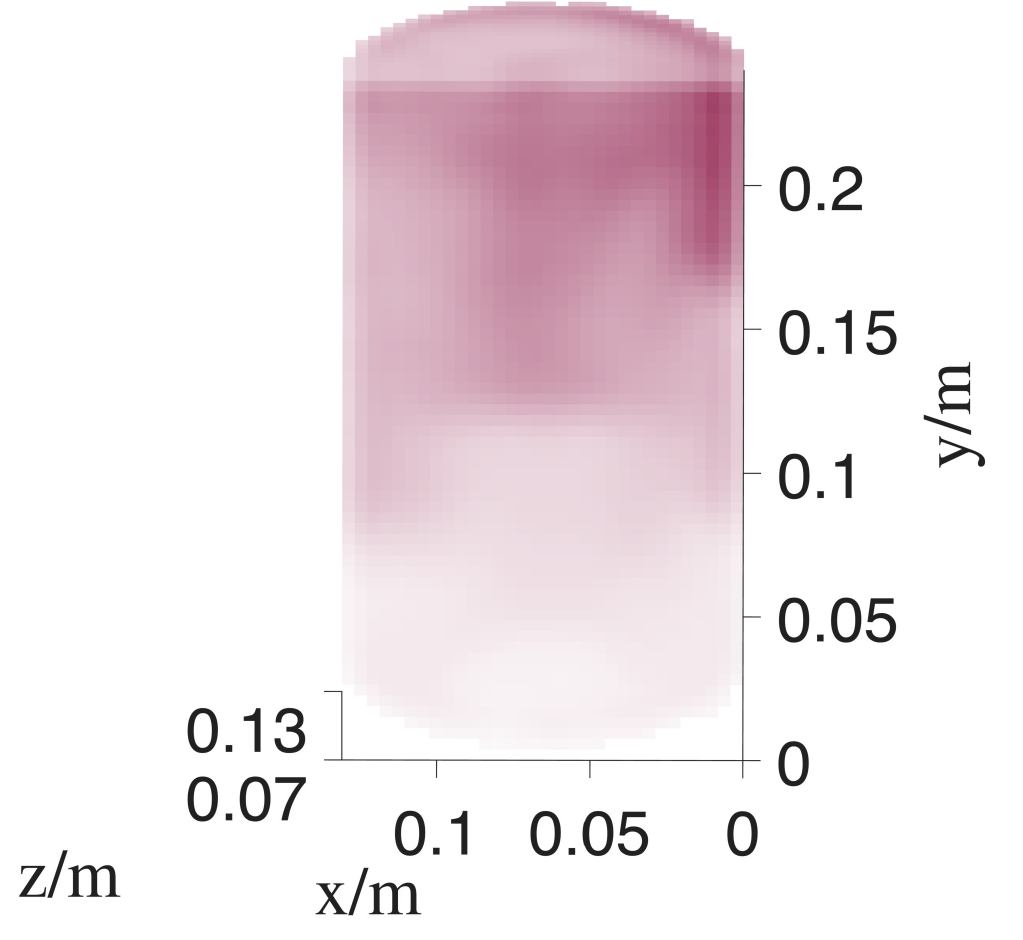}    \includegraphics[height=0.10\textheight]{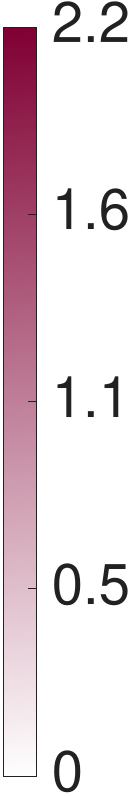}
    \caption{Input area (northwest) and output area (southeast) of the CSTR model (a) and longitudinal section of concentrations $C_B$ [\unit{\mol\per\l}] without diffusion (b) and with diffusion (c) at $t=t_0+\qty{20}{\s}$.}
    \label{fig:STR_VDV}
\end{figure}
Only very low concentrations of B reach a $y$-height of the reactor of less than \qty{0.1}{\m}, meaning that only a small amount of B passes beyond the upper region with high b-FTE values, which B passes through in the direction of the output area (see fig.~\ref{fig:STR_FTE}). This suggests that high b-FTE values may also pose a barrier in a CSTR.

\section{Conclusion and Outlook}
Our numerical studies using DG, ABC flow, and STR reactor models link mixing with chemical reactions via the transfer operator. These simulations provide insight into the reactor models to help us understand where high concentrations and reaction fronts are present. A comparison of the backward time FTE fields with reaction fronts has shown that, across all models, high product concentrations are present in boxes with high FTEs. However, depending on the initial distribution reactions may also occur in regions with low b-FTE values.

Different product yields are achieved depending on the reactor model because the mixing process influences the distribution and interaction of the reactants within the reactor.
For the simple chemical reaction, the Damköhler numbers confirmed that transport is faster than the reaction—across all reactor models considered. Transport dominates the reaction most strongly in the STR model. The results reflect the simulated total concentrations of the target product in the respective reactor models.
Therefore, the duration of a reaction process also depends on the underlying flow field of the respective reactor.
In the CSTR model, the low Damköhler numbers could explain the different concentrations of the target- and byproducts.
We also found that, in all models, the reaction process occurs faster with diffusion and leads to higher desired concentrations, except for the closed STR, where the reaction overall is slightly faster without diffusion.

The transfer operator framework can be used as a time- and cost-effective tool for investigating chemical reaction processes. This includes open systems and complex reactions, such as van de Vusse's competitive consecutive reaction in a CSTR, can be simulated and investigated using the transfer operator method. Selecting specific input and output areas allows one to view and trace the distribution or flow of a product within the reactor. We found that high b-FTE values could reduce the effectiveness of a CSTR.
Shifting inflow and outflow regions demonstrate the most effective method for maximizing the desired product.
The transfer operator method can be used to quickly and easily adjust the mixing parameters and initial distribution to ensure optimal settings before a real reaction process.
The method can be used to conduct in silico reaction experiments, testing for example different initial distributions and provides a simple model basis for optimization.

Additional effects, such as the presence of catalysts, could be taken into account in the proposed framework.
In future work, we aim to compare our simulations with experimental data. Further, we plan to use a network-based method to model chemical reactions. That framework is grid-free and has been applied to model advective-diffusive transport of scalar quantities \cite{kluenker2025}. Thus it will be  applicable to complex reactor geometries and cases where only sparse Lagrangian tracer trajectories are available.
\begin{acknowledgments}
This project has been supported by the Deutsche Forschungsgemeinschaft (DFG, German Research Foundation) -- SFB 1615 -- 503850735.
\end{acknowledgments}
\section*{Data Availability Statement}

The numerical data for the stirred tank reactor is available at \url{https://doi.org/10.15480/882.16844} \cite{weiland2026data}. The MATLAB code to reproduce the results shown in this paper  will be
archived on Zenodo upon publication of this manuscript.

%that support the findings of this study are openly available at \url{https://doi.org/...}, Ref.\.....

\appendix*
\section{}
To demonstrate the validity of our approach, we consider the numerical solution of the two-dimensional  advection-diffusion-reaction equation for the simple reaction $A + B \longrightarrow C$ in the double gyre flow. Let $c_A(\mathbf{x},t)$, $c_B(\mathbf{x},t)$, and $c_C(\mathbf{x},t)$
denote the concentrations of species $A$, $B$, and $C$, respectively.
Their transport and reaction is governed by the following system of partial differential equations
\begin{align*}
    \frac{\partial c_A}{\partial t}
        + \mathbf{u} \cdot \nabla c_A
        &= \kappa \, \nabla^2 c_A - k \, c_A \, c_B,\\
    \frac{\partial c_B}{\partial t}
        + \mathbf{u} \cdot \nabla c_B
        &= \kappa \, \nabla^2 c_B - k \, c_A \, c_B,\\
    \frac{\partial c_C}{\partial t}
        + \mathbf{u} \cdot \nabla c_C
        &= \kappa \, \nabla^2 c_C + k \, c_A \, c_B,
\end{align*}
where $\kappa > 0$ is the (common) diffusion coefficient and $k > 0$ is the
reaction rate constant (here $k=0.1$). The advecting velocity field $\mathbf{u}=(\dot{x},\dot{y})^\top$ is that of the double gyre flow (equation \eqref{eq:doublegyre}). We choose the same initial conditions on the domain $[0,2]\times [0,1]$ as described in section \ref{sec:mixreactDG}, i.e.
$$
c_A(x,y,0) =
    \begin{cases}
        1 & x \leq 1, \\
        0 & x > 1,
    \end{cases} , \qquad
    c_B(x,y,0) =
    \begin{cases}
        0 & x \leq 1, \\
        1 & x > 1,
    \end{cases}, 
    $$
    and  $c_C(x,y,0) = 0$.
Zero-flux (Neumann) boundary conditions are applied to all species on every wall. The system is numerically solved by a high-order finite-volume scheme, the code will be published alongside the manuscript. The resulting concentrations of $C$ at $t=4\unit{\s}$ are shown in fig.~\ref{fig:DG_reaction_ADR}.
\begin{figure}[!htb]
    \centering
    \makebox[0.3\linewidth][l]{\text{a)}}
    \makebox[0.3\linewidth][l]{\text{b)}}
    \makebox[0.3\linewidth][l]{\text{c)}} \\   \includegraphics[height=0.06\textheight]{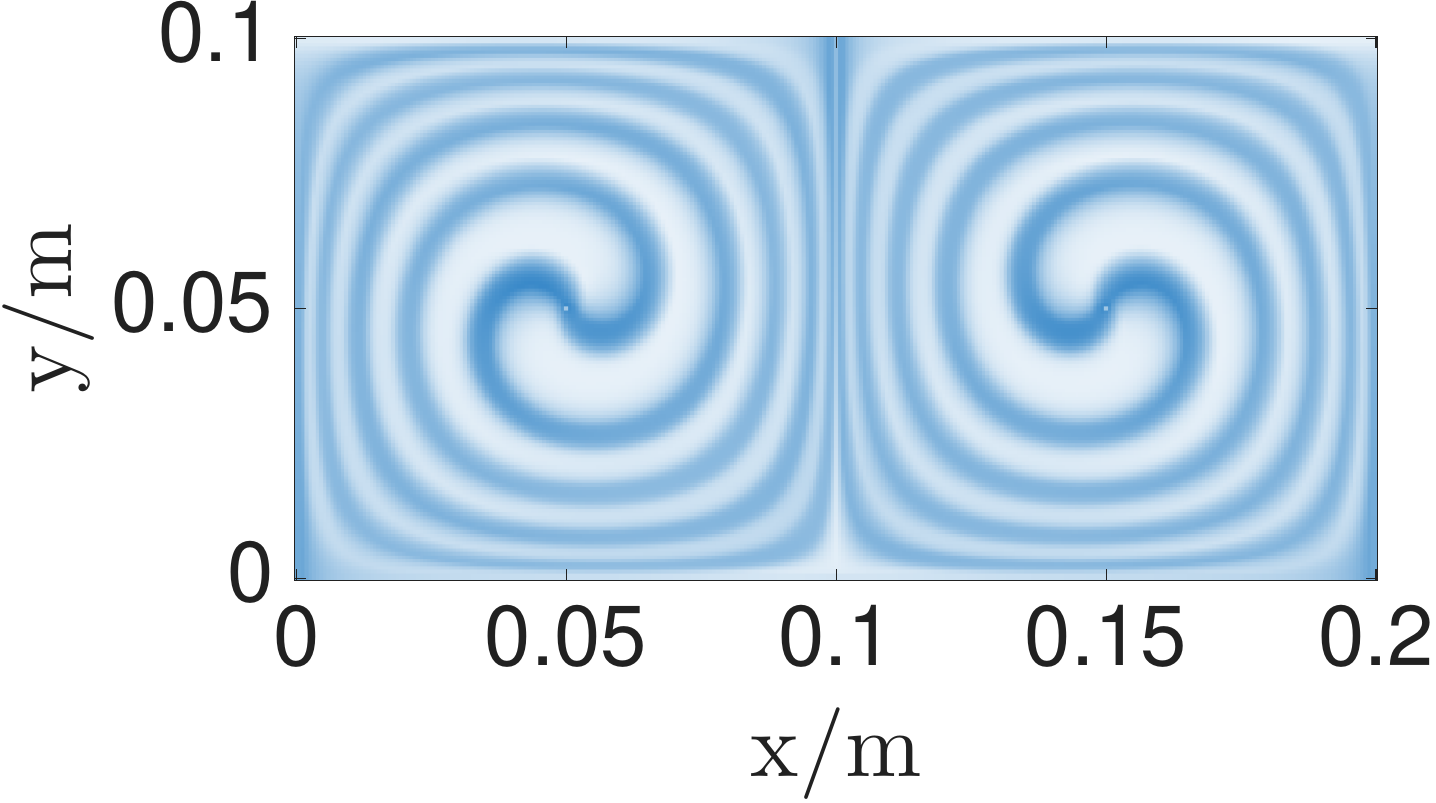} \includegraphics[height=0.06\textheight]{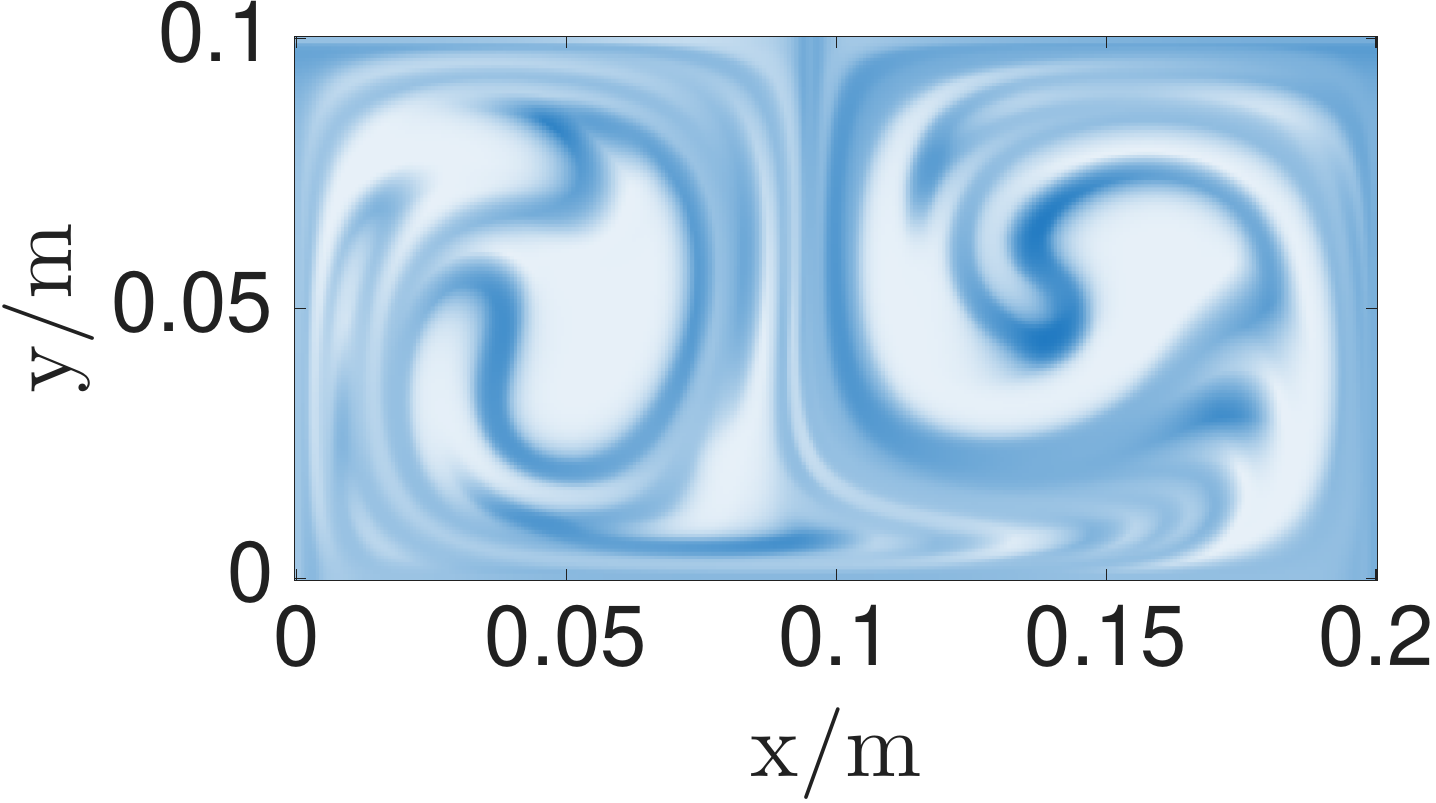}    \includegraphics[height=0.06\textheight]{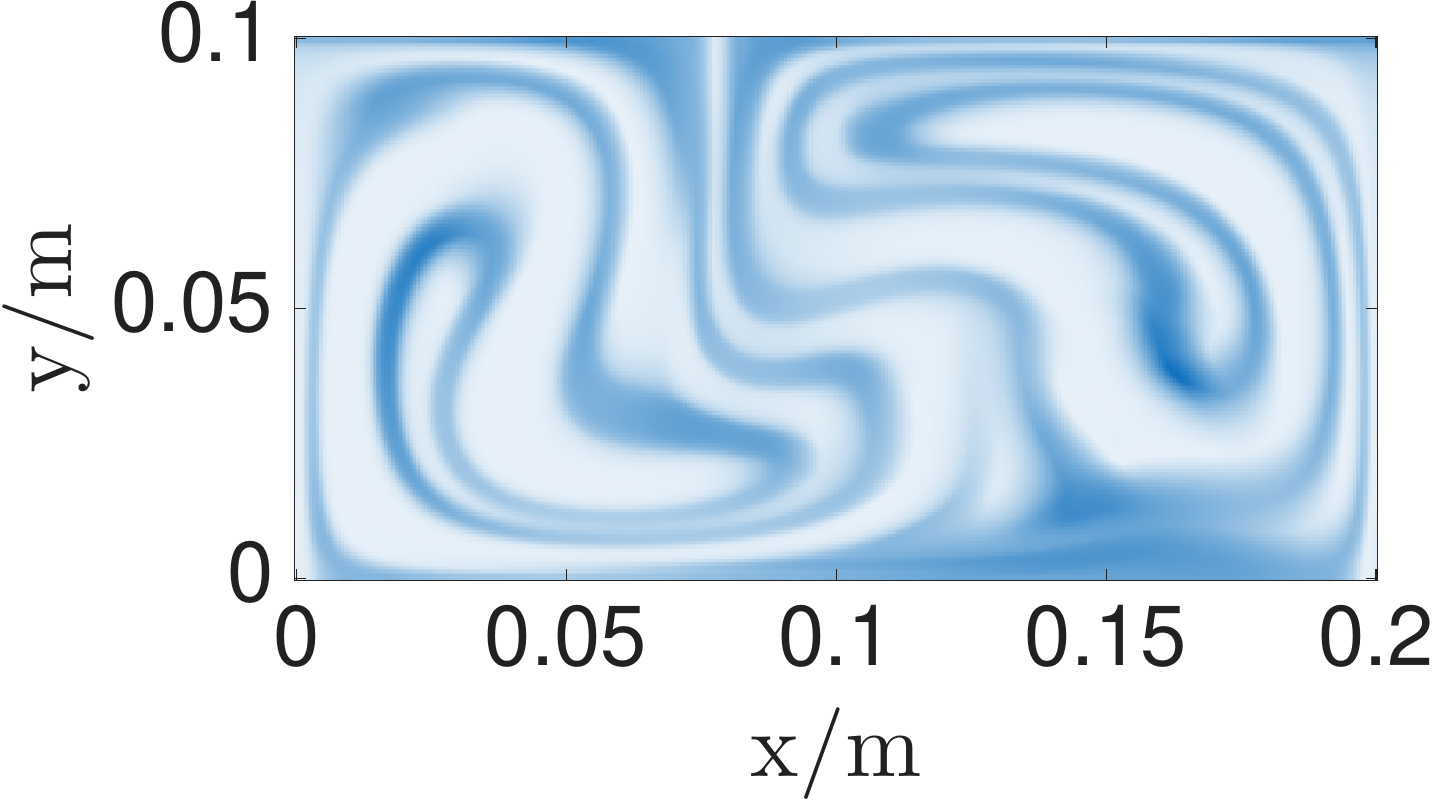}    \includegraphics[height=0.06\textheight]{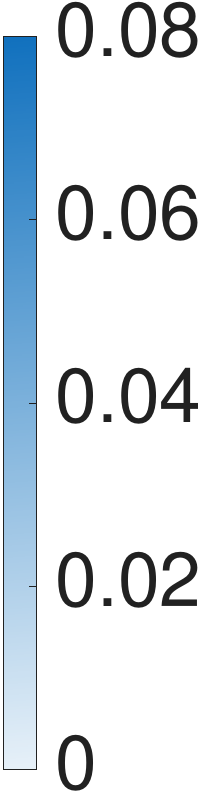} \\
    \makebox[0.3\linewidth][l]{\text{d)}}
    \makebox[0.3\linewidth][l]{\text{e)}}
    \makebox[0.3\linewidth][l]{\text{f)}} \\
    \includegraphics[height=0.06\textheight]{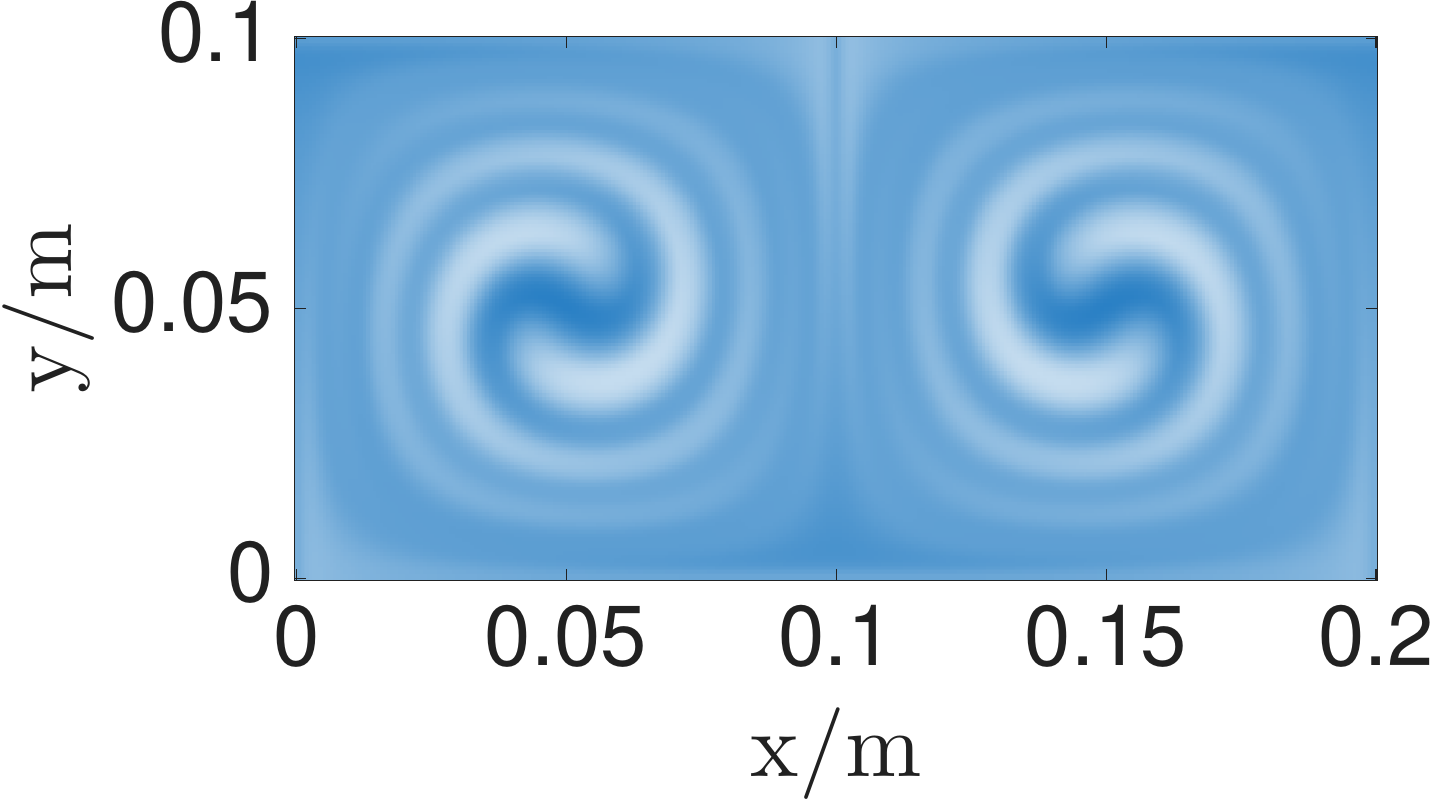}
    \includegraphics[height=0.06\textheight]{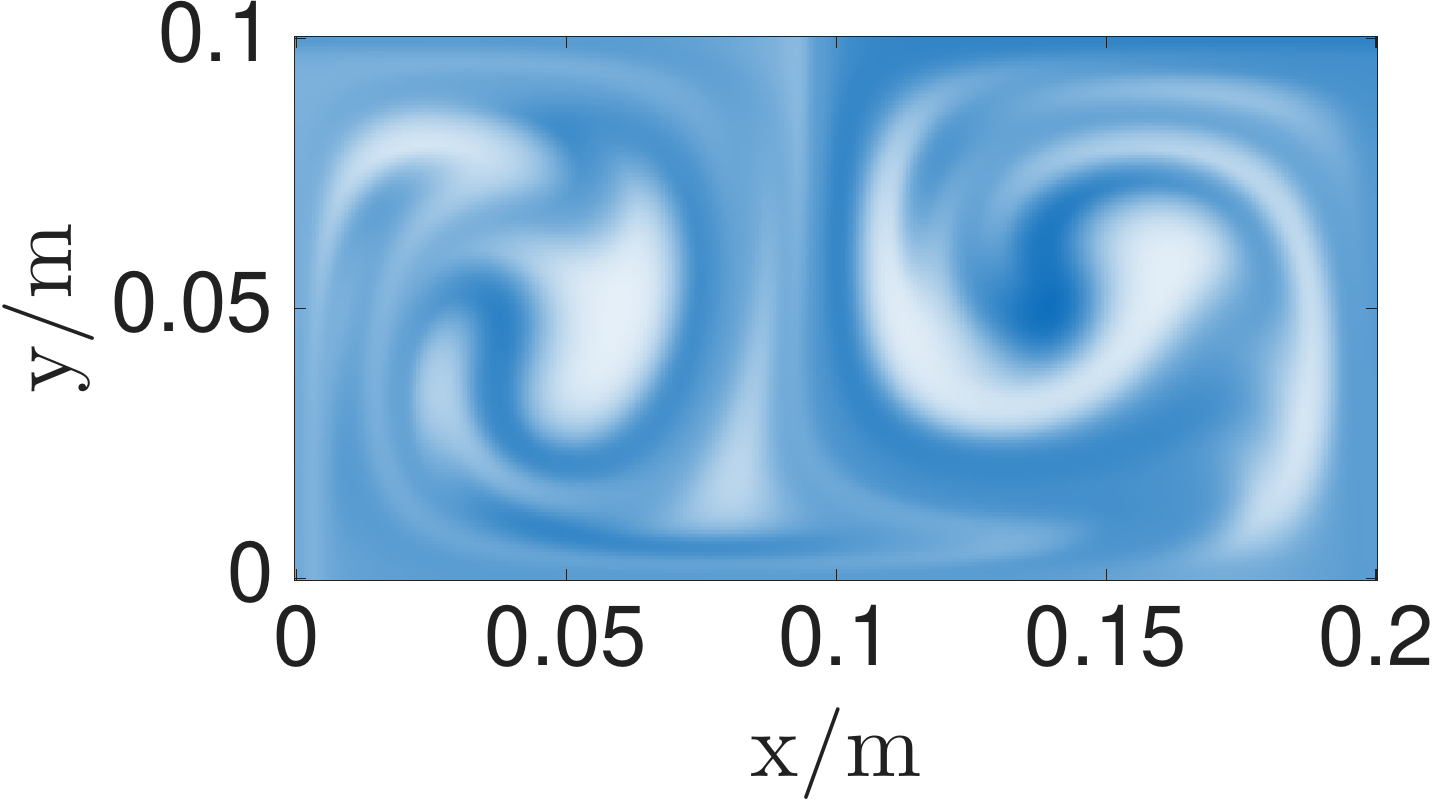}
    \includegraphics[height=0.06\textheight]{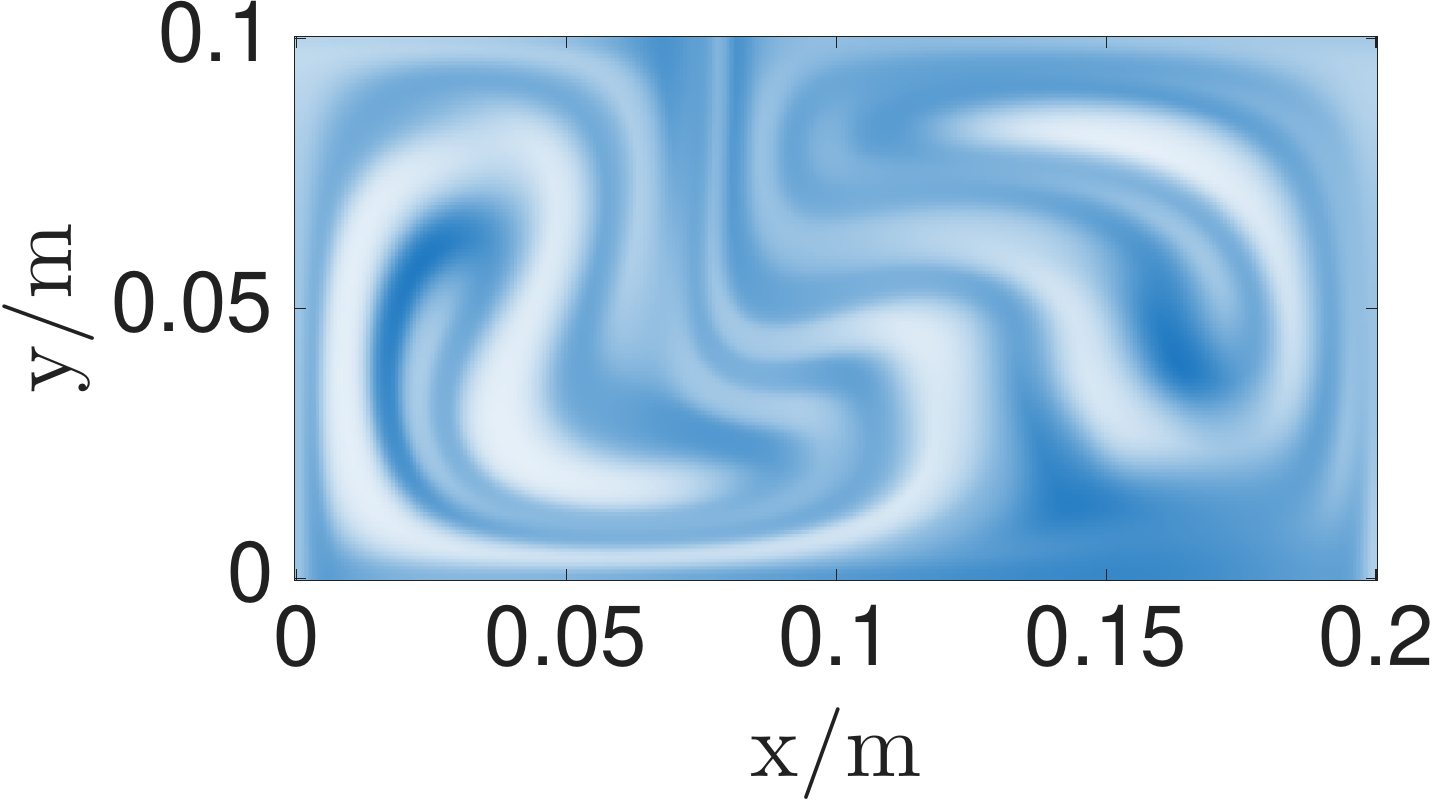}
    \includegraphics[height=0.06\textheight]{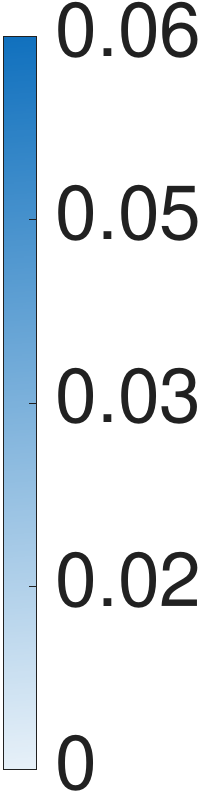}
    \caption{Concentrations $\bm{C}_C$ [\unit{\mol\per\l}] at $t=4\unit{\s}$ for the DG advection-diffusion-reaction system with diffusion coefficient $\kappa=\num{5e-4}$ (a-c) and $\kappa=\num{1e-6}$ (d-f) for $\delta=0$ (a,d), $\delta=0.1$ (b,e) and $\delta=0.5$ (c,f).}
    \label{fig:DG_reaction_ADR}
\end{figure}

%\nocite{*}
%\section*{References}
\bibliography{references}%
\end{document}